\documentclass[11pt,letter, margin=1in]{article}
\usepackage[utf8]{inputenc}
\usepackage{amssymb}
\usepackage{amsmath}
\usepackage{amsthm}
\usepackage{amsfonts}
\usepackage[langfont=typewriter,funcfont=slant]{complexity}
\usepackage[left=1in,bottom=1in,top=1in,right=1in]{geometry}
\usepackage[colorlinks=true, linkcolor=blue, citecolor=blue]{hyperref}
\usepackage{framed}
\usepackage{algorithm}
\usepackage[noend]{algpseudocode}
\usepackage{soul}
\usepackage{thmtools,thm-restate}
\usepackage{csquotes}
\usepackage{booktabs}       
\usepackage{nicefrac}       
\usepackage{microtype} 
\usepackage[shortlabels]{enumitem}
\usepackage{bbm}
 \usepackage{setspace}
 \allowdisplaybreaks
\usepackage[round]{natbib}
\usepackage{multirow,array}
\usepackage[disable]{todonotes} 
\presetkeys{todonotes}{inline}{}

\makeatletter
\if@todonotes@disabled

\else

\fi
\makeatother

\usepackage{commath} 

\usepackage{bbm}
\usepackage{enumitem}

\newtheorem{corollary}{Corollary}
\newtheorem{proposition}{Proposition}
\newtheorem{lemma}{Lemma}
\allowdisplaybreaks

\title{The Ordinary Least Eigenvalues}
\author{Yassine Sbai Sassi \footnote{Department of Economics, University of California - Berkeley, e-mail: yassine@berkeley.edu. I am thankful to Bryan Graham, Michael Jansson, Demian Pouzo and Kévin Dano for helpful conversations. All errors are mine.}}
\date{March 2023}

\begin{document}

\maketitle

\begin{abstract}
    
    We propose a rate optimal estimator for the linear regression model on network data with interacted (unobservable) individual effects. The estimator achieves a faster rate of convergence $N$  compared to the standard 
 estimator' rate $\sqrt{N}$ rate 
    and is efficient in cases that we discuss. We  observe that the individual effects  alter the eigenvalue distribution of the data's matrix representation in significant and distinctive ways.  We subsequently offer a correction for the \textit{ordinary least squares}' objective function to attenuate the statistical noise that arises due to the individual effects, and in some cases, completely eliminate it. The new estimator is asymptotically normal and we provide a valid estimator for its asymptotic covariance matrix. While this paper only considers models accounting for first-order interactions between individual effects, our estimation procedure is naturally extendable to higher-order interactions and more general specifications of the error terms. 
\end{abstract}
\vspace{1cm}
\section*{Introduction}

Linear regression models with individual-specific effects are widely used to fit data with network structures. Such linear models were used to explain trade flows between countries (\cite{AndersonWincoop2003}, \cite{Fally2015}), to fit matched employer-employee  data (\cite{AKM}, \cite{Bonhommeetal2019}), or to study teacher effect on student performance (\cite{Jacksonetal2014}), to mention  a few examples.  In applications, these linear regression models are most often used with a particular specification of the individual-specific effect. A popular specification consists of including the individual effects as additive error terms. This is the approach taken for instance  in \cite{AKM}  and \cite{Jacksonetal2014}. Broadly, three types of estimators are used under this specification. The two way fixed effects estimator (\cite{AKM}) exploits the additivity of the model to eliminate the individual effects. After double differencing, the initial model is turned into a regular linear regression model (free of the individual effects), and estimators are obtained by least squares on the transformed model. When the data is non bipartite with a number $N$ of agents (respectively, when it is bipartite, with $N$ and $M$ agents on each side), the two way fixed estimator of the slope parameters converges  at the optimal rate of $N$ (resp. $\sqrt{NM}$). The two way fixed estimator comes with a significant caveat: the slope parameters on any agent-specific observable covariates disappear in the double differencing process, in the same way as the individual effects. Those can be recovered in a second stage by ordinary least squares if we further assume the individual effects to be exogenous with respect to the additive observable. The second stage OLS estimators for the slope parameters on the additive covariates converges at a $\sqrt{N}$ rate.

A second approach appeals to the standard OLS estimator (e.g. \cite{Rose2004}, \cite{FafchampsGubet2007}). In the dydadic linear regression setting, the OLS estimator is in general $\sqrt{N}$ consistent for all the parameters. Given that for some covariates the two way fixed effects estimator can provide $N$-consistent estimators, the OLS estimator is severely inefficient. Other approaches consist of estimating a fixed effects model, by regressing the output variable on the covariates, individual indicators and interaction of individual interactions. As such, this procedure suffers from the incidental parameter problem, since the number of parameters to be estimated grows faster than the sample size. The incidental parameter problem is usually resolved through a dimension reduction, for instance by ``grouping" individuals into a small number of categories that share the same individual effect, or simimilarly by assuming the individual effects are drawn from some discrete distribution (\cite{BonhommeManresa2015}, \cite{Bonhommeetal2019}).

This paper proposes an $N$-consistent estimator for the slope parameters under an exogeneity assumption on the individual (unobserved) effects. We  exploit the matrix structure of network data and  identify the individual effects' footprint on the spectrum of the output matrix. We then correct for the unobservables'effect on the spectrum. The correction proposed in this paper consists of tweaking the standard minimization problem that yields the OLS estimator in a way that mechanically attenuates the effect of the unobservables on the explained variable's spectrum.

This paper contributes to the literature on the dyadic linear regression model in a few ways. First, we provide a rate optimal estimator for \textit{all} slope parameters. In the process, we provide valid estimators for the individual effects and overcome the incidental parameter issue without any appeal to discretization. This is an improvement over existing estimation procedures that are either rate inferior (e.g. OLS), or that only provide rate optimal estimators for a subset of slope parameters (e.g. the two-way fixed effects). This comes at the cost of the exogeneity assumption the paper imposes on the  individual effects.  Second, to the best of my knowledge, this is the first time tools from the random matrix literature are used for inference. These tools allow for the analysis of a new set of estimators that are based on a matrix representation of the regression model. These new estimators are severely non linear and are defined as extrema of globally ill-behaved objective functions. We show that in a simple sub-specification, our estimator is semi-parametrically efficient, unlike OLS or any other linear estimator. This suggests that achieving semi parametric efficiency in  models with dyadic dependence might require the recourse to non standard estimators.

The next section introduces the setup and lays out the main intuitions leading up to  the definition of the new estimator. Section \ref{AsymSection} discusses the estimator's theoretical properties and numerical implementation. Section \ref{BiasCorrection} proposes estimators for the asymptotic bias and covariance matrix. Finally, section \ref{Simulations} shows the results from Monte Carlo simulations. All proof are differed to the end of the paper  (section \ref{Proofs}).

\section{The eigenvalue corrected OLS}
Consider the model:
\begin{equation} \label{the model}
    Y_{ij}=\sum_{l=1}^L \beta_{0,l} X_{ij,l}+\gamma(A_i+A_j)+  \delta A_i \times A_j+V_{ij}
\end{equation}
for all $i\neq j$, where $A_i$'s are i.i.d centered random variables with finite fourth moments. The $V_{ij}$'s are i.i.d centered square integrable random variables with $V_{ij}=V_{ji}$,  $\beta_0:=(\beta_{0,1},.., \beta_{0,L})$ is the parameter of interest and  $\gamma \geq 0$ and $\delta \in \{ -1, +1\}$ are unknown nuisance parameters.\footnote{$\gamma$ is set to be positive because the model \ref{the model} can also be re-expressed as$Y_{ij}= X_{ij}\beta +(-\gamma)((-A_i)+(-A_j))+  \delta\times  (-A_i) \times (-A_j)+V_{ij}$. The sign of $\gamma$ is not identified.} The covariates $X$ are such that for all $i,j,l$ such that $i\neq j$: $X_{ij,l}= X_{ji, l}=\phi(X_i, X_j, W_{ij})$, for some (unknown) function $\phi$, i.i.d random variables $X_i$ and $i.i.d.$ variables $W_{ij}$. By convention, $X_{ii,l}=0$ for all $i$ and $l$ by convention.

The model in equation (\ref{the model}) can also be re-expressed:
\begin{equation}\label{the model modified}
\begin{split}
Y_{ij}&=\sum_{l=1}^L \beta_{0,l} X_{ij,l}-\delta \gamma^2+ \delta (A_i+\gamma)(A_j+\gamma)+V_{ij}\\
\end{split}
\end{equation}
which reduces the study of the model \ref{the model} to that of the model:
\begin{equation} \label{the model of interest}
\begin{split}
    Y_{ij}&=\sum_{l=1}^L \mu_{0,l} X_{ij,l}+\delta U_i U_j+V_{ij}\\
\end{split}
\end{equation}
where the errors $U_i:=\gamma+A_i$ are no longer assumed to be centered. All the slope parameters remain unchanged as you move from model 
(\ref{the model}) to (\ref{the model modified}) (or (\ref{the model of interest})), only the intercept is altered by the correction term``$-\delta \gamma^2$" in equation (\ref{the model modified}). Therefore, any ``good" estimators for the parameters of the model (\ref{the model of interest}) also provide good estimators for the parameters in models (\ref{the model}) and (\ref{the model modified}), except perhaps for their intercepts. For reasons that will soon become clear, the rest of the paper focuses on the model (\ref{the model of interest}). This paper focuses on the estimation of the slope parameter and does not propose an improved estimator for the intercept.

\begin{figure}
\includegraphics[width=\textwidth,height=\textheight,keepaspectratio]{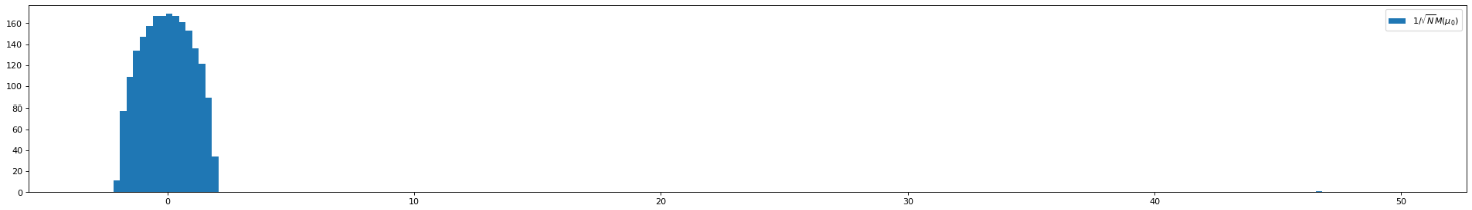}
\caption{\textit{ A histogram for $ \frac{1}{\sqrt{N}} M(\mu_0)$'s eigenvalues; $
\sigma_u=\sigma_v=1$ , $N=2000$}}
\label{Mmu0}
\end{figure}

Let $N$ be the sample size (number of nodes or agents $i$). Denote $Y$ and $V$ the $N\times N$ matrices with entries $Y_{ij}$, $V_{ij}$ and $X_l$  the matrix with entries $X_{ij,l}$ for every $l=1..L$. $Y$ and $X_l$'s diagonal entries are equal to  zero. $V$'s $i$th diagonal term is equal to $\delta(E(U_1^2)-U_i^2)$. Finally, stack the individual random effects into a vector denoted $U$. This allows for the formulation of model \ref{the model of interest} in a compact matrix form :
\begin{equation}\label{matrix form}
    Y=\sum_{l=1}^L \mu_{0,l} X_{l}+ \delta UU'+V- \delta E(U_1^2)I_N
\end{equation}
Looking at model (\ref{the model of interest}) through the lens of the matrix formulation (\ref{matrix form}) allows for novel interpretations of classical estimators. It also gives access to potentially interesting estimators that are based on standard matrix functions. To illustrate, consider the ordinary least squares estimator on model \ref{the model of interest},  defined by
$$ \hat{\mu}_{OLS}:= \arg \min_{\mu \in \mathbb{R}^L} \sum_{i,j}\left(Y_{ij}-\sum_{l=1}^L \mu_{l} X_{ij,l}\right)^2$$
Under the formulation (\ref{matrix form}), $ \hat{\mu}_{OLS}$ can also be expressed:
\begin{equation}\label{OLS}
    \begin{split}
        \hat{\mu}_{OLS}&=\arg \min_{\mu \in \mathbb{R}^L}  Trace\left( \left(Y-\sum_{l=1}^L \mu_{l} X_{l}\right)^2\right)\\
        &=\arg \min_{\mu \in \mathbb{R}^L}  \sum_{i=1}^N \left(\lambda_i \left(Y-\sum_{l=1}^L \mu_{l} X_{l}\right)^2\right)\\
        &=:\arg \min_{\mu \in \mathbb{R}^L}  \sum_{i=1}^N \lambda_i \left(M(\mu)^2\right)\\
    \end{split}
\end{equation}
where, for any $N\times N$ matrix $M$, $Trace(M)$ denotes $M$'s trace,  $\lambda_1(M)\geq \lambda_2(M)\geq \dots \lambda_N(M)$ are $M$'s eigenvalues ranked from largest to smallest and $M( . )$ is the matrix valued function that takes an element $\mu \in \mathbb{R}^L$ and returns the matrix
\begin{equation}\label{matrixM}
    M(\mu):=Y-\sum_{l=1}^L \mu_{l} X_{l} = \sum_{l=1}^L (\mu_{0,l} - \mu_l) X_{l}+ \delta UU'+V- \delta E(U_1^2)I_N
\end{equation}
Equation (\ref{OLS}) indicates that the OLS estimator can also be defined as a minimizer of the average squared eigenvalues of the matrix $M(\mu)$. Let's examine the distribution of $M(\mu)$'s eigenvalues for values of $\mu$ that are ``close" to the true value $\mu_0$, assuming $\delta=1$ (the treatment for $\delta=-1$ is similar). Begin with the value $\mu=\mu_0$, that is, let's look at the distribution of the eigenvalues of the matrix $UU'+V$.
\footnote{ We ignore the effect of the matrix $E(U_1^2)I_N$ in the discussion that follows. $E(U_1^2)I_N$  simply shifts all eigenvalues by the same quantity $E(U_1^2)$. The shift size  will turn out to be of a low order of magnitude  compared to the bulk of $UU'+V$'s eigenvalues and its effect will be negligible anyways.} Figure \ref{Mmu0} shows the histogram of the eigenvalues of the simulated matrix $\frac{1}{\sqrt{N}}\left( UU'+V\right)$, where the $U$'s and $V$'s are i.i.d standard normal and the sample size is set to $N=2000$. 

\begin{figure}
\centering
\includegraphics[scale=1]{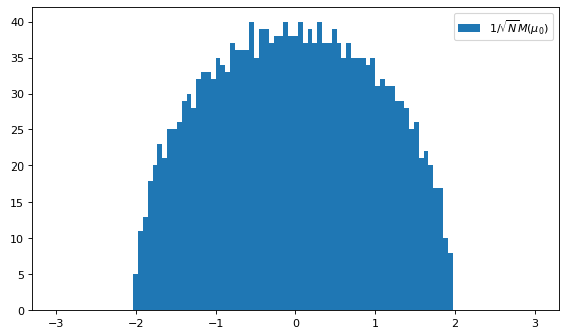}
\caption{ \textit{A zoom into the semi circle (the left block in Figure \ref{Mmu0})}}
\label{Mmu0Zoomed}
\end{figure}

The histogram in figure \ref{Mmu0} shows two distinct parts: to the left, a block of eigenvalues concentrated between values $\sim -2$ and $\sim +2$, and a single eigenvalue, further to the right, at around value $\sim 46$. After proper rescaling (and ignoring the single eigenvalue to the left for the rescaled histogram to fit on a page) the block of eigenvalues to the left has the shape of a semi-circle as shown in figure \ref{Mmu0Zoomed} .

To rationalize the shape of the histogram \ref{Mmu0}, let's examine the eigenvalues of each of the terms composing $M(\mu_0)$. The matrix $UU'$ is of rank 1, its unique non null eigenvalue is equal to $U'U=\sum_i U_i^2$ which is of the same order as $N E(U_1^2)$ when $N$ is large enough.

Figure \ref{V} shows the histogram of $V$'s eigenvalues. The two histograms in \ref{Mmu0Zoomed} and \ref{V} are seemingly identical. Only $M(\mu_0)$'s outlier eigenvalue (the one approximately equal to 46) is absent from $V$'s histogram. This should come as no surprise: the matrix $M(\mu_0)$ is a rank 1 deformation of $V$. The impact of rank 1 deformations on the eigenvalues of the original matrix ($V$ here) is well studied (e.g. \cite{Bunch1978}). Because $UU'$s unique eigenvalue is positive, modifying $V$ through $UU'$ shifts all of $V$'s eigenvalues upwards such that $V$'s eigenvalues are interlaced with $V+UU'$'s, that is, for $i=2, ..., N$: 
\[\lambda_{i}(V)  \leq \lambda_i\left(V+UU'\right)\leq \lambda_{i-1} (V) \]
and 
\[ \lambda_{1} (V) \leq \lambda_1\left(V+UU'\right)\]
Provided that $V$'s  eigenvalues (rescaled by $\frac{1}{\sqrt{N}}$) are concentrated roughly between -2 and 2, then the inequalities above predict that  $V+UU'$'s $N-1$ smallest eigenvalues will be only shifted by a small amount, which explains why the figures \ref{Mmu0Zoomed} and \ref{V} are not visually distinguishible.
\begin{figure}
\centering
\includegraphics[scale=1]{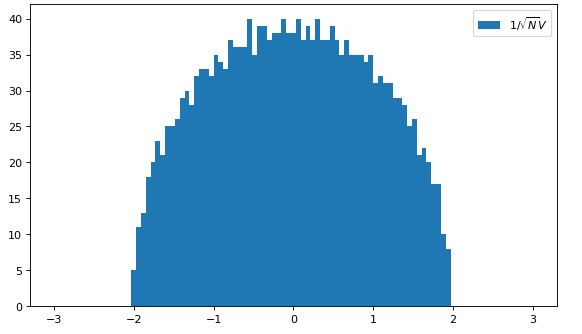}
\caption{\textit{A histogram for $ \frac{1}{\sqrt{N}} V$'s eigenvalues; $\sigma_v=1$ , $N=2000$} }
\label{V}
\end{figure}

The semi-circle in figure \ref{Mmu0Zoomed} is reminiscent of Weigner's semi-circle law in the random matrix literature (see for instance \cite{FlorentAntti2018}). Weigner's law states that the empirical distribution of the eigenvalues of a random symmetric matrix with centered square integrable entries ``converges" (in a sense that is made precise below) to a distribution with a semi-circular probability density function . Formally:\\

\textbf{Theorem [\cite{FlorentAntti2018}].} \textit{Let $W_N$ be a sequence of $N\times N$ symmetric matrices with i.i.d centered square integrable entries below the diagonal with $E(W_{N, 12}^2)=\sigma^2$, and i.i.d centered square integrable entries on the diagonal. Let $Z_N$ be a rescaling of $W_N$: $Z_N:=\frac{1}{\sqrt{N}}W_N$. Let $\mathcal{S}$ be the probability distribution characterized by the the probability density function $f$:
$$
f(x) := \left\{
    \begin{array}{ll}
        \frac{1}{2\pi \sigma^2 }\sqrt{4\sigma^2-x^2} & \mbox{if } x \in [-2\sigma^2 , 2\sigma^2] \\
        0 & \mbox{otherwise}
    \end{array}
\right.
$$
Then for any measurable set $I$:
$$ \frac{\sum_{i=1}^N \mathbbm{1}\{ \lambda_i(Z_N)\in I\}}{N}\rightarrow_p \mathcal{S}(I) \\$$
}

In particular, \cite{Furedi1981} show that $V$'s largest eigenvalue is of order $\sqrt{N}$ with probability approaching 1 as $N$ grows.

These observations combined suggest the following rough interpretation of the histogram \ref{Mmu0}:  $M(\mu_0)$'s $N-1$ smallest eigenvalues are of order $\sqrt{N}$ and are "very close" to $V$'s eigenvalues, whereas the largest eigenvalue is due to the $UU'$ deformation and is of order $N$. 

Let's extend these intuitions to values of $\mu$ that are different from the true parameter $\mu_0$. If $\mu$ is too far from $\mu_0$, then the term $ \sum_{l=1}^L (\mu_{0,l} - \mu_l) X_{l}$ in equation (\ref{matrixM}) can become dominant and dwarf the contributions of $V$ and $UU'$ in $M(\mu)$'s eigenvalue distribution. In the other extreme, when the candidate $\mu$ is``very close" to $\mu_0$, then the contribution of the covariates' term becomes negligible and we obtain a histogram that is similar to the one in figure \ref{Mmu0}. 

The values of $\mu$ that are abberantly far from $\mu_0$ lead to the eigenvalues of $ \sum_{l=1}^L (\mu_{0,l} - \mu_l) X_{l}$ being of a higher than order $\sqrt{N}$. Subsequently, they are easy to eliminate as they produce a histogram that is grossly different from the one in figure \ref{Mmu0}. However, this rough discrimination strategy will be ineffective for values of $\mu$ that return a term  $\sum_{l=1}^L (\mu_{0,l} - \mu_l) X_{l}$ of order $\sqrt{N}$ or lower. In any case, for model (\ref{the model}), the OLS estimator is known to be $\sqrt{N}$ consistent in general (See for instance \cite{Menzel2021} or section 4 in \cite{Graham2020}). The following lemma shows that any $\sqrt{N}$ estimator is in fact ``close enough" for our purposes.\footnote{Because the intercept is shifted by $-\delta \gamma^2$ when we move from the original model (\ref{the model}) to model (\ref{the model of interest}), the OLS estimator of the intercept would need to be corrected to account for the shift. That is done in proposition \ref{OLSAdj} and relegated to the appendix.}
\begin{lemma}\label{eigenvalues of X}
    For any $N\times N$ matrix $X_N$ with entries $X_{ij}$ that can be represented as: $X_{ij}=_d \psi(X_i, X_j, W_{ij})$ for some $i.i.d.$ and finite dimensional $X_i$'s and some i.i.d $W_{ij}'$s that are i.i.d. and  independent from the $X_i$'s. Assume $X_{ij}$ has at least 4 finite moments. We have that
    $$ \max_i |\lambda_i(X_N)|=O_p(N)$$
    \begin{proof}
        Refer to subsection \ref{ProofLemma1}.
    \end{proof}
 \end{lemma}
The lemma implies that any initial estimator $\tilde{\mu}$ that is $\sqrt{N}$ consistent - like the OLS estimator - would yield a covariates' term such that $\sum_{l=1}^L (\mu_{0,l} - \tilde{\mu}_l) X_{l}=O_p(\sqrt{N})$. It would   produce an eigenvalue histogram for $M(\tilde{\mu})$ that is similar to figure \ref{Mmu0} with one outlier eigenvalue of order $N$, due to the rank 1 modification $UU'$, and a cloud of eigenvalues that are of a smaller order $\sqrt{N}$ but that need not form a semi-circle this time.

Provided that the candidate $\mu$ is close enough to $\mu_0$, the largest eigenvalue of $M(\mu)$ is, at least up to a first order approximation, closely tied to the error term $UU'$. Notice that, absent the $UU'$ from the model \ref{the model of interest} (or the random effects  $A_i$ and $A_j$ from the model (\ref{the model})), we would be back to the standard linear regression model with i.i.d. and exogenous noise 
$V_{ij}$. In that case, we know that OLS is efficient, and since the sample size is $\frac{N(N-1)}{2}$, the rate of convergence of the OLS estimator would be $N$, rather than $\sqrt{N}$ under models (\ref{the model}) or (\ref{the model of interest}).

An appealing idea is then to modify the objective function in the matrix form definition of the OLS in (\ref{OLS}) to remove the contribution of the random effects. Following the intuition laid down so far, this can for instance be done by removing $M(\mu)$'s largest eigenvalue from the sum of squared errors before minimizing. The new estimator would be a solution to the minimization problem
\begin{equation}\label{Oracle} \min_{\mu \in \mathbb{R}^L}  \sum_{i=2}^N \lambda_i \left(M(\mu)^2\right)  
\end{equation}
First, let's show that the problem (\ref{Oracle}) admits at least a solution. Let $f_N$ the function $$f_N: \mu \rightarrow \left( \sum_{i\neq j} X_{ij}'X_{ij}-\sum_{i\neq j, k\neq i,j} \nu_i({\mu}) \nu_j({\mu}) X_{jk}'X_{ik} \right)^{-1} \left( \sum_{i\neq j}X_{ij}'Y_{ij} -\sum_{i\neq j, k\neq i,j} \nu_i({\mu}) \nu_j({\mu}) X_{jk}'Y_{ik} \right)$$

\begin{lemma}\label{Existence}
    Assume $E(X_{12}'X_{12})$ is invertible. With probability approaching 1, the problem \ref{Oracle} admits a solution for $N$ large enough. Moreover, $\mu^*$ is a  minimizer of (\ref{Oracle}) if and only if it is a solution to the fixed point problem:
\begin{equation}\label{FixedPoint}
    {\mu}=f_N(\mu)
\end{equation}
where $\nu(\mu)$ is the normalized $\left(||\nu(\mu)||_2=1\right)$ eigenvector of $M(\mu)$ corresponding to $M(\mu)'s$ largest eigenvalue.
\end{lemma}
\begin{proof}
   See section \ref{ExistenceProof}. 
\end{proof}
The condition on $E(X_{12}'X_{12})$ is standard in the classical least squares theory and insures that in the population, none of the regressors is a linear combination of the others (see for instance \cite{Wooldridge2010}, chapter 4).

The optimization problem (\ref{Oracle}) involves functions that are in general not smooth. It is not solvable in closed form.  In addition to guaranteeing the existence of a solution, lemma \ref{Existence} provides a practical tool to study the behavior of estimators obtained through the optimization problem (\ref{Oracle}). Intuitively, equation (\ref{FixedPoint}) is a first order condition of a minimization problem that is equivalent to (\ref{Oracle}). Let $\mu^* \in \arg \min_{\mu \in \mathbb{R}^L}  \sum_{i=2}^N \lambda_i \left(M(\mu)\right)^2 $ and note
{\small
 \begin{equation*}
 \begin{split}
\mu^* \in \arg \min_{\mu \in \mathbb{R}^L}&   \sum_{i=2}^N \lambda_i \left(M(\mu)\right)^2 \\
&\iff  \mu^* \in arg\min_\mu \sum_{i\neq j} \left(Y_{ij}-\sum_{l=1}^L \mu_{l} X_{ij,l}\right)^2-\max_{ \nu: ||\nu||=1}  \nu'M(\mu)^2\nu\\
    & \Rightarrow \mu^* \in arg\min_\mu \sum_{i\neq j} \left(Y_{ij}-\sum_{l=1}^L \mu_{l} X_{ij,l}\right)^2-\nu({\mu}^*)'M(\mu)^2\nu({\mu}^*); \mbox{ where } \nu({\mu}) \in \arg \max_{ \nu: ||\nu||=1}  \nu'M(\mu)^2\nu\\
    & \Rightarrow \mu^* \in arg\min_\mu \left(Y_{ij}-\sum_{l=1}^L \mu_{l} X_{ij,l}\right)^2- \sum_{i\neq j,k\neq i,j}\nu_i({\mu}^*) \nu_j({\mu}^*)\left(Y_{ik}-\sum_{l=1}^L \mu_{l} X_{ik,l}\right)\left(Y_{kj}-\sum_{l=1}^L \mu_{l} X_{kj,l}\right)
    \end{split}
\end{equation*}}  

The last equality allows for the expression of ${\mu}^*$ as the minimizer of a smooth and convex function over $\mathbb{R}^L$ (in fact, strictly convex with probability 1, when $N$ is large enough): $$\mu \rightarrow  \sum_{i\neq j} \left(Y_{ij}-\sum_{l=1}^L \mu_{l} X_{ij,l}\right)^2- \sum_{i\neq j,k\neq i,j}\nu_i({\mu}^*) \nu_j ({\mu}^*) \left(Y_{ik}-\sum_{l=1}^L \mu_{l} X_{ik,l}\right)\left(Y_{kj}-\sum_{l=1}^L \mu_{l} X_{kj,l}\right)$$
the first order condition results in the fixed point problem (\ref{FixedPoint}). The proof in section \ref{ExistenceProof} closely follows this sketch.

\begin{figure}\label{Objective}
\centering
\includegraphics[scale=1]{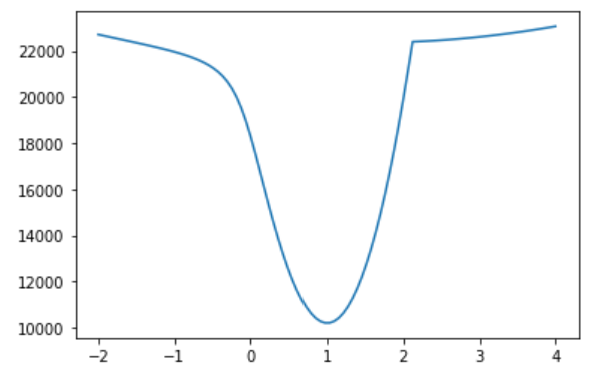}
\caption{\textit{The graph of the function $f_N$ for the model $Y_{ij}=\mu+U_iU_j+V_{ij}$; $\sigma_U=\sigma_V=1$ , $N=100$. The values of $\mu$ are on the X-axis, and the corresponding $f_N(\mu)$ is on the Y-axis.}}
\label{V}
\end{figure}

Lemma \ref{Existence} does not guarantee the uniqueness of the solution to the minimization problem (\ref{Oracle}). The iteration process just described, when it converges, could converge to one of many potential  fixed point of (\ref{FixedPoint}) (solutions to (\ref{Oracle})). Additionally, the iteration process could be explosive, leading the iterations to diverge rather than approach one of the fixed points. The function $f_N$ is generally ill-behaved. In general, it is neither convex, nor quasi-convex, nor differentiable.  Figure \ref{Objective} illustrates $f_N$'s behavior for the simplest model nested in model (\ref{the model}): $Y_{ij}=\mu_0 +U_iU_j+V_{ij}$ for $\sigma_U=\sigma_V=1$, $\mu_0=1$ and for $N=100$. In this example, $f_N$ is convex between $\approx -0.5$ and $\approx 2$, it has a point of inflexion, smoothly switching convexity at $\approx -0.5$. $f_N$  is not derivable at $\approx 2$. However, $f_N$ has a unique minimum (on the interval displayed in figure \ref{Objective}), that is close to the true parameter $\mu=\mu_0=1$. The figure also points to the direction that the results in the sequel will follow: I show that with high probability,  $f_N$ is well behaved in a shrinking neighborhood of $\mu_0$ . $\mu_0$ being unknown, knowledge of a \textit{good enough} first stage estimator will be essential throughout the paper. In particular, we study  the estimator defined defined in (\ref{Oracle}) by studying single successive iterations on the fixed point problem (\ref{FixedPoint}).  It turns out that when the iteration process is initiated with a \textit{good} first stage estimator, e.g. OLS, it converges to a fixed point or a minimizer (formal statements are presented in corollary \ref{Contraction} in the following section). In fact, the main reason why the errors $U$ are required to be exogenous throughout the paper  is that without exogeneity, no valid first stage estimators are available to the best of my knowledge.

The estimator(s) studied in this paper are obtained by iterating equation (\ref{FixedPoint}), that is, by plugging  some ``reasonable" initial candidate estimator in the right hand side of (\ref{FixedPoint}) to obtain what we show is a more precise estimator on the left hand side, then iterating this process as needed until the true fixed point distribution is achieved.  We show that when $\gamma$ in model (\ref{the model}) is equal to zero, a single iteration starting with the ordinary least squares (or any $\sqrt{N}$-consistent first stage) is enough to achieve full efficiency. This is reminiscent of ``one-step theorems" in standard cross-section models where, knowing that efficiency is achieved by an estimator minimizing some objective function, full efficiency can also be achieved through a single iteration of a gradient descent algorithm applied to the objective function provided that the algorithm is initiated with an appropriate first stage estimator (see  \cite{NeweyMcFadden1994}   section 3.4 for a detailed discussion). As in the standard cross-section setting,  iterating beyond the first iteration has no first order effect.

For non-null $\gamma$ however, any  $\sqrt{N}$-consistent first stage can only return a $\sqrt{N}-$ consistent estimator following any finite number of iterations, even though the asymptotic variance of the generated estimators decays exponentially in the number of iterations. For this case, the next section lays out a general method to obtain a rate $N$ estimator.

\section{The estimator's asymptotic behavior}\label{AsymSection}

The first result examines a single iteration of the fixed point problem (\ref{FixedPoint}).
\begin{proposition}\label{MainTheorem}
Consider the model \ref{the model of interest}:
$$ Y_{ij}=\sum_{k=1}^K \mu_{0,k} X_{ij,k}+  \delta U_i U_j+V_{ij}=X_{ij}\mu_0+\delta U_iU_j+V_{ij}$$
where $\delta\in \{-1, +1\}$. Assume that the $U$'s have at least 4 finite moments, that $Var(U)=\sigma_U^2\neq 0$ and that the $V$'s have at least 2 finite moments.
Given a first stage estimator $ \Tilde{\mu}$ such that $ ||\Tilde{\mu}-\mu_0||=O_p\left(\frac{1}{\sqrt{N}}\right)$,
the single iteration estimator
\begin{equation}\label{Definition}
    \hat{\mu}:=\left( \sum_{i\neq j} X_{ij}'X_{ij}-\sum_{i\neq j, k\neq i,j} \nu_i(\tilde{\mu}) \nu_j(\tilde{\mu}) X_{jk}'X_{ik} \right)^{-1} \left( \sum_{i\neq j} X_{ij}'Y_{ij} -\sum_{i\neq j, k\neq i,j} \nu_i(\tilde{\mu}) \nu_j(\tilde{\mu}) X_{jk}'Y_{ik} \right)
\end{equation} 
satisfies
\begin{equation}\label{SingleIteration}
\begin{split}
    \sqrt{N}(\hat{\mu}-\mu_0)& =K \sqrt{N}(\Tilde{\mu}-\mu_{0})+ O_p\left(\frac{1}{\sqrt{N}}\right)
    \end{split}
\end{equation}
for
$$ K:= \frac{E(U_1)^2}{E(U_1^2)} \left(  E(X_{12}X_{12}')-\frac{E(U_1)^2}{E(U_1^2)}E(X_{12}X_{23}')\right)^{-1}  \left(E(X_{12}X_{23}')-\frac{E(U_1)^2 }{E(U_1^2)} E(X_{12})E(X_{12}')\right) $$
\end{proposition}
A detailed proof is  presented in section \ref{MainTheoremProof}. Proposition \ref{Keig} shows that $\left(  E(X_{12}X_{12}')-\frac{E(U_1)^2}{E(U_1^2)}E(X_{12}X_{23}')\right)$ is invertible, insuring that $K$ is well defined.

Equation (\ref{SingleIteration}) describes how the distribution of the single iteration estimator relates to the first stage estimator's. An immediate corollary of proposition (\ref{MainTheorem}) is that the single iteration estimator $\hat{\mu}$ is consistent and converges  to $\mu_0$ at least as at a rate of ${\sqrt{N}}$. Also, up to a first order approximation, the single stage estimator depends linearly on the initial $\tilde{\mu}$.

Whether the iteration process improves the quality of estimation depends on the matrix $K$. When $E(U_1)=0$ in proposition \ref{MainTheorem} (i.e. when $\gamma=0$ in model (\ref{the model})), the matrix $K$ is null and equation (\ref{SingleIteration}) becomes $$\hat{\mu}-\mu_0=O_p\left( \frac{1}{N}\right).$$
After a single iteration, we are able to achieve the optimal rate of convergence $N$. Unfortunately, proposition \ref{MainTheorem} does not provide the asymptotic distribution of $\hat{\mu}$ or the effect iterations have beyond the first iteration. To answer both these questions, we need to \textit{zoom} into the $O_p\left( \frac{1}{\sqrt{N}}\right)$ term in equation (\ref{SingleIteration}) and determine how it depends on the first stage estimator $\tilde{\mu}$ and/or how it behaves asymptotically. The next proposition and its proof in section \ref{MainTheoremProof}  address this case.

\begin{proposition}\label{MainTheorem2}
   In addition to the assumptions in proposition \ref{MainTheorem}, assume $E(U_i)=0$, then
  \begin{align}\label{EfficientCase}
    N(\hat{\mu}-\mu_0)\rightarrow_d \mathcal{N}\left( 0, 2\sigma_v^2 E(X_{12}X_{12}')^{-1} \right)
    \end{align}
\end{proposition}

A ``one step theorem" applies, one iteration is enough to achieve full efficiency. The argument proving the efficiency of $\hat{\mu}$ in proposition \ref{MainTheorem2} is simple: consider the alternative model $ Y_{ij}=\sum_{l=1}^L \mu_{0,l} X_{ij,l}+  V_{ij}=X_{ij}\mu_0+V_{ij}$ is $i.i.d.$ with $i.i.d.$ errors $V_{ij}$. In this model, the ordinary least squares estimator is known to be efficient and asymptotically normal, with asymptotic covariance matrix $2\sigma_v^2 E(X_{12}X_{12}')^{-1}$ - the same asymptotic distribution as in (\ref{EfficientCase}) (see for instance \cite{Chamberlain1987} or \cite{Newey1990}). Given that our model of interest (\ref{the model of interest}) is noisier than the alternative model, the following corollary holds.

\begin{corollary}
Under the assumptions of proposition \ref{MainTheorem}, when $E(U_i)=0$, the single iteration estimator defined in (\ref{SingleIteration}) is semi-parametrically efficient.
\end{corollary} 

When $K\neq 0$, the one step theorem no longer applies. After any finite number of iterations, the new estimator is still $\sqrt{N}$-consistent. To understand the role of $K$ when $K\neq 0$, consider the simple case where we have a single regressor ($L=1$). $K$ becomes a scalar and  when $|K|<1$, $\hat{\mu}$ is to a first order closer to $\mu_0$ than $\tilde{\mu}$. If the first stage estimator is asymptotically normal (the standard \textit{ordinary least squares} estimator for example) with an asymptotic variance of $\tilde{\sigma}^2$, then $\hat{\mu}$ is normally distributed with variance $K^2 \tilde{\sigma}^2 <\tilde{\sigma}^2$. Moreover, as we iterate, the variance decays exponentially in the number of iterations. Conversely, if $|K|>1$, iterations produce noisier estimators, and the variance explodes exponentially with the number of iterations. Finally, if $|K|=1$, then the new estimator is asymptotically equivalent to the first stage estimator, iteration is neither useful nor harmful.

Simplify further, and assume that the single regressor is in fact just a constant $X_{ij}=1$, that is, we are interested in estimating the mean of $Y_{ij}$. The constant $K$ becomes $K=\frac{E(U_1)^2}{E(U_1^2)}$ which is positive and  strictly smaller than 1 (since by assumption $\sigma_U^2>0$), the iterations improve estimation quality. 

When $L>1$, $K$ is a matrix. Rather than comparing $K$ to 1, the relevant comparison is now between $K$ and $I_L$ - the identity matrix of dimension $L$ - in the partial order on symmetric matrices. When $K^2> I_L$, that is, when  $K$'s eigenvalues are all larger than 1 in absolute value, the successive iterations follow an explosive path of covariance matrices. The conclusions are similar to the univariate setting in the two cases: $K^2< I_L$ or $K^2=I_L$.  In the multivariate case however, these three cases are not exhaustive, since $>$ here is only a partial order. Fortunately, the next proposition shows that the only possibile case, given our assumptions,  is in fact $0<K<I_L$.

\begin{proposition}\label{Keig}
  Under the assumptions of proposition \ref{MainTheorem}, the matrix $\left(  E(X_{12}X_{12}')-\frac{E(U_1)^2}{E(U_1^2)}E(X_{12}X_{23}')\right)$ is definite positive and all the eigenvalues of the matrix   
$$ K= \frac{E(U_1)^2}{E(U_1^2)}\left(  E(X_{12}X_{12}')-\frac{E(U_1)^2}{E(U_1^2)}E(X_{12}X_{23}')\right)^{-1}   \left(E(X_{12}X_{23}')-\frac{E(U_1)^2 }{E(U_1^2)} E(X_{12})E(X_{12}')\right)   $$
  are positive and strictly smaller than 1.
\end{proposition}

\begin{proof}
    cf. section \ref{KeigProof}
\end{proof}

Together, the propositions \ref{MainTheorem} and \ref{Keig} imply that given a $\sqrt{N}$-consistent initial estimator and a fixed $\epsilon>0$, we  iterate the process described in the equation (\ref{Definition}) to obtain a new  $\sqrt{N}$-consistent estimator with a variance that is smaller than $\epsilon$ (or $\epsilon I_L$ in the multivariate case). This strongly suggests that an estimator with a faster than $\sqrt{N}$ rate of convergence exists. In fact, using a simple trick,  the propositions \ref{Prop2} and \ref{Keig} provide a rate $N$ (a rate optimal) estimator.\footnote{A similar idea is used for instance to construct a ``Generalized Jackknife estimator" (e.g. \cite{PowellStockStoker1989},\cite{CattaneoCrumpJansson2013}). In the context of proposition \ref{MainTheorem}, the term $\sqrt{N}(\tilde{\mu}-\mu_0)$ is eliminated by taking the convex combination, in the same fashion that the bias is removed in the generalized jacknife by taking a convex combination of estimator with the same bias.  }

Another corollary of proposition \ref{MainTheorem} is that, if $f_N$ has a fixed point $\hat{\mu}^*$ that is $\sqrt{N}$-consistent, then equation (\ref{SingleIteration}) yields:
$$(I-K)\sqrt{N}(\hat{\mu}^*-\mu_0)=O_p\left(\frac{1}{\sqrt{N}}\right) $$
so $\hat{\mu}^*$ is in fact $N$-consistent. Proposition \ref{MainTheorem} is  silent about the exact  asymptotic distribution of $\hat{\mu}^*$ and about its existence.  

To establish the existence of a fixed point, notice that  equation (\ref{SingleIteration}) has the flavor of Taylor expansion, where the matrix $K$ would represent a gradient. Because the matrix $K$ has a spectral radius that is smaller than 1 (proposition \ref{Keig}), then $f_N$ must be contracting in a local sense. Then (a variation on) the Banach fixed point theorem should prove existence. This intuition is the main idea for the proof for the next corollary. 

\begin{corollary}\label{Contraction}
    Let $\Tilde{\mu}$ be an estimator such that $\tilde{\mu}-\mu_0=O_p\left( \frac{1}{\sqrt{N}}\right)$. 
    Fix $\kappa \in (\lambda_1(K), 1)$ and some $C>0$. With probability approaching 1:
    \begin{enumerate}
        \item The function $f_N$ in equation (\ref{FixedPoint}) is differentiable in the closed ball $B({\mu}_0, \frac{C}{\sqrt{N}})$ centered at ${\mu}_0$ and with radius $\frac{C}{\sqrt{N}}$.
        \item $\sup_{\mu \in B({\mu}_0, \frac{C}{\sqrt{N}})} ||f_N'(\mu)||\leq \kappa < 1$
\end{enumerate}
Moreover, define the sequence $\hat{\mu}_m$   by: $\hat{\mu}_0:=\Tilde{\mu} $ and $\hat{\mu}_{m+1}:=f_N(\hat{\mu}_{m})$, and $\hat{\mu}^*:=\limsup_{m} \hat{\mu}_{m}$. Then $\hat{\mu}^*-\mu_0 =O_p\left( \frac{1}{\sqrt{N}}\right)$ and with probability approaching 1 $\hat{\mu}^*= \lim_{m\rightarrow +\infty} \hat{\mu}_{m}  $ and $\hat{\mu}^*$ is a solution to (\ref{Oracle}).
\end{corollary}

\begin{proof}
    Cf. Section \ref{ContractionProof}
    .
\end{proof}

So $ \hat{\mu}^*$ exists with probability approaching 1 and is rate optimal. It is left to determine its asymptotic distribution. We need to compute a higher order term in the expansion (\ref{SingleIteration}) of proposition \ref{MainTheorem}. That is the purpose of proposition \ref{N rate version}.
\begin{proposition}\label{N rate version}
    Under the assumptions of proposition \ref{MainTheorem}, if the first stage estimator is such that $\Tilde{\mu}-\mu_0=O_p\left(\frac{1}{N} \right)$, then
    \begin{align}\label{SingleIteration2}
    N(\hat{\mu}-\mu_0)&=K N (\Tilde{\mu}-\mu_{0}) +R_N + O_p\left(\frac{1}{\sqrt{N}}\right)
\end{align}
with
\begin{align*}
    R_N& \rightarrow_d  2 \delta \frac{E(U_1) E(U_1^3)}{E(U_1^2)}\left(E(X_{12}X_{12}')- \frac{E(U_1)^2}{E(U_1^2)}E(X_{12}X_{32}')\right)^{-1} E(X_{12})\\
&\; \; \; + \left(E(X_{12}X_{12}')- \frac{E(U_1)^2}{E(U_1^2)}E(X_{12}X_{32}')\right)^{-1}  \mathcal{N}\left( 0,\sigma_V^2 \Sigma \right)
\end{align*}
for 
\begin{align*}
    \Sigma &:= \bigg( 2 E(X_{12 }X_{12 }')+10\frac{E(U_1)^4}{ E(U_1^2)^2} E(X_{12})E(X_{12})'-4\frac{E(U_1)^2}{ E(U_1^2)} E(X_{12}X_{23}')\bigg) 
\end{align*}
\end{proposition}

\begin{proof}
    C.f. section \ref{Proof N rate version}
\end{proof}

Because of the presence of the residual $R_N$ in equation (\ref{SingleIteration2}), the new expansion is fundamentally different from the previous one ((\ref{SingleIteration}) in proposition \ref{MainTheorem}). The effect of an iteration on the estimation quality is now ambiguous and depends on how the first stage estimator $\tilde{\mu}$ relates to the residual $R_N$. Even if they were independent, it is not clear whether iteration improves estimation. Unfortunately, even though proposition \ref{MainTheorem2} provides the asymptotic distribution of $R_N$, that is not enough to fully characterize the distribution of the single iteration estimator. For that, we would need the joint distribution of the first stage $\Tilde{\mu}$ and $R_N$, which is challenging even for a single iteration. However, we can see that because $K$ has a smaller than one spectrum (by proposition \ref{Keig}), as we iterate, the contribution of the initial (first stage or input) estimator fades away. Intuitively, starting with some $N-$ consistent first stage $\tilde{\mu}(=: \hat{\mu}_0)$, from (\ref{SingleIteration2}):
\begin{align*}
N(\hat{\mu}_m-\mu_0)&\approx K^m N(\Tilde{\mu}-\mu_0)+\sum_{i=0}^{m-1} K^i R_N  \\
&\approx K^m N(\Tilde{\mu}-\mu_0) + (I_L-K^{m})(I_L-K)^{-1}R_N\\
& \approx (I_L-K)^{-1}R_N; \mbox{ when } m \mbox{ is lage.}
\end{align*}

So the limit  distribution (when $m$ approaches infinity) should not depend on the initial estimator $\tilde{\mu}$. Corollary \ref{asymResult} formalizes these thoughts.

\begin{corollary}\label{asymResult}
 Let $\Tilde{\mu}$ be a $\sqrt{N}$ consistent estimator.    
Define the sequence $\hat{\mu}_m$: $\hat{\mu}_0:=\Tilde{\mu} $ and $\hat{\mu}_{m+1}:=f_N(\hat{\mu}_{m})$ for all $m\geq 0$, and let $\hat{\mu}^*:=\limsup_{m} \hat{\mu}_{m}$. Then 
    \begin{align*}
    N(\hat{\mu}^*-\mu_0)&=(I-K)^{-1 }R_N + O_p\left(\frac{1}{\sqrt{N}}\right)
\end{align*}
and with  probability  approaching 1 $\hat{\mu}^*$ is a solution to (\ref{Oracle}).
Therefore 
\begin{equation}\label{asymDist}
\begin{split}
N(\hat{\mu}^*-\mu_0)&\rightarrow_d  2 \delta \frac{E(U_1) E(U_1^3)}{E(U_1^2)}(I-K)^{-1 }\left(E(X_{12}X_{12}')- \frac{E(U_1)^2}{E(U_1^2)}E(X_{12}X_{32}')\right)^{-1} E(X_{12})\\
&\; \; \; + (I-K)^{-1 }\left(E(X_{12}X_{12}')- \frac{E(U_1)^2}{E(U_1^2)}E(X_{12}X_{32}')\right)^{-1}  \mathcal{N}\left( 0,\sigma_V^2 \Sigma \right)
\end{split}
\end{equation}
for 
\begin{align*}
    \Sigma &:= \bigg( 2 E(X_{12 }X_{12 }')+10\frac{E(U_1)^4}{ E(U_1^2)^2} E(X_{12})E(X_{12})'-4\frac{E(U_1)^2}{ E(U_1^2)} E(X_{12}X_{23}')\bigg) 
\end{align*}

\end{corollary}
\begin{proof}
    Immediately follows from proposition \ref{MainTheorem2} and corollary \ref{Contraction}.
\end{proof}

Notice that $\hat{\mu}^*$ is asymptotically biased. In the next section, we offer a correction to this bias by proposing a consistent estimator for the bias term.

The corollary show that if we initiate a sequence $\hat{\mu}_0:=\tilde{\mu}$, for some initial $\sqrt{N}$-consistent estimator $\tilde{\mu}$, and then we iterate ``infinitely many" $\hat{\mu}_{m+1}:=f_N(\hat{\mu}_m)$ as in corollary \ref{Contraction}, then with high probability $\hat{\mu}_m$ approaches a fixed point $\hat{\mu}^*$. As is standard in numerical optimization methods, ``infinitely many" repetitions can in practice be read as ``sufficiently many repetitions". None of the results so far in this paper provides any guidance regarding how many repetitions are enough. In fact, one of proposition \ref{MainTheorem}'s corollaries can be concerning: equation (\ref{SingleIteration}) establishes that if we initiate  with a $\sqrt{N}$ consistent estimator, then we can only hope the iteration process to return $\sqrt{N}$-consistent estimators if we stop after a finite number of iterations. Therefore, from equation (\ref{MainEq}), we get a sense of what a lower bound on the number of iterations should be, and it is rather massive. The number of iterations should be a diverging function of the sample size $N$ for us to have any hope to escape the $\sqrt{N}$ rate of convergence. How fast the number of iterations grows with $N$ will have an effect on the rate of convergence of the final estimator, but it is hard to tell what the proper order of magnitude is. It is even less clear what the rate of convergence would be if the number of iterations is indexed on some stoppage criterion on the value of the objective function, as is usually the case in standard numerical optimization algorithms.  In simulations, the question of the number of iterations does not seem to be problematic. The standard optimization methods deliver distributions that are in line with the predictions of the asymptotic results presented so far, in particular the asymptotic distribution of $\hat{\mu}^*$ in corollary \ref{asymResult}.

Fortunately, the propositions \ref{MainTheorem} and \ref{MainTheorem2} can be put to use differently to extract  an estimator that is asymptotically equivalent to the minimizer $\hat{\mu}^*$. The alternative estimator requires exactly 2 iterations over the function $f_N$ and is therefore numerically more efficient. Using the alternative estimator, we can circumvent the concerns we highlighted around the number of iterations that are sufficient to achieve the desired asymptotic distribution.

\subsection*{An equivalent estimator} 

First, assume that the matrix $K$ is observed. Let $\Tilde{\mu}$ be an initial $\sqrt{N}$ consistent estimator and let $\hat{\mu}_1$ be the estimator returned in the equation (\ref{Definition}) after a single iteration. Write $\check{\mu}_1^*:=G\hat{\mu}_1+(I_L-G)\tilde{\mu}$, for some fixed $L\times L$ matrix $G$ and where $I_L$ is the identity matrix of dimension $L$. We will choose the matrix $G$ so that $\check{\mu}_1^*$ converges at rate $N$. Write
\begin{align*}
    \sqrt{N}(\check{\mu}_1^*- \mu_0)&= \sqrt{N}G(\hat{\mu}_1- \mu_0)+\sqrt{N}(I_L-G)(\tilde{\mu}- \mu_0)\\
    &=(GK+I_L - G) \sqrt{N} (\tilde{\mu}- \mu_0) + O_p\left(\frac{1}{\sqrt{N}}\right)\\
    &=\left(I_L - G(I_L-K)\right) \sqrt{N} (\tilde{\mu}- \mu_0) + O_p\left(\frac{1}{\sqrt{N}}\right)
\end{align*}
choosing $G$ such that $I_L - G(I_L-K)=0$ - i.e. $G=(I_L-K)^{-1}$ - yields a rate $N$ estimator. Note that by the proposition \ref{Keig}, $I_L-K$ is invertible and $G$ is well defined.

In practice, the matrix $K$ is not observed. Instead, it needs to be estimated and plugged in to generate an estimator for $G$. Assume we have a consistent estimator $\hat{K}$ for $K$. Define $\hat{G}:=(I_L-\hat{K})^{-1}$ and $\check{\mu}_1:= \hat{G}\hat{\mu}+(I_L-\hat{G})\tilde{\mu}$. As for $\check{\mu}^*$
\begin{equation}\label{mucheck}
\begin{split}
    \sqrt{N}(\check{\mu}_1- \mu_0)&= \sqrt{N}\hat{G}(\hat{\mu}- \mu_0)+\sqrt{N}(I_L-\hat{G})(\tilde{\mu}- \mu_0)\\
    &=\left(I_L - (I_L-\hat{K})^{-1}(I_L-K)\right) \sqrt{N} (\tilde{\mu}- \mu_0) + O_p\left(\frac{1}{\sqrt{N}}\right)\\
    &=(I_L-\hat{K})^{-1}\left(K-\hat{K}\right)\sqrt{N} (\tilde{\mu}- \mu_0) + O_p\left(\frac{1}{\sqrt{N}}\right)
    \end{split}
\end{equation}
If $\hat{K}$ is a $\sqrt{N}$ - consistent estimator for $K$, that is, if $\hat{K}-K=O_p\left(\frac{1}{\sqrt{N}}\right)$, then the new estimator $\check{\mu}$ is rate optimal. The following proposition offers an example of a $\sqrt{N}$ consistent estimator for $K$.

\begin{proposition}\label{Kmatrix}
    Let $\tilde{\mu}$ be a $\sqrt{N}$-consistent estimator for $\mu_0$. Define: 
    \begin{equation*}
        \begin{split}
        \hat{K}:&=\frac{\left(  \sum_i \nu_i(\Tilde{\mu})\right)^2}{N}\left(  \frac{\sum_{i=1 \leq N/2} X_{2i, 2i+1}X_{2i, 2i+1}'}{N/2}-\frac{\left(  \sum_i \nu_i(\Tilde{\mu})\right)^2}{N}\frac{\sum_{i=1 \leq N/3} X_{3i, 3i+1}X_{3i+1, 3i+2}'}{N/3}\right)^{-1} \\
        &\times \left(\frac{\sum_{i=1 \leq N/3} X_{3i, 3i+1}X_{3i+1, 3i+2}'}{N/3}-\frac{\left(  \sum_i \nu_i(\Tilde{\mu})\right)^2}{N} \left( \frac{\sum_{i=1 \leq N/2} X_{2i, 2i+1}}{N/2}\right)\left( \frac{\sum_{i=1 \leq N/2} X_{2i, 2i+1}}{N/2}\right)'\right) \\
    \end{split}
    \end{equation*}
    Then
    $$ \hat{K}-K=O_p\left(\frac{1}{\sqrt{N}}\right)$$
\end{proposition}

\begin{proof}
    c.f. section \ref{KmatrixProof}
\end{proof}

Proposition \ref{Kmatrix} allows for the construction of an estimator that is rate optimal. However, studying the asymptotic distribution of $\check{\mu}_1$ defined in (\ref{mucheck}) is challenging. It requires that we determine the joint asymptotic distribution of $\hat{K}$, $\sqrt{N}(\tilde{\mu}-\mu_0)$ and the residual of order $O_p\left(\frac{1}{\sqrt{N}}\right)$ in equation (\ref{mucheck}). However, as for the study of the fixed point $\hat{\mu}^*$, as we iterate, the effect of first stage estimator fades away. Rather than iterating here again, we use the same linear combination trick that allows us again to achieve the ``infinite iterations" distribution using one iteration only. 

Let $\hat{\mu}_2:=f_N(\check{\mu}_1)$ and define $\check{\mu}_2:= \hat{G}\hat{\mu}_2+(I_L-\hat{G})\check{\mu}_1$. Following the steps in equation (\ref{mucheck}), 
\begin{equation}\label{mucheck2}
\begin{split}
    {N}(\check{\mu}_2- \mu_0)&=(I_L-\hat{K})^{-1}\left(K-\hat{K}\right){N} (\check{\mu}_1- \mu_0) +(I_L-\hat{K})^{-1} R_N+ O_p\left(\frac{1}{\sqrt{N}}\right)\\
    &=(I_L-K)^{-1} R_N+(I_L-\hat{K})^{-1}\left(K-\hat{K}\right){N} (\check{\mu}_1- \mu_0) +(K-\hat{K})^{-1} R_N+ O_p\left(\frac{1}{\sqrt{N}}\right)\\
    &=(I_L-K)^{-1} R_N+ O_p\left(\frac{1}{\sqrt{N}}\right)\\
    \end{split}
\end{equation}
the last equality is a consequence of proposition \ref{Kmatrix}. This proves that $\check{\mu}_2$ is asymptotically equivalent to $\hat{\mu}^*$, the fixed point studied through corollary \ref{asymResult}.

To summerize, the alternative estimation procedure follows these steps:
\begin{enumerate}
    \item Compute a $\sqrt{N}$ consistent estimator $\tilde{\mu}$ (e.g. OLS with the correction in appendix \ref{OLSAdj}) and $\hat{K}$ a $\sqrt{N}$ consistent  estimator for $K$ (e.g. proposition \ref{Kmatrix}).
    \item Run one iteration to get $\hat{\mu}_1:=f_N(\tilde{\mu})$
    \item Compute  $\check{\mu}_1:= (I_L-\hat{K})^{-1}\hat{\mu}_1+(I_L-(I_L-\hat{K})^{-1})\tilde{\mu}$
    \item Iterate on $\check{\mu}_1$ to get $\hat{\mu}_2:=f_N(\check{\mu}_1)$
    \item Compute  $\check{\mu}_2:= (I_L-\hat{K})^{-1}\hat{\mu}_2+(I_L-(I_L-\hat{K})^{-1})\check{\mu}_1$
\end{enumerate}

\begin{corollary}\label{alternative}
    \begin{equation}\label{alternativeDist}
        {N}(\check{\mu}_2- \mu_0)=(I_L-K)^{-1} R_N+ O_p\left(\frac{1}{\sqrt{N}}\right)
    \end{equation}
\end{corollary}

\begin{proof}
    See the steps leading to equation (\ref{mucheck}).
\end{proof}

\section{Inference and bias correction}
To be able to do inference on the (asymptotically equivalent) estimators presented in the previous section. We need to
\begin{enumerate}
    \item correct for the bias term $2 \delta  \frac{E(U_1) E(U_1^3)}{E(U_1^2)}(I-K)^{-1 }\left(E(X_{12}X_{12}')- \frac{E(U_1)^2}{E(U_1^2)}E(X_{12}X_{32}')\right)^{-1} E(X_{12})$ in equations (\ref{asymDist}) and (\ref{alternativeDist})
    \item provide a consistent estimator for the covariance matrix.
\end{enumerate} 
{\small
    $$\sigma_V^2 (I-K)^{-1 }\left(E(X_{12}X_{12}')- \frac{E(U_1)^2}{E(U_1^2)}E(X_{12}X_{32}')\right)^{-1} \Sigma \left(E(X_{12}X_{12}')- \frac{E(U_1)^2}{E(U_1^2)}E(X_{12}X_{32}')\right)^{-1}(I-K)^{-1 } $$}
    (c.f. proposition \ref{asymResult})

To provide a consistent estimator for the bias term, we can use $\hat{K}$ defined in proposition \ref{Kmatrix} as a consistent estimator for $K$. The matrices $ E(X_{12}X_{12}')$, $E(X_{12}X_{32}')$ and the vector $E(X_{12})$ can be estimated through their sample analogues. $\delta$ and the moments of $U$ are what is left to be estimated. Assume that $\delta=1$, section 2 explained how the eigenvector corresponding to the largest eigenvalue of $M(\tilde{\mu})$ is a good approximation to the normalized vector $\frac{U}{|| U||_2}$. Moreover, the largest eigenvalue informs about $U'U$, the norm of the vector $U$. Combining both, we can recover an  estimator for $U$. When $\delta$ is -1, then we reason in terms of the largest eigenvalue in absolute value, and its corresponding eigenvalue. The difference when $\delta=-1$ is that the corresponding eigenvalue in fact estimates $-\frac{U}{||U||_2}$ rather than $\frac{U}{||U||_2}$  and a sign correction is necessary. The sign of $\delta$ is, with probability approaching 1, the sign of the largest eigenvalue in absolute value. These ideas are formalized through lemma  \ref{BiasCorrection}.

\begin{lemma}\label{BiasCorrection}
    Under the conditions and notation of proposition \ref{MainTheorem}, denote 
    $$
\hat{\delta} := \left\{
    \begin{array}{ll}
        sign(\lambda_1(\Tilde{\mu})) & \mbox{if } \lambda_1(\Tilde{\mu}):=\max_i |\lambda_{i}(\Tilde{\mu})| \\
        sign(\lambda_N(\Tilde{\mu})) & \mbox{if } |\lambda_N(\Tilde{\mu})|=\max_i |\lambda_{i}(\Tilde{\mu})|
    \end{array}
\right.
$$
and $\hat{U}_i=\hat{\delta} \sqrt{\max_i |\lambda_i(\tilde{\mu})|} \nu_i(\tilde{\mu})$. We have
    \begin{enumerate}
        \item $\mathbb{P}(\hat{\delta}=\delta)$ converges to 1 as $N$ grows to $+\infty$,
        \item $\frac{\sum_{i} U_i -\hat{U}_i}{N}=O_p\left(\frac{1}{\sqrt{N}} \right)$,
         \item $\frac{\sum_{i} U_i^2 -\hat{U}_i^2}{N}=O_p\left(\frac{1}{N} \right)$,
        \item $\frac{\sum_{i} U_i^3 -\hat{U}_i^3}{N}=O_p\left(\frac{1}{\sqrt{N}} \right)$.
    \end{enumerate}
\end{lemma}

\begin{proof}
    Cf section \ref{BiasCorrectionProof}.
\end{proof}

Lemma \ref{BiasCorrection} leads to the following consistent estimator for the bias term.

\begin{corollary}
    The estimator $$2 \hat{\delta}  \frac{\sum_i \hat{U}_i \sum_i \hat{U}_i^3}{N \sum_i \hat{U}_i^2}(I-\hat{K})^{-1 }\left(\frac{1}{N(N-1)}\sum_{i \neq j}X_{ij}'X_{ij}- \frac{\left(\sum_i \hat{U}_i\right)^2 }{N \sum_i \hat{U}_i^2}\frac{1}{N(N-1)}\sum_{i\neq j, k\neq i, j}X_{ij}'X_{jk}\right)^{-1} \sum_{i \neq j}\frac{1}{N(N-1)} X_{ij}$$

    is consistent for the bias term $2 \delta  \frac{E(U_1) E(U_1^3)}{E(U_1^2)}(I-K)^{-1 }\left(E(X_{12}X_{12}')- \frac{E(U_1)^2}{E(U_1^2)}E(X_{12}X_{32}')\right)^{-1} E(X_{12})$.
\end{corollary}

Finally, to obtain an estimator for the covariance matrix, it is left to provide a consistent estimator for the variance $\sigma_V^2$. Let's go back to the model (\ref{the model modified})
$$  Y_{ij}=\sum_{l=1}^L \beta_{0,l} X_{ij,l}+ \delta U_i  U_j+V_{ij} $$
Denote $\epsilon_{ij}:=  \delta U_i  U_j- \delta E(U_1)^2+V_{ij} $. First, observe that 
$$ \sigma_\epsilon^2:=Var(\epsilon_{ij})=E(U_1^2)^2-E(U_1)^4+\sigma_V^2 $$
Therefore, given the estimators for $E(U_1)$ and $E(U_1^2)$ provided in lemma \ref{BiasCorrection}, any  estimator $\hat{\sigma}_\epsilon^2$ that is consistent for $\sigma_\epsilon^2$, we obtain a consistent estimator $\hat{\sigma}_V^2:= \hat{\sigma}_\epsilon^2- \left( \frac{\hat{U}_i^2}{N}\right)^2+ \left( \frac{\hat{U}_i}{N}\right)^4$, where $\hat{U}_i$ are defined in lemma \ref{BiasCorrection}. Notice that the sub-sample of the data $\{ (Y_{2k-1, 2k}, X_{2k-1, 2k} ), k=1...[N/2]\}$ is i.i.d. with i.i.d residuals $\epsilon_{2k-1, 2k}$. 

Using the i.i.d. sub-sample, standard estimators for $\sigma_\epsilon^2$ are available from the cross-sectional OLS literature. Following for instance \cite{Wooldridge2010} (Section 4.2.2.), let $\hat{\epsilon}_{2k-1,2k}$ for $k=1...[N/2]$ be the residuals OLS fit of $Y_{2k-1, 2k}$ on $X_{2k-1, 2k}$  (and an intercept, when $X_{2k-1, 2k}$ does not include an intercept). A consistent estimator for $\epsilon$ is
\begin{equation} \label{SigmaEpsilon}
    \hat{\sigma}_\epsilon^2:=\frac{1}{[N/2]}\sum_{k=1}^{[N/2]} \hat{\epsilon}_{2k-1,2k}
\end{equation}
To summarize:
\begin{corollary}\label{SigmaV}
    Define $$ \hat{\sigma}_V^2:= \hat{\sigma}_\epsilon^2- \left( \frac{\hat{U}_i^2}{N}\right)^2+ \left( \frac{\hat{U}_i}{N}\right)^4 $$
    where $\hat{\sigma}_\epsilon^2$ is defined in equation (\ref{SigmaEpsilon}) and $\hat{U}_i$ is defined in lemma \ref{BiasCorrection}. We have
    $$\hat{\sigma}_V^2 \rightarrow_p {\sigma}_V^2 $$
\end{corollary}
\begin{proof}
    Follows from the consistency of $\hat{\sigma}_\epsilon^2$ for ${\sigma}_\epsilon^2$ (\cite{Wooldridge2010}), the consistency of $\frac{\hat{U}_i^2}{N}$ and $ \frac{\hat{U}_i}{N}$ for $E(U_1^2)$ and $E(U_1)$ (lemma \ref{BiasCorrection}), and the observation that $ \sigma_\epsilon^2:=Var(\epsilon_{ij})=E(U_1^2)^2-E(U_1)^4+\sigma_V^2 $.
\end{proof}

\section{Simulation study}\label{Simulations}
\begin{figure}[H]
\centering
\includegraphics[scale=0.5]{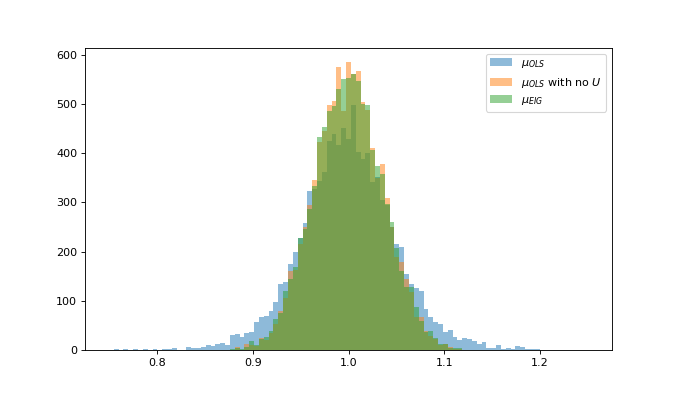}
\caption{\textit{OLS (blue) and eigenvalue-corrected (green) estimators for the slope parameter $\beta_{0,1}$ in the model $ Y_{ij}:= \beta_{0, 1}+ \beta_{0,2} (X_{i}+X_{j}) + A_i A_j +V_{ij}$, and the ``oracle" OLS estimator (orange) for the slope parameter $\beta_{0,1}$ in the model $ Y_{ij}:= \beta_{0, 1}+ \beta_{0,2} (X_{i}+X_{j}) +V_{ij}$.} }
\label{Hist1}
\end{figure}

\begin{figure}[H]
\centering
\includegraphics[scale=0.5]{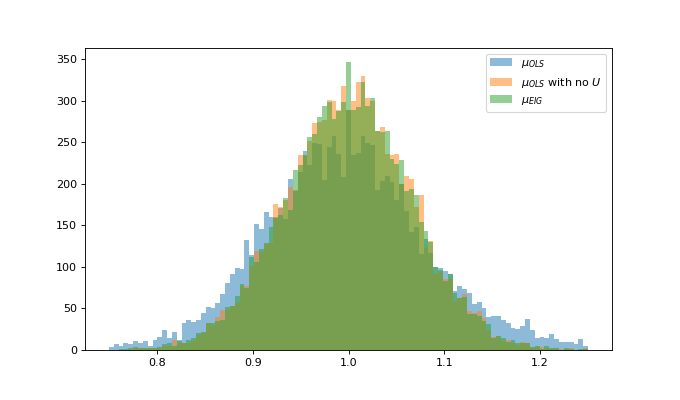}
\caption{\textit{OLS (blue) and eigenvalue-corrected (green) estimators for the slope parameter $\beta_{0,1}$ in the model $ Y_{ij}:= \beta_{0, 1}+ \beta_{0,2} X_{i}X_{j} + A_i A_j +V_{ij}$, and the ``oracle" OLS estimator (orange) for the slope parameter $\beta_{0,1}$ in the model $ Y_{ij}:= \beta_{0, 1}+ \beta_{0,2} (X_{i}+X_{j}) +V_{ij}$. }}
\label{Hist2}
\end{figure}

\begin{figure}[H]
\centering
\includegraphics[scale=0.5]{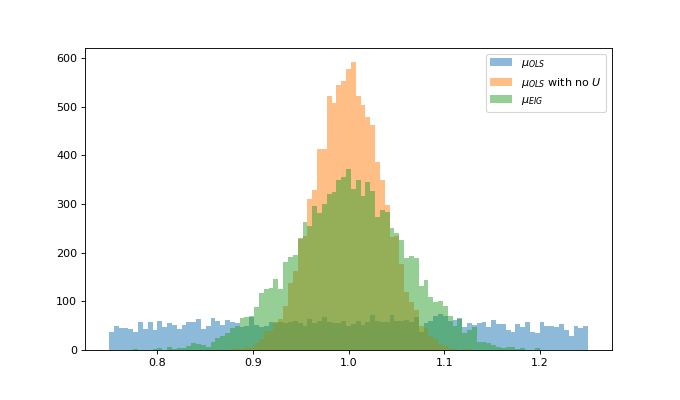}
\caption{\textit{OLS (blue) and eigenvalue-corrected (green) estimators for the slope parameter $\beta_{0,1}$ in the model $ Y_{ij}:= \beta_{0, 1}+ \beta_{0,2} (X_{i}+X_{j}) +A_i+A_j+ A_i A_j +V_{ij}$, and the ``oracle" OLS estimator (orange) for the slope parameter $\beta_{0,1}$ in the model $ Y_{ij}:= \beta_{0, 1}+ \beta_{0,2} (X_{i}+X_{j}) +V_{ij}$.} }
\label{Hist3}
\end{figure}

\begin{figure}[H]
\centering
\includegraphics[scale=0.5]{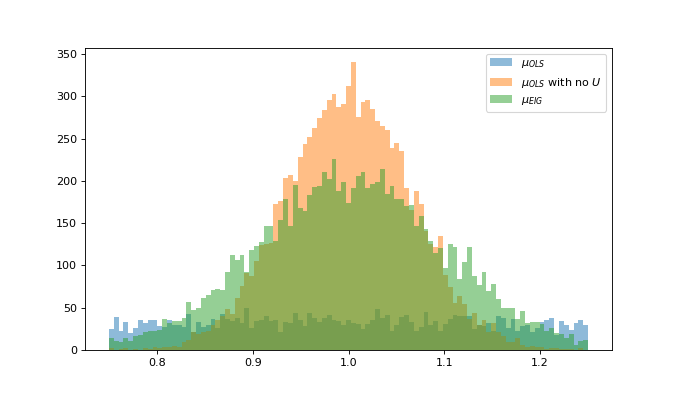}
\caption{\textit{OLS (blue) and eigenvalue-corrected (green) estimators for the slope parameter $\beta_{0,1}$ in the model $ Y_{ij}:= \beta_{0, 1}+ \beta_{0,2} X_{i}X_{j}+A_i+A_j+ A_i A_j +V_{ij}$, and the ``oracle" OLS estimator (orange) for the slope parameter $\beta_{0,1}$ in the model $ Y_{ij}:= \beta_{0, 1}+ \beta_{0,2} (X_{i}+X_{j}) +V_{ij}$.} }
\label{Hist4}
\end{figure}

I run $S=10000$ simulations on each of the 4 following designs, with a network of $N=100$ nodes in each simulation.
\begin{enumerate}
\item An intercept and an additive regressor , with $\gamma=0$ $$ Y_{ij}:= \beta_{0, 1}+ \beta_{0,2} (X_{i}+X_{j}) + A_i A_j +V_{ij} $$
    \item An intercept and a multiplicative regressor, with $\gamma=0$
    $$ Y_{ij}:= \beta_{0, 1}+ \beta_{0,2} X_{i}X_{j}  + A_i A_j +V_{ij} $$
    \item An intercept and an additive regressor, with $\gamma=1$ $$ Y_{ij}:= \beta_{0, 1}+ \beta_{0,2} (X_{i}+X_{j}) +A_i +A_j + A_i A_j +V_{ij} $$
    \item An intercept and a multiplicative regressor, with $\gamma=1$
    $$ Y_{ij}:= \beta_{0, 1}+ \beta_{0,2} X_{i}X_{j} +A_i +A_j + A_i A_j +V_{ij} $$
\end{enumerate}
for each of the two designs $X\sim Unif(0,1)$, $\beta_{0,1}=\beta_{0,2}=E(A_1^2)=E(V_{12}^2)=1$.\footnote{I also generated simulations with $X\sim \mathcal{N}(0,1)$ or $X\sim 1+ \mathcal{N}(0,1)$ and the outcomes are similar.} The histograms for the estimated slope parameters $\beta_{0,2}$ are in the figures \ref{Hist1} to \ref{Hist4}. In each graph, we show the histogram for the OLS estimator (in bleu) on the original model (\ref{the model}) as a benchmark,  the estimator $\hat{\mu}_{EIG}$ defined in this paper in green. The OLS estimator is semi-parametrically efficient in the model without individual effects, $Y_{ij}:=\beta_{0,1}+\beta_{0,2}X_{ij}+V_{ij}$ as a ``gold standard" in orange, it is also estimated for each of the simulations and displayed in orange in the figures \ref{Hist1} to \ref{Hist4} as an oracle estimator. The estimators for the intercepts are not shown since the slope parameter are our concern in this paper. As discussed in the introduction, our estimator is $N-$ consistent for $\beta_{0,1}-\delta \gamma=1-1=0$ rather than for $\beta_{0,1}=1$. The term $\delta \gamma $ can't be estimated at a higher rate  than $\sqrt{N}$. Any estimator for $\beta_{0,1}$ based on our estimator and an estimated correction for $\delta \gamma $ would only yield a $\sqrt{N}$-consistent estimator, even though  $\beta_{0,1}-\delta \gamma$ is estimated at rate $N$.

\begin{figure}[H]
\centering
\includegraphics[scale=0.5]{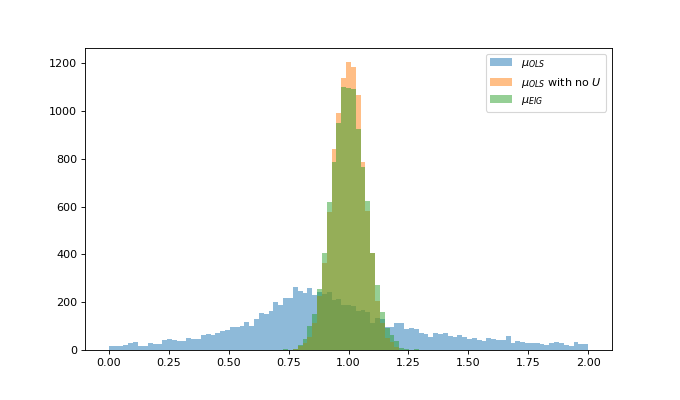}
\caption{\textit{OLS (blue) and eigenvalue-corrected (green) estimators for the slope parameter $\beta_{0,1}$ in the model $ Y_{ij}:= \beta_{0, 1}+ \beta_{0,2} X_{i}X_{j}+ 10\times A_i A_j +V_{ij}$, and the ``oracle" OLS estimator (orange) for the slope parameter $\beta_{0,1}$ in the model $ Y_{ij}:= \beta_{0, 1}+ \beta_{0,2} (X_{i}+X_{j}) +V_{ij}$. Note the 10 factor multiplying the $A_i A_j$ term to amplify the skew of the OLS estimator.}}
\label{Hist5}
\end{figure}

The first two histograms (figures \ref{Hist1} and \ref{Hist2}) confirm the result in proposition \ref{MainTheorem2}. The histogram for the eigenvalue-corrected estimator (in green) is  close to the oracle (orange). On both histogram, the OLS estimator (blue) seems to have a larger variance. In fact, the OLS estimator has a non standard asymptotic distribution (cf. \cite{Menzel2021}) and its distribution is slightly skewed to the left. The skew is not visible in figures \ref{Hist1} and \ref{Hist2}, because the variance of $A$ is not large enough (see figure \ref{Hist5} for a version of figure \ref{Hist2} with a $Var(A)=100$ and where the skew is now obvious on the OLS estimator, whereas the eigenvalue corrected estimator is unaffected).

Figures \ref{Hist3} and \ref{Hist4} show that the histogram of the OLS estimator (blue) is much less concentrated than the eigenvalue-corrected estimator (green). This reflects the prediction of corollary \ref{alternative}. The eigenvector corrected estimator is itself less efficient than the oracle (orange), but is rate optimal.

\section{Extensions}
The idea that this paper explores finds its origins in a simple observation: the matrix $ UU' + V$ has a single order $N$ eigenvalue that is due to the interaction between (re-centered) individual effects $U_i \times U_j$. A natural extension, when the model has higher order interactions, as in  the model
$$Y_{ij}=X_{ij}\mu_0+U_iU_j+U_i^2U_j^2+ V_{ij}$$
for instance. Then, similarly, the matrix of residuals would have two largest eigenvalues of order $N$ and $N-2$ eigenvalues of order at most $\sqrt{N}$. An estimator of the same flavor as the eigenvector corrected estimator discussed in this paper would consist of removing the two largest eigenvalues from the objective function in equation (\ref{Objective}), rather than just the largest eigenvalue.

More generally, if the model of interest is instead
$$ Y_{ij}=X_{ij}\mu_0+g(U_i,U_j)+ V_{ij}$$
for some unknown but well-behaved and symmetric function $g$, under some smoothness assumption, $g$ has a spectral representation:
$$g(x,y)=\sum_{k=1}^{+\infty}c_k \phi_k(x) \phi_k(y)$$
for some $\mathcal{L}^2$-orthonormal basis $(\phi_k)$ and a decaying sequence $c_k$ of real scalars. Because $c_k$ vanishes to 0, correcting for a certain number of first terms in the expansion of $g$ by removing the corresponding number of largest eigenvalues  seems like an interesting extension of the estimator the current paper concerned itself with.

\newpage

\bibliographystyle{plainnat}
\bibliography{Biblio.bib}

\newpage

\section{Proofs and intermediary results}\label{Proofs}

This section details the proofs of all the results in the paper. It begins by showing how the OLS estimator of the intercept (in mode (\ref{the model})) can be adjusted to obtain a $\sqrt{N}$- consistent estimator of the modified estimator (in model \ref{the model modified}). Then we provide the  technical ingredients (propositions \ref{Prop2} and  \ref{Prop4}) that our main results heavily rely on.

\subsection{Adjustment to the intercept}

\begin{proposition}\label{OLSAdj}
    Under model (\ref{the model}) and under the assumptions of theorem \ref{MainTheorem}, $\gamma\geq 0$, $\delta\in \{-1, 1\}$, $\sigma_U^2=E(A_1^2)\neq 0$. Let $\tilde{\beta}_1$ be a $\sqrt{N}$-consistent estimator of the intercept $\beta_0$ in equation (\ref{the model}). Then $\mu_{0,1}$, the intercept in the modified model (\ref{the model modified}) is equal to: $\mu_{0,1}= \beta_{0,1}-\delta \gamma^2 $. Define 
    \begin{align*}
\epsilon_{ij}:&=\gamma(A_i+A_j)+\delta A_{i} A_j +V_{ij}\\
a:&=E(\epsilon_{12}\epsilon_{23})=\gamma^2 E(A_i^2)\\
b: &=E(\epsilon_{12}\epsilon_{23}\epsilon_{31})=3\delta \gamma^2 E(A_i^2)^2+\delta E(A_i^2)^3
\end{align*}
Then $ |\beta|=3\delta \gamma^2 E(A_i^2)^2+\delta E(A_i^2)^3 $ and $E(A_i^2)$ is the unique real root  of the polynomial $P(x; a, |b|):=x^3+3a x -|b|$ .
    Denote 
    \begin{align*}
    \hat{\epsilon}_{ij}:&=Y_{ij}-\sum_{l=1}^L X_{ij, l}\tilde{\beta}_{l}=\sum_{l=1}^L X_{ij, l}(\beta_{0,l}-\tilde{\beta}_{l})+\epsilon_{ij}\\
    \hat{a}:&=\frac{1}{N^3}\sum_{i\neq j \neq k } \hat{\epsilon}_{ij} \hat{\epsilon}_{ik}\\
    \hat{b} :&=\frac{1}{N^3}\sum_{i\neq j \neq k } \hat{\epsilon}_{ij} \hat{\epsilon}_{ik}\hat{\epsilon}_{jk}\\
    \Tilde{\delta}:&=sign(\hat{b})
    \end{align*}
    
    Let $\hat{\sigma}_U^2$ be a real root of the polynomial $P(x; \hat{a}, \hat{b})$ and define $\hat{\gamma}^2:=\frac{\hat{a}}{\hat{\sigma}_U^2} $. We have
    \begin{align*}
        |\hat{\sigma}_U^2-E(A_1^2)|&= O_p\left( \frac{1}{\sqrt{N}}\right)\\
        |\hat{\gamma}^2-\gamma^2|&=O_p\left( \frac{1}{\sqrt{N}}\right)\\
        \mu_{0,1}-\Tilde{\mu}_1&=O_p\left( \frac{1}{\sqrt{N}}\right)
    \end{align*}
    where $\Tilde{\mu}_1:=\hat{\beta}_0-\hat{\delta}\hat{\gamma}^2$
\end{proposition}

\begin{proof}
    That $P$ has a unique real solution whenever $a\geq 0$ results from the observation that $\lim _{x\rightarrow -\infty} P(x; a, |b|)= -\infty$, $\lim _{x\rightarrow +\infty} P(x; a, |b|)= +\infty$  and $P(., a, b)$ is strictly increasing when $a\geq 0$.
    
    Observe that
    \begin{align*}
        |\hat{a}-a|&= O_p\left( \frac{1}{\sqrt{N}}\right)\\
        |\hat{b}-b|&= O_p\left( \frac{1}{\sqrt{N}}\right)
    \end{align*}
    The roots of a polynomial being continuous in its coefficients (e.g. \cite{HarrisMartin1987}), the continuous mapping theorem proves the consistency of $\hat{\sigma}_U^2$.

    Moreover, note that $\delta =sign(b)$, that $b\neq 0$, and that

    In addition, by the mean  value theorem, for some $(\Bar{x}, \Bar{a}, \Bar{b})$ between $(E(U_1^2), a, b)$ and $(\hat{\sigma}_U^2,  \hat{\alpha}, \hat{\beta})$
 \begin{align*}
    0=P(\hat{\sigma}_U^2; \hat{a}, \hat{b})&=P(E(A_1^2); a, b)+\frac{\partial P}{\partial x} (\Bar{x}, \Bar{a}, \Bar{b})(\hat{\sigma}_U^2- E(A_1^2))+\frac{\partial P}{\partial a } (\hat{a}- a)+\frac{\partial P}{\partial |b| } (|\hat{b}|-|b| )\\
    &=(3\Bar{x}^2 +3\Bar{a} ) (\hat{\sigma}_U^2- E(A_1^2)) +3\Bar{x} (\hat{a}-a)-(|\hat{b}|-|b| )
\end{align*}
because $\big||\hat{b}|-|b|\big|\leq |\hat{b}-b|$ , then $|\hat{b}|-|b|= O_p\left( \frac{1}{\sqrt{N}}\right)$, implying:
$$ \hat{\sigma}_U^2- E(A_1^2)=\frac{-1}{(3\Bar{x}^2 +3\Bar{a} )} \left(3\Bar{x} (\hat{a}-a)-(|\hat{b}|-|b| )  \right)=O_p\left( \frac{1}{\sqrt{N}}\right) $$
Finally
\begin{align*}
    \hat{\gamma}^2-\gamma^2&=	\frac{\hat{a}}{\hat{\sigma}_U^2}- \frac{a}{E(A_1^2)}\\
    &=\frac{\hat{a}E(A_1^2)-a\hat{\sigma}_U^2}{\hat{\sigma}_U^2 E(A_1^2) }\\
    &=\frac{(\hat{a}-a)E(A_1^2)+a(E(A_1^2)-\hat{\sigma}_U^2)}{\hat{\sigma}_U^2 E(A_1^2) }\\
    &=O_p\left( \frac{1}{\sqrt{N}}\right)
\end{align*}
\end{proof}
\subsection{On the distribution of the largest eigenvalue}

\begin{proposition}\label{Prop2}
Let $A=(a_{ij})$ be a matrix such that:
$$a_{ij}=U_i U_j +V_{ij}  \mbox{ for all } i\neq j$$
and $a_{ii}=0$ for all $i$, where the $V_{ij}$'s for $i\neq j $ are $i.i.d.$ mean 0 random variables with variance $\sigma_v$ and $V_{ij}=V_{ji}$, and the diagonal entries of $V$ given by $V_{ii}=E(U_1)^2-U_i^2$.\\
The $U$'s  are also i.i.d but not necessarily centered.
Let $\lambda_1(A)>\lambda_2(A)....$ be $A$'s eigenvalues. Then:
$$ \lambda_1(A)= U'U+\frac{U'VU}{U'U}+\frac{U'V^2U}{(U'U)^2}-E(U_1)^2+o_p(1)$$
\end{proposition}
\begin{proof}
The proof draws from \cite{Furedi1981}. In all what follows,``with high probability (w.h.p.)" means``with probability approaching 1 as $N$ grows". Write
$$A= U U' +V -E(U_1)^2I_N $$ 
define $\Tilde{A}:=A+E(U_1)^2I_N$ and decompose $U$ into $U=v+r$ such that $r'v=0$ and $\Tilde{A}v=\lambda_1 v$. We first show that, with high probability, $r$ is bounded.

Define:
$$ S:= \Tilde{A} U= (U'U)U+  V U=\lambda_1 v+Ar$$
define:
$$L:=E(S|U)=(U'U)U$$
therefore:
$$L_i= (U'U)U_i $$
Notice  
$$|| Ar ||^2=r'\Tilde{A}'\Tilde{A}r \leq  \lambda_2(\Tilde{A}'\Tilde{A})\times || r ||^2 = max_{i>1} |\lambda_i(\Tilde{A})|^2 \times || r ||^2$$
where the inequality results from the Courant-Fisher theorem (equation (11) in \cite{Furedi1981} ) and the second equality results from: $\Tilde{A}'\Tilde{A}=\Tilde{A}^2 $. Therefore:
$$|| \Tilde{A}r ||\leq max_{i>1} |\lambda_i(\Tilde{A})| \times || r ||$$
By a standard result on rank 1 modifications (e.g. \cite{Bunch1978}), for all $i>1$
$$ \lambda_i(V)\leq  \lambda_i(\Tilde{A}) \leq \lambda_{i-1}(V)$$
So :
$$max_{i>1} |\lambda_i(\Tilde{A})| \leq \max\{ |\lambda_N(V)|, \lambda_1(V)\} $$
By theorem 2 in \cite{Furedi1981} , almost surely:
$$\max\{ |\lambda_N(V)|, \lambda_1(V)\}=2\sigma_v \sqrt{N}+o\left(N^\frac{1}{2}\right)$$
so with high probability, for $N$ large enough:
\begin{equation}\label{Bound-1}
||\Tilde{A}r||\leq \max\{ |\lambda_N(V)|, \lambda_1(V)\} \leq 3\sigma_v \sqrt{N} ||r||\end{equation}
Thus:
\begin{equation}
    \begin{split}
        || \Tilde{A}r-(U'U)r||\geq (U'U)||r||-||\Tilde{A}r||\geq (U'U-\max\{|\lambda_N(V)|, \lambda_1(V) \})||r||
    \end{split}
\end{equation}
implying:
\begin{equation}\label{Bound0}
||r||^2 \leq \frac{|| \Tilde{A}r-(U'U)r||^2}{(U'U-\max\{|\lambda_N(V)|, \lambda_1(V) \})^2}\leq \frac{||S-L||^2}{(U'U-\max\{|\lambda_N(V)|, \lambda_1(V) \})^2} \end{equation}

With high probability:
\begin{equation}\label{Bound1}
    (U'U-\max\{|\lambda_N(V)|, \lambda_1(V) \})^2\geq \frac{\sigma_u^4}{2}N^2
\end{equation}
The second inequality is a result of Pythagorean theorem.

To show that $r$ is bounded w.h.p., it is left to show that $||S-L||^2$ also grows as $N^2$. I use Chebychev's inequality on $||S-L||^2$:
\begin{equation*}
    \begin{split}
        E(||S-L||^2|U)&=E\left(\sum_i (S_i-L_i)^2\bigg|U\right)\\
        &=E\left(\sum_i \left(\sum_j V_{ij}U_j\right)^2\bigg|U\right)\\
        &=E\left(\sum_i U_i^2 V_{ii}^2+\sum_i \left(\sum_{j\neq i} V_{ij}U_j\right)^2+2\sum_i U_i V_{ii} \sum_{j\neq i} V_{ij} U_j\bigg|U\right)\\
        &=\sum_i U_i^2V_{ii}^2+\sigma_v^2 \sum_i \sum_{j\neq i}U_j^2\\
        &=\sum_i U_i^2(E(U_1)^2-U_i^2)^2+ \sigma_v^2 (N-1)\sum_i U_i^2 
    \end{split}
\end{equation*}
so 
\begin{equation}\label{Mean}
    \frac{E(||S-L||^2|U)}{N^2}\rightarrow \sigma_v^2 E(U_1^2) \mbox{ almost surely.} 
\end{equation} 
Also: \begin{align*}
        Var(||S-L||^2|U)&=Var\left(\sum_i \left(\sum_j V_{ij}U_j\right)^2\bigg|U\right)\\
        &=Var\left(\sum_i U_i^2 V_{ii}^2+\sum_i \left(\sum_{j\neq i} V_{ij}U_j\right)^2+2\sum_i U_i V_{ii} \sum_{j\neq i} V_{ij} U_j\bigg|U\right)\\
        &=Var\left(\sum_i \left[ \left(\sum_{j\neq i} V_{ij}U_j\right)^2+2 U_i V_{ii} \sum_{j\neq i} V_{ij} U_j \right]\bigg|U\right)\\
        &=\sum_{i,l} Cov\left( \left(\sum_{j\neq i} V_{ij}U_j\right)^2+2 U_i V_{ii} \sum_{j\neq i} V_{ij} U_j,  \left(\sum_{j\neq l} V_{lj}U_j\right)^2+2 U_l V_{ll} \sum_{j\neq l} V_{lj} U_j \bigg| U\right)\\
        &=\sum_{i,l} Cov\left(\left(\sum_{j\neq i} V_{ij}U_j\right)^2, \left(\sum_{j\neq l} V_{lj}U_j\right)^2  \bigg| U\right)\\
        &+4\sum_{i,l}U_i V_{ii}U_lV_{ll} Cov\left(\sum_{j\neq i} V_{ij} U_j,\sum_{j\neq l} V_{lj} U_j \bigg| U\right)\\
        &+4\sum_{i,l}U_i V_{ii}Cov\left(\left(\sum_{j\neq i} V_{ij}U_j\right)^2,\sum_{j\neq l} V_{lj} U_j \bigg| U\right)\\
    \end{align*}
Hence:
\begin{equation*}
    \begin{split}
        Var(||S-L||^2|U)&=\sum_{i,l} \sum_{j_1,j_2\neq i}\sum_{k_1,k_2\neq l}U_{j_1}U_{j_2}U_{k_1}U_{k_2}Cov\left(V_{ij_1}V_{ij_2}, V_{lk_1}V_{lk_2} \right)\\
        &+4\sum_{i,l}\sum_{j\neq i}\sum_{k\neq l}U_i V_{ii}U_lV_{ll}U_j U_k Cov(V_{ij}, V_{lk})\\
        &+4\sum_{i,l}\sum_{j_1,j_2\neq i}\sum_{k\neq l}U_i V_{ii}U_{j_1}U_{j_2}U_{k}Cov\left(V_{ij_1} V_{ij_2} ,V_{lk} \right)\\
        &= 2\sigma_v^4\sum_{i}\sum_{j,k\neq i, k\neq j}U_{j}^2U_{k}^2+Var(V_{ij}^2)\sum_{i}\sum_{j\neq i}U_j^4+Var(V_{ij}^2)\sum_{i}\sum_{j\neq i}U_j^2U_i^2\\
        &+4\sigma_v^2\sum_{i}\sum_{j\neq i} U_i^2 U_j^2 V_{ii}^2+4\sigma_v^2\sum_{i}\sum_{j\neq i} U_i V_{ii}U_jV_{jj}U_j U_i\\
        &+4E(V_{12}^3)\sum_{i}\sum_{j\neq i}U_i^4V_{ii}+4E(V_{12}^3)\sum_{i}\sum_{j\neq i}U_i^2 V_{ii}U_{j}^2\\
        \end{split}
\end{equation*}

so there exists a constant $c_1\geq 0$ such that
$$\frac{Var(||S-L||^2|U)}{N^3}\rightarrow c_1 \mbox{ almost surely.} $$
By Chebychev's inequality:
\begin{equation}\label{Cheby}
    \begin{split}
    \mathbb{P}\left(\bigg|||S-L||^2-E(||S-L||^2|U)\bigg|\geq \sqrt{Var(||S-L||^2|U)}N^{1/3} \right)\leq \frac{1}{N^{2/3}}
    \end{split}
\end{equation}  
By (\ref{Mean}) and (\ref{Cheby}), with high probability:
\begin{equation}\label{Bound2} ||S-L||^2 \leq 2N^2 E(U_1^2) \sigma_v^2 \end{equation} 
Combining (\ref{Bound0}),(\ref{Bound1}) and (\ref{Bound2}), with high probability:
\begin{equation}\label{Bound3}
    ||r||^2\leq \frac{4E(U_1^2)}{\sigma_v^2}
\end{equation}
Now note that:
\begin{equation}
    \begin{split}
        \frac{\sum_i S_i^2}{\sum_i S_i U_i}&=\frac{S'S}{S'U}=\frac{\lambda_1^2 ||v||^2+||\Tilde{A}r||^2}{\lambda_1 ||v||^2+r'\Tilde{A}r}=\lambda_1 + \frac{||\Tilde{A}r||^2-\lambda_1 r'\Tilde{A}r}{\sum_i S_i U_i}
    \end{split}
\end{equation} 
let's now show that $\bigg| \frac{||\Tilde{A}r||^2-\lambda_1 r\Tilde{A}r}{\sum_i S_i U_i}\bigg|=O\left(\frac{1}{\sqrt{N}}\right)$. From (\ref{Bound-1}), w.h.p.:
$$||\Tilde{A}r||^2 \leq 9\sigma_v^2 N ||r||^2$$
then by (\ref{Bound3})
$$||\Tilde{A}r||^2 \leq 9\sigma_v^2 \frac{4E(U_1^2)}{\sigma_v^2} N=36 E(U_1^2) N $$
then:
$$ |r'\Tilde{A}r|\leq ||r||\times  ||\Tilde{A}r|| \leq \frac{2 \sqrt{E(U_1^2)}}{\sigma_v}\times 6 \sqrt{E(U_1^2)} \sqrt{N}=12\frac{E(U_1^2)}{\sigma_v}\sqrt{N}$$
To bound $\lambda_1(\Tilde{A})$, note that $\Tilde{A}v=\lambda_1 v$. So $|\lambda_1(\Tilde{A})| |v_i|=|\sum_{j\neq i} a_{ij}v_j -E(U_1)^2 v_i|\leq \max_j |v_j| (E(U_1)^2+\sum_{j\neq i} |a_{ij}| $. Taking a max over the $i$'s: $|\lambda_1(\Tilde{A})| \max_i|v_i|\leq \max_j |v_j| \times \max_i \sum_{j} |a_{ij}| $, therefore:  $|\lambda_1| \leq E(U_1)^2+ \max_i \sum_{j\neq i} |a_{ij}| $. For any $\eta>0$, Markov's inequality shows that $\max_i \sum_{j} |a_{ij}|=o_p(N^{1+\eta})$ \\
Finally:
$$ \sum_{i} S_i U_i =S'U=(U'U)^2+U'VU=(\sum_i U_i^2)^2+\sum_{i\neq j} U_i U_j V_{ij} +\sum_{i} U_i^2(U_i^2-\sigma_u^2) $$
so, almost surely,
\begin{equation}
\frac{1}{N^2}\sum_i S_i U_i =E(U_1^2)^2+o_p(1)
\end{equation}
implying that:
\begin{equation} \label{lambda_1}
    \begin{split}
     \lambda_1 & = \frac{\sum_i S_i^2}{\sum_i S_i U_i}+o_p\left(1\right)\\
&=\frac{(U'U)^3+2(U'U) U'VU+U'V^2U}{(U'U)^2+U'VU}+o_p\left(1\right)\\
&=U'U+\frac{U'VU}{U'U}+\frac{(U'U)U'V^2U-(U'VU)^2}{(U'U)((U'U)^2+U'VU)}+o_p\left(1\right)   
    \end{split}
\end{equation}
Note that, by the CLT $ U'VU=O_p({N})$, and note that $U'V^2U= O_p(N^2)$ so
$$\lambda_1(\Tilde{A})= U'U+\frac{U'VU}{U'U}+\frac{U'V^2U}{(U'U)^2}+o_p(1)$$
or
$$\lambda_1({A})=\lambda_1(\Tilde{A})-E(U_1)^2= U'U+\frac{U'VU}{U'U}+\frac{U'V^2U}{(U'U)^2}-E(U_1)^2+o_p(1)$$
\end{proof}

\begin{proposition}\label{Prop4}
Fix some vector $\mu_0\in \mathbb{R}^L$. For all $\mu$, denote $M(\mu)$ the matrix:
$$ M(\mu):=X(\mu_0-\mu) +UU'+V-E(U_1^2)I_N$$
where $U$ and $V$ are defined as in Proposition (\ref{Prop2}), and $X$ is a linear function of the vector $(\mu_0-\mu)$: $X=\sum_{l=1}^L (\mu_{0,l}-\mu_l)X_l $, with $L$ a fixed, known number, $X_l$ are symmetric matrices with zeros on the diagonal and such that $\lambda_1(X):=\max_{l=1..L}\lambda_1(X_l)=O_p(N)$.\\
Let $\lambda_1(\mu)>\lambda_2(\mu)...>\lambda_N(\mu)$ be the eignevalues of $M(\mu)$, then:
$$\lambda_1(\Tilde{\mu})=U'U+ \frac{\sum_{k}(\mu_{0,k}-\Tilde{\mu}_k)U'X_kU}{U'U}+O_p(1) $$
Moreover, define $v(\mu)$ and $r(\mu)$ the vectors such that:
\begin{enumerate}
    \item $U=v(\mu)+r(\mu)$
    \item $v(\mu)'r(\mu)=0$
    \item $M(\mu)v(\mu)=\lambda_1(\mu)v(\mu)$
\end{enumerate}
Let $\Tilde{\mu}$ be an estimator for $\mu_0$ such that $||\Tilde{\mu}-\mu_0||=O_p\left(\frac{1}{\sqrt{N}}\right)$.Then
\begin{enumerate}
    \item $$||U-v(\Tilde{\mu}) ||=O_p(1) $$
    \item 
    $$ ||U||^2-||v(\Tilde{\mu})||^2=||r(\Tilde{\mu})||^2= O_p(1
)$$

\item for any $l,l'=1.. K$:
$$v(\Tilde{\mu})'X_{l'}X_l v(\Tilde{\mu})= U' X_{l'}X_l U+ O_p\left(N^2\sqrt{N}\right)$$
\end{enumerate}

\end{proposition}

\begin{proof}
Note that $$ ||M(\tilde{\mu})r(\tilde{\mu})||\leq |\lambda_2(\tilde{\mu})| \times || r(\tilde{\mu})||$$
and:
for all $i=2...N$
$$ \lambda_{i}(M(\tilde{\mu})-UU')\leq \lambda_{i}(M(\tilde{\mu}))\leq  \lambda_{i-1}(M(\tilde{\mu})-UU')$$
and by Weyl's inequalities:
$$ -||\mu_0 -\tilde{\mu}||\times |\lambda_1(X)| +\lambda_i(V-E(U_1^2)I_N)\leq \lambda_{i}(M(\tilde{\mu})-UU')\leq ||\mu_0 -\tilde{\mu}||\times |\lambda_1(X)| +\lambda_i(V-E(U_1^2)I_N)$$
so:$$ |\lambda_2(M(\tilde{\mu})|\leq \max\{\lambda_1(V), |\lambda_N(V)|\}+E(U_1^2)+||\mu_0 -\tilde{\mu}||\times |\lambda_1(X)|$$
by Theorem 2 in \cite{Furedi1981}, almost surely:
$$\max\{ |\lambda_N(V)|, \lambda_1(V)\}=2\sigma_v \sqrt{N}+o\left(N^\frac{1}{2}\right)$$
so:
$$ |\lambda_2(M(\tilde{\mu})|=O_p(\sqrt{N})$$
as in the proof of proposition (\ref{Prop2}), with high probability:
$$ ||M(\tilde{\mu})r(\tilde{\mu})-(U'U)r(\tilde{\mu})||\geq (U'U)||r(\tilde{\mu})||-||M(\tilde{\mu})r(\tilde{\mu})||\geq ((U'U)-|\lambda_2(M(\tilde{\mu}))|)||r(\tilde{\mu})||$$
so with high probability:
\begin{align*}
||r(\tilde{\mu})||^2&\leq \frac{||M(\tilde{\mu})r(\tilde{\mu})-(U'U)r(\tilde{\mu})||^2}{(U'U-|\lambda_2(M(\tilde{\mu}))|)^2}\\
&\leq \frac{||M(\tilde{\mu})U-(U'U)U||^2}{(U'U-|\lambda_2(M(\tilde{\mu}))|)^2} \\
&\leq \frac{\left(||M(\tilde{\mu})U-M(\mu_0)U|| +||M(\mu_0)U- (U'U)U||\right)^2}{(U'U-|\lambda_2(M(\tilde{\mu}))|)^2} \\ 
&= \frac{\left(||\sum_l (\mu_{0,l}-\tilde{\mu}_l)X_l U|| +||M(\mu_0)U- (U'U)U||\right)^2}{(U'U-|\lambda_2(M(\tilde{\mu}))|)^2} \\
&\leq \frac{\left(\sum_l |\mu_{0,l}-\tilde{\mu}_l| \times |\lambda_1(X_l)| \times  ||U|| +||VU-E(U_1^2) U||\right)^2}{(U'U-|\lambda_2(M(\tilde{\mu}))|)^2} \\
&=\frac{\left(\sum_l |\mu_{0,l}-\tilde{\mu}_l| \times |\lambda_1(X_l)| \times  ||U|| +||S-L||+E(U_1^2) ||U||\right)^2}{(U'U-|\lambda_2(M(\tilde{\mu}))|)^2} \\
\end{align*}
where $S$ and $L$ are defined in the proof for equation (\ref{Prop2}). By equation (\ref{Bound2}),
with high probability:
\begin{equation*} ||S-L|| \leq \sqrt{2} N \sqrt{E(U_1^2)} \sigma_v \end{equation*} 
so
$$ ||r(\tilde{\mu})||= O_p(1)$$
which proves the first result:
$$||U-v(\Tilde{\mu}) ||=O_p(1) $$
Also, as in equation (\ref{lambda_1}):
{\small
\begin{equation*}
    \begin{split}
     \lambda_1 (\Tilde{\mu})& = \frac{U'M(\Tilde{\mu})'M(\Tilde{\mu})U}{U'M(\Tilde{\mu})U}+o_p\left(1\right)\\
     &=\frac{\sum_{k=1}^K \sum_{l=1}^K(\mu_{0,l}-\tilde{\mu}_l)(\mu_{0,k}-\tilde{\mu}_l)U'X_k X_l U+(U'U)\sum_{k=1}^K (\mu_{0,k}-\tilde{\mu}_l)U'X_kU+\sum_{k=1}^K (\mu_{0,k}-\tilde{\mu}_l)U'X_kVU}{\sum_{k}(\mu_{0,k}-\Tilde{\mu}_k)U'X_kU+(U'U)^2+ U'VU-E(U_i^2) U'U}\\
     &+\frac{-E(U_1^2)\sum_{k=1}^K (\mu_{0,k}-\tilde{\mu}_l)U'X_kU}{\sum_{k}(\mu_{0,k}-\Tilde{\mu}_k)U'X_kU+(U'U)^2+ U'VU-E(U_i^2) U'U}\\
     &+\frac{(U'U)\sum_{l=1}^K(\mu_{0,l}-\tilde{\mu}_l)U'X_l U+(U'U)^3+(U'U)U'VU-E(U_1^2)(U'U)^2}{\sum_{k}(\mu_{0,k}-\Tilde{\mu}_k)U'X_kU+(U'U)^2+ U'VU-E(U_i^2) U'U}\\
     &+\frac{\sum_{l=1}^K(\mu_{0,l}-\tilde{\mu}_l)U'V X_l U+(U'U)U'VU+U'V^2U-E(U_1^2)U'VU}{\sum_{k}(\mu_{0,k}-\Tilde{\mu}_k)U'X_kU+(U'U)^2+ U'VU-E(U_i^2) U'U}\\
     & +\frac{-E(U_1^2) \sum_{l=1}^K(\mu_{0,l}-\tilde{\mu}_l)U'X_l U-E(U_1^2)(U'U)^2-E(U_1^2)U'VU+E(U_1^2)^2U'U}{\sum_{k}(\mu_{0,k}-\Tilde{\mu}_k)U'X_kU+(U'U)^2+ U'VU-E(U_i^2) U'U}+o_p(1)\\
     &=\frac{(U'U)\left(\sum_{k}(\mu_{0,k}-\Tilde{\mu}_k)U'X_kU+(U'U)^2+ U'VU-E(U_i^2) U'U \right)}{\sum_{k}(\mu_{0,k}-\Tilde{\mu}_k)U'X_kU+(U'U)^2+ U'VU-E(U_i^2) U'U}\\
     &+\frac{\left(\sum_{k}(\mu_{0,k}-\Tilde{\mu}_k)U'X_kU /U'U\right)\left(\sum_{k}(\mu_{0,k}-\Tilde{\mu}_k)U'X_kU+(U'U)^2+ U'VU-E(U_i^2) U'U \right)}{\sum_{k}(\mu_{0,k}-\Tilde{\mu}_k)U'X_kU+(U'U)^2+ U'VU-E(U_i^2) U'U}\\
     &+\frac{\sum_{k=1}^K \sum_{l=1}^K(\mu_{0,l}-\tilde{\mu}_l)(\mu_{0,k}-\tilde{\mu}_l)U'X_k X_l U+(U'U)(U'VU)}{\sum_{k}(\mu_{0,k}-\Tilde{\mu}_k)U'X_kU+(U'U)^2+ U'VU-E(U_i^2) U'U}\\
     &\frac{-E(U_1^2)(U'U)^2-\left(\sum_{k}(\mu_{0,k}-\Tilde{\mu}_k)U'X_kU \right)^2/U'U+U'V^2U}{\sum_{k}(\mu_{0,k}-\Tilde{\mu}_k)U'X_kU+(U'U)^2+ U'VU-E(U_i^2) U'U}+O_p\left(\frac{1}{\sqrt{N}} \right)\\
     &=U'U+ \frac{\sum_{k}(\mu_{0,k}-\Tilde{\mu}_k)U'X_kU}{U'U}\\
     &+\sum_{k=1}^K \sum_{l=1}^K(\mu_{0,l}-\tilde{\mu}_l)(\mu_{0,k}-\tilde{\mu}_l)\frac{U'X_k X_l U}{(U'U)^2}+\frac{U'VU}{U'U} -E(U_1^2)-\frac{\left(\sum_{k}(\mu_{0,k}-\Tilde{\mu}_k)U'X_kU \right)^2}{(U'U)^3}+\frac{U'V^2U}{(U'U)^2}+O_p\left(\frac{1}{\sqrt{N}} \right)
     \end{split}
\end{equation*}
}
as desired.\\

For the third part of the proposition, note that:$$ M(\Tilde{\mu})U=\sum_{l=1}^K(\mu_{0,l}-\tilde{\mu}_l)X_l U+(U'U)U+VU-E(U_1^2)U$$
so
$$M(\Tilde{\mu})U-\lambda_1(M(\Tilde{\mu}))U=VU+(U'U-\lambda_1(M(\tilde{\mu})))U +\sum_{l=1}^K(\mu_{0,l}-\tilde{\mu}_l)X_l U-E(U_1^2)U$$
remember:
$$M(\Tilde{\mu})U=M(\Tilde{\mu})r+\lambda_1(M(\Tilde{\mu})) v$$
hence:
{\small
\begin{equation}\label{MainEq}
\begin{split}
\lambda_1(M(\Tilde{\mu}))(U-v(\Tilde{\mu}))&=M(\Tilde{\mu})r-VU+(\lambda_1(M(\tilde{\mu}))-U'U)U +\sum_{l=1}^L(\tilde{\mu}_l-\mu_{0,l})X_l U+E(U_1^2)U\\
&=M(\Tilde{\mu})r-VU+O_p(1)U +\sum_{l=1}^L(\tilde{\mu}_l-\mu_{0,l})X_l U+E(U_1^2)U- \frac{\sum_{k}(\Tilde{\mu}_k-\mu_{0,k})U'X_kU}{U'U}U
\end{split}
\end{equation}
}
fix some $l$ in $1..L$, multiplying both sides by $\iota'X_l'$:
{\small
\begin{align*}
\lambda_1(M(\Tilde{\mu}))\iota'X_l'(v(\Tilde{\mu})-U)&=-\iota'X_l'M(\Tilde{\mu})r(\Tilde{\mu})+\iota'X_l'VU+(U'U-\lambda_1(M(\tilde{\mu})))\iota'X_l'U\\
&+\sum_{l'=1}^L(\mu_{0,l'}-\tilde{\mu}_{l'})\iota'X_l'X_{l'}U-E(U_i^2)\iota'X_l'U\\
\end{align*} }

The the proposition's second result, remember that $r(\Tilde{\mu})$ and $v(\Tilde{\mu})$ are orthogonal and that $U=v(\Tilde{\mu})+r(\Tilde{\mu})$, so by the Pythagorean theorem:

$$ ||U||^2=||v(\Tilde{\mu})||^2+||r(\Tilde{\mu})||^2 $$

as desired.\\

Finally, remember that
$$\lambda_1(M(\Tilde{\mu}))(v(\Tilde{\mu})-U)=-M(\Tilde{\mu})r+VU+(U'U-\lambda_1(M(\tilde{\mu})))U +\sum_{l=1}^L(\mu_{0,l}-\tilde{\mu}_l)X_l U-E(U_1^2)U=:\Delta$$
so 
$$  X_lv(\Tilde{\mu})=X_lU+ \frac{1}{\lambda_1(M(\Tilde{\mu}))}X_l \Delta$$
and
$$v(\Tilde{\mu})' X_{l'} X_lv(\Tilde{\mu})=U' X_{l'} X_l U+\frac{1}{\lambda_1(M(\Tilde{\mu}))^2}\Delta' X_{l'}X_l \Delta+ \frac{1}{\lambda_1(M(\Tilde{\mu}))}\Delta' X_{l'}X_lU +  \frac{1}{\lambda_1(M(\Tilde{\mu}))}UX_{l'}X_l\Delta'$$
Note that 
$|| \Delta||= O_p(N) $, $ \lambda_1(M(\Tilde{\mu}))= O_p(N)$ and $||X_l \Delta||=O_p(N^2)$ since by assumption:$\lambda_{\max}(X_l), \lambda_{\min}(X_l)=O_p(N)$ (by lemma \ref{eigenvalues of X}), also $$ ||X_l U|| \leq \lambda_{\max}(X_l)||U||=O_p(N\sqrt{N})$$
so
$$v(\Tilde{\mu})' X_{l'} X_lv(\Tilde{\mu})= U' X_{l'} X_l U+ O_p(N^2\sqrt{N})
$$

\end{proof}

\subsection{Proof of Lemma \ref{eigenvalues of X}} \label{ProofLemma1}
Assume $X$ satisfies the lemma's assumptions. Let $\lambda$ be one of $X$'s eigenvalues and let $x$ be a corresponding eigenvector. Then 
\begin{align*}
    |\lambda| ||x||_2&=||\lambda x||_2\\
    &=||Xx||_2\\
    &\leq ||X||_2 ||x||_2
\end{align*}
where $||.||_2$ designates the Euclidean norm for vectors and the spectral norm for matrices. Hence
$$ |\lambda|\leq ||X||_2$$
but we know that the spectral normal is smaller than the Forbenius norm for any matrix. Therefore:
$$ |\lambda| \leq \sqrt{\sum_{i,j} X_{ij}^2}$$
It is left to show that $\sum_{i,j} X_{ij}^2=O_p(N^2)$. 
Decompose $$\sum_{ij}X_{ij}^2=\sum_{ij} E(X_{ij}^2|X_i, X_j)+\sum_{ij}X_{ij}^2-E(X_{ij}^2|X_i, X_j)$$
First, by a U-statistic law of large numbers (e.g. theorem 3.1.3. in \cite{Ustattheory}), $\sum_{ij} E(X_{ij}^2|X_i, X_j)= O_p(N^2)$. For the second term in the decomposition, it is enough to note:
$$ Var\left(\frac{1}{N^2}\sum_{ij}X_{ij}^2  \bigg| (X_i)_{i=1}^\infty \right)\rightarrow 0, \; \mbox{ almost surely.}$$

\subsection{Proof of lemma \ref{Existence}}\label{ExistenceProof}
\begin{proof}

First, note that, given the assumption $E(X_{12}'X_{12})$ is invertible. By a standard law of large numbers, the matrix $\frac{1}{N}\sum_{k=1}^N X_{2k,2k+1}'X_{2k,2k+1}$ converges almost surely to  $E(X_{12}'X_{12})$, then with probability 1, $\frac{1}{N}\sum_{k=1}^N X_{2k,2k+1}'X_{2k,2k+1}$ is invertible for $N$ large enough. Under the condition that $\frac{1}{N}\sum_{k=1}^N X_{2k,2k+1}'X_{2k,2k+1}$ is invertible:

    Write{\small
\begin{equation*}
    \begin{split}
    \arg \min_{\mu \in \mathbb{R}^L}  \sum_{i=2}^N \lambda_i \left(M(\mu)\right)^2 &=arg\min_\mu \sum_{i\neq j} \left(Y_{ij}-X_{ij}\mu \right)^2-\max_{ \nu: ||\nu||=1}  \nu'M(\mu)^2\nu\\
    &=arg\min_\mu \min_{ \nu: ||\nu||=1} \sum_{i\neq j} \left(Y_{ij}-X_{ij}\mu \right)^2- \sum_{i\neq j,k\neq i,j}\nu_i\nu_j\left(Y_{ik}-X_{ik}\mu\right)\left(Y_{kj}-X_{kj}\mu\right)
    \end{split} 
    \end{equation*}}
    For a fixed $\nu$ in the unit sphere, the function that associates each $\mu$ to $\sum_{i\neq j} \left(Y_{ij}-X_{ij}\mu \right)^2- \sum_{i\neq j,k\neq i,j}\nu_i\nu_j\left(Y_{ik}-X_{ik}\mu\right)\left(Y_{kj}-X_{kj}\mu\right)$ is twice continuously differentiable with a Hessian equal to:
    $$ H:=2\left( \sum_{ij} X_{ij}'X_{ij}-\sum_{i\neq j,k\neq i,j}\nu_i \nu_j X_{ik}'X_{jk}\right)$$
    let's show that $H$ is definite positive. Fix $\alpha\neq 0$ in $\mathbb{R}^L$, denote: $x_{ij}:=\sqrt{2}  X_{ij} \alpha$ and $X$ the matrix with entries $x_{ij}$. Because $X$ is symmetric, represent $\nu$ in an orthonormal basis of eigenvector of $X$: $\nu=\sum_{i=1}^N m_i e_i$, where $e_i$ is a normalized eigenvector of $X$.
    Note
    \begin{align*}
        \alpha'H\alpha &= \sum_{ij} x_{ij}^2-\sum_{i\neq j,k\neq i,j}\nu_i \nu_j x_{ik}x_{jk}\\
        &=Trace(X^2)-(X\nu)'(X\nu)+\sum_{i\neq k} \nu_i^2 x_{ik}^2\\
        &=\sum_{i=1}^N \lambda_i(X)^2-\sum_{i=1}^N m_i^2 \lambda_i(X)^2+\sum_{i\neq k} \nu_i^2 x_{ik}^2 \\
        &=\sum_{i=1}^N (1-m_i^2) \lambda_i(X)^2+\sum_{i\neq k} \nu_i^2 x_{ik}^2>0
    \end{align*}
    since $\sum_{i=1}^N (1-m_i^2) \lambda_i(X)^2 =0$ implies that $X$ is of rank at most 1 and $\nu$ is its unique eigenvector (up to a normalization) corresponding to a non null eigenvalue, if $X$ is rank 1. Along with $\nu_i x_{ik}=0$, this implies that $X=0$, so $X_{ij}\alpha=0$ for all $i,j$. Therefore $\alpha' \frac{1}{N}\sum_{k=1}^N X_{2k,2k+1}'X_{2k,2k+1} \alpha=0$ and the matrix $\frac{1}{N}\sum_{k=1}^N X_{2k,2k+1}'X_{2k,2k+1}$ is not invertible; a contradiction. 
    
    This proves that, almost surely, when $N$ is large enough, $H(\nu)$ is definite positive for all $\nu$.\footnote{In fact, we have shown that almost surely, for $N$ large enough, $\min_\nu \lambda_N(H(\nu))>0$.}

    For any fixed $\nu$, the function  $\sum_{i\neq j} \left(Y_{ij}-X_{ij}\mu \right)^2- \sum_{i\neq j,k\neq i,j}\nu_i\nu_j\left(Y_{ik}-X_{ik}\mu\right)\left(Y_{kj}-X_{kj}\mu\right)$ is minimized at $\mu^*(\nu)$ that is continuous in $\nu$. So the problem of minimizing $$ \sum_{i\neq j} \left(Y_{ij}-\sum_{l=1}^L \mu_{l} X_{ij,l}\right)^2- \sum_{i\neq j,k\neq i,j}\nu_i(\hat{\mu}^*) \nu_j(\hat{\mu}^*)\left(Y_{ik}-\sum_{l=1}^L \mu_{l} X_{ik,l}\right)\left(Y_{kj}-\sum_{l=1}^L \mu_{l} X_{kj,l}\right) $$
    on the unit circle admits a solution (minimizing a continuous function on a compact).

    So let $(\mu^*, \nu^*)$ be a minimizer of the function $\sum_{i\neq j} \left(Y_{ij}-\sum_{l=1}^L \mu_{l} X_{ij,l}\right)^2-  \nu'M(\mu)^2\nu$, then:
    $$\mu^* =\arg\min  \left(Y_{ij}-\sum_{l=1}^L \mu_{l} X_{ij,l}\right)^2-  \nu(\mu^*)'M(\mu)^2\nu(\mu^*) $$
    and 
    $$ \nu^*=\arg \max_{||\nu||_2=1} \nu'M(\mu^*)^2\nu$$
    taking a first order condition for $\mu$, we get that $\mu^*$ is a fixed point of $f_N$.

    Conversely, let $\mu^*$ be a fixed point of $f_N$. Then $\mu^*$ satisfies the first order condition for the minimization of the function:
    $$\mu \rightarrow  \sum_{i\neq j} \left(Y_{ij}-X_{ij}\mu \right)^2- \sum_{i\neq j,k\neq i,j}\nu_i(\mu^*)\nu_j(\mu^*)\left(Y_{ik}-X_{ik}\mu\right)\left(Y_{kj}-X_{kj}\mu\right)$$
we have shown that this function is strictly convex with probability approaching 1. Therefore $\mu^*$ is a minimizer, implying that $\mu^*$ minimizes the initial objective function $\mu \rightarrow \sum_{i=2}^N \lambda_i \left(M(\mu)\right)^2$

\end{proof}

\subsection{ Proof of propositions \ref{MainTheorem} and \ref{MainTheorem2}}\label{MainTheoremProof}
\begin{proof}
Note that the function $f_N$ is symmetric as a function of the data, that is $f_N(Y_N, X_N; \mu)=f_N(- Y_N, - X_N; \mu)=f_N(\delta Y_N,\delta X_N; \mu)$ for all $\mu$ and for any sequence of data $(Y_N, X_N)$. Therefore, an iteration using $f_N(Y_N, X_N; .)$ produces the exact same effect as an iteration using the function $f_N(\delta Y_N, \delta X_N; .)$. In other words, given a first stage estimator $\Tilde{\mu}$, the estimator $\hat{\mu}$ is numerically the same whether it is computed on the model
$$ Y_{ij}= X_{ij}\mu_0+\delta U_i U_j +V_{ij}$$
or
$$ (\delta Y_{ij})= (\delta X_{ij})\mu_0+ U_i U_j + \delta  V_{ij}$$

To ease notation, I will  prove the proposition for the case $\delta=1$. The result for any $\delta \in \{-1,1\}$ is easily derived through the previous observation.

First, note that:
\begin{align*}
    (\hat{\mu}-\mu_0) =(1+o_p(1))& \left( \sum_{i\neq j} X_{ij}'X_{ij}-\sum_{j=1}^N\left(\sum_{i=1}^N \nu_i(\Tilde{\mu}) X_{ij}\right)'\left(\sum_{i=1}^N\nu_i(\Tilde{\mu}) X_{ij}\right) \right)^{-1}\\
    &\times \left(\sum_{i\neq j}X_{ij}'\left(Y_{ij}-\sum_{l=1}^K \mu_{0,l} X_{ij,l}\right)-\sum_{i\neq j,k\neq i,j}\nu_i(\Tilde{\mu})\nu_j(\Tilde{\mu})X_{jk}'\left(Y_{ik}-\sum_{l=1}^K \mu_{0,l} X_{ik,l}\right)\right)
\end{align*}

the $(l,l')$ entry of the matrix $\sum_{j=1}^N\left(\sum_{i=1}^N \nu_i(\Tilde{\mu})X_{ij}\right)'\left(\sum_{i=1}^N\nu_i(\Tilde{\mu}) X_{ij}\right) $ is given by:
\begin{align*}
\sum_{i,j,k}\nu_i(\Tilde{\mu})\nu_k (\Tilde{\mu}) X_{ij,l}X_{kj,l'}&=\frac{1}{|| v||^2}\sum_{i,j,k}v_i(\Tilde{\mu})v_k (\Tilde{\mu})X_{ij,l}X_{kj,l'}\\
&= \frac{1}{|| v||^2}\sum_{i,j,k}U_iU_k X_{ij,l}X_{kj,l'}+O_p(N\sqrt{N})
\end{align*}
where the last inequality results from proposition \ref{Prop4}. Observe:
\begin{align*}
\sum_{i,j,k}U_iU_k X_{ij,l}X_{kj,l'}
&=\sum_{ik}(U_i- E(U_1))(U_k- E(U_1)) \sum_{j}  X_{ij,l}X_{kj,l'} \\
&+E(U_1)\sum_{ik}(U_i-E(U_1) )\sum_{j}  X_{ij,l}X_{kj,l'} +E(U_1)\sum_{ik}(U_k-E(U_1) ) \sum_{j}  X_{ij,l}X_{kj,l'}\\
&+E(U_1)^2\sum_{i,j,k}  X_{ij,l}X_{kj,l'}\\
\end{align*}
notice that:
\begin{align*}
Var\left(\sum_{ik}(U_i- E(U_1))(U_k- E(U_1)) \sum_{j}  X_{ij,l}X_{kj,l'}|X\right)&= \sum_{i\neq k}\left( \sum_{j} X_{ij,l}X_{kj,l'} \right)^2 Var(U)^2\\
&+\sum_{i\neq k}\left(\sum_{j} X_{ij,l}X_{ij,l'} \right)\left(\sum_{j} X_{kj,l}X_{kj,l'} \right)Var(U)^2\\
&+\sum_{i\neq k}\left( \sum_{j} X_{ij,l}X_{kj,l'} \right)\left( \sum_{j} X_{kj,l}X_{ij,l'} \right) Var(U)^2\\
&+\sum_i E((U_i-E(U_1))^4)\left( \sum_{j} X_{ij,l}X_{ij,l'} \right)^2\\
&-\left(\sum_{i} \sum_{j} X_{ij,l}X_{ij,l'}\right)^2 Var(U)^2
\end{align*}
so $$ \sum_{ik}(U_i- E(U_1))(U_k- E(U_1)) \sum_{j}  X_{ij,l}X_{kj,l'} =O_p(N^2)$$
likewise:
$$Var\left(\sum_{ik}U_i \sum_{j}  X_{ij,l}X_{kj,l'}|X\right)=\sum_{i}\left(\sum_{j,k}  X_{ij,l}X_{kj,l'} \right)^2 Var(U_1)^2$$
so $$ \sum_{ik}(U_i-E(U_1) ) \sum_{j}  X_{ij,l}X_{kj,l'}=O_p(N^2 \sqrt{N})$$
hence:
\begin{align*}
\sum_{i,j,k}U_iU_k X_{ij,l}X_{kj,l'}&=E(U_1)^2\sum_{i,j,k}  X_{ij,l}X_{kj,l'}+O_p(N^2 \sqrt{N})\\
\end{align*}
Using a central limit theorem
$$ \frac{1}{N^3}\sum_{i,j,k}U_iU_k X_{ij,l}X_{kj,l'}=E(U_1)^2E(X_{12,l}X_{32, l'})+O_p\left(\frac{1}{\sqrt{N}}\right)$$
and, by proposition \ref{Prop4}:
$$||U-v(\Tilde{\mu})||=O_p(1) $$
implying
$$\bigg| ||U||-||v(\Tilde{\mu})|| \bigg|\leq ||U-v(\Tilde{\mu})||= O_p(1) $$
hence
\begin{align*}
    \frac{||v(\Tilde{\mu})||^2}{N}=E(U_1^2)+O_p\left(\frac{1}{\sqrt{N}}\right)
\end{align*}
so

$$ \frac{1}{N^2 ||v(\Tilde{\mu})||^2}\sum_{i,j,k}U_iU_k X_{ij,l}X_{kj,l'}=\frac{E(U_1)^2}{E(U_1^2)}E(X_{12,l}X_{32, l'})+O_p\left(\frac{1}{\sqrt{N}}\right)$$
and
\begin{align*}
 \frac{1}{N^2}\sum_{i,j,k}\nu_i(\Tilde{\mu})\nu_k (\Tilde{\mu}) X_{ij,l}X_{kj,l'}&=\frac{E(U_1)^2}{E(U_1^2)}E(X_{12,l}X_{32, l'})+O_p\left(\frac{1}{\sqrt{N}}\right)
\end{align*}
implying:
\begin{align*}
\frac{1}{N^2}\sum_{j=1}^N\left(\sum_{i=1}^N \nu_i(\Tilde{\mu}) X_{ij}\right)'\left(\sum_{i=1}^N\nu_i(\Tilde{\mu}) X_{ij}\right)=\frac{E(U_1)^2}{E(U_1^2)}E(X_{12}'X_{32})+O_p\left(\frac{1}{\sqrt{N}}\right)
\end{align*}
and $\hat{\mu}-\mu_0$ has the same distribution as 
\begin{align*}
&\left(E(X_{12}'X_{12})- \frac{E(U_1)^2}{E(U_1^2)}E(X_{12}'X_{32})\right)^{-1}\\
&\times \frac{1}{N^2}\left(\sum_{i\neq j}X_{ij}'\left(Y_{ij}-\sum_{l=1}^K \mu_{0,l} X_{ij,l}\right)-\sum_{i\neq j,k\neq i,j}\nu_i(\Tilde{\mu})\nu_j(\Tilde{\mu})X_{jk}'\left(Y_{ik}-\sum_{l=1}^K \mu_{0,l} X_{ik,l}\right)\right)\\
&=\left(E(X_{12}'X_{12})- \frac{E(U_1)^2}{E(U_1^2)}E(X_{12}'X_{32})\right)^{-1}\\
&\times \frac{1}{N^2}\left(\sum_{i\neq j}X_{ij}'(U_iU_j+V_{ij})-\sum_{i\neq j,k\neq i,j}\nu_i(\Tilde{\mu})\nu_j(\Tilde{\mu})X_{jk}' (U_iU_k+V_{ik}) \right)\\
\end{align*}

Now, define:
\begin{equation*}
\begin{split}
\hat{\mu}^*:=\mu_0+&\left(E(X_{12}'X_{12})- \frac{E(U_1)^2}{E(U_1^2)}E(X_{12}'X_{32})\right)^{-1}\\
&\times \frac{1}{N^2}\left(\sum_{i\neq j}X_{ij}'(U_iU_j+V_{ij})-\sum_{i\neq j,k\neq i,j}\frac{U_i}{||U||}\frac{U_j}{||U||}X_{jk}' (U_iU_k+V_{ik}) \right) 
\end{split}
\end{equation*}

The proof proceeds in two steps. First,  find the asymptotic distribution of $N(\hat{\mu}^*-\mu_0)$. Second, determine the asymptotic distribution of:
\begin{align*}
   \hat{\mu}^*-\mu_0-&\left(E(X_{12}'X_{12})-\frac{E(U_1)^2}{E(U_1^2)}E(X_{12}'X_{32})\right)^{-1}\\
&\times \frac{1}{N^2}\left(\sum_{i\neq j}X_{ij}'(U_iU_j+V_{ij})-\sum_{i\neq j,k\neq i,j}\nu_i(\Tilde{\mu})\nu_j(\Tilde{\mu})X_{jk}' (U_iU_k+V_{ik}) \right)\\
\end{align*} 
combining the results of both steps allow to conclude.\\

\textbf{Step 1:} 
I will begin by assuming that $L=1$, then generalize to an arbitrary but known $L$.\\

Let's determine the asymptotic distribution of $\hat{\mu}^*-\mu_0$, that is, of:
$$\sum_{i\neq j}X_{ij}(U_iU_j+V_{ij})-\sum_{i\neq j,k\neq i,j}\frac{U_i}{||U||}\frac{U_j}{||U||}X_{jk} (U_iU_k+V_{ik})  $$
First, note:
\begin{equation}\label{Eq1Step1}
\begin{split}
    \sum_{i\neq j}X_{ij}U_iU_j-\sum_{i\neq j,k\neq i,j}\frac{U_i}{||U||}\frac{U_j}{||U||}X_{jk} U_iU_k &=\sum_{i\neq j}X_{ij}U_iU_j-\sum_{i,j,k\neq i,j}\frac{U_i}{||U||}\frac{U_j}{||U||}X_{jk} U_iU_k+ \sum_{i,k\neq i}\frac{U_i^3}{||U||^2}X_{ik} U_k\\
    &= \sum_{i\neq j}X_{ij}U_iU_j-\sum_{i,j,k\neq j}\frac{U_i}{||U||}\frac{U_j}{||U||}X_{jk} U_iU_k\\
    &+ \sum_{i,k\neq i}\frac{U_i^3}{||U||^2}X_{ik} U_k+\sum_{i,j}\frac{U_i^3}{||U||^2}X_{ij} U_j\\
    &=2 \sum_{i,j}\frac{U_i^3}{||U||^2}X_{ij} U_j\\
    &=N \left(2 \frac{E(U_1^3)E(U_1)}{E(U_1^2)}E(X_{12})+o_p\left(\frac{1}{\sqrt{N}}\right)\right)
    \end{split}
\end{equation}

second:

\begin{align*}
   \sum_{i\neq j}X_{ij}V_{ij}-\sum_{i\neq j,k\neq i,j}\frac{U_i}{||U||}\frac{U_j}{||U||}X_{jk} V_{ik} &= \sum_{i \neq j} V_{ij} \left(X_{ij} -\frac{U_i}{||U||^2}\sum_{k\neq i,j}U_k X_{jk}\right)
\end{align*}
note that
\begin{equation*}
\begin{split}
    Var\bigg( \sum_{i \neq j} V_{ij} & \bigg(X_{ij} -\frac{U_i}{||U||^2}\sum_{k\neq i,j}U_k X_{jk}\bigg)\bigg| U,X\bigg)\\
    &= Var\bigg( \sum_{i < j} V_{ij} \left(2X_{ij} -\frac{U_i}{||U||^2}\sum_{k\neq i,j}U_k X_{jk}-\frac{U_j}{||U||^2}\sum_{k\neq i,j}U_k X_{ik}\right)\bigg| U,X\bigg)\\ &=\sigma_V^2 \sum_{i<j} \left(2X_{ij} -\frac{U_i}{||U||^2}\sum_{k\neq i,j}U_k X_{jk}-\frac{U_j}{||U||^2}\sum_{k\neq i,j}U_k X_{ik}\right)^2\\
    &=\sigma_V^2N^2 \left( 2 E(X_{12}^2)-3 \frac{E(U_1)^2}{E(U_1^2)}E(X_{12}X_{23})+\frac{E(U_1)^4}{E(U_1^2)^2}E(X_{12})^2+o_p(1)\right) \; \mbox{ ; almost surely}
\end{split}
\end{equation*}

by a standard CLT:
\begin{align*}
   &\frac{1}{N}\left(\sum_{i\neq j}X_{ij}V_{ij}-\sum_{i\neq j,k\neq i,j}\frac{U_i}{||U||}\frac{U_j}{||U||}X_{jk} V_{ik} \right)\\
   &\rightarrow_d \mathcal{N}\left(0, \sigma_V^2 \left( 2 E(X_{12}^2)-3 \frac{E(U_1)^2}{E(U_1^2)}E(X_{12}X_{23})+\frac{E(U_1)^4}{E(U_1^2)^2}E(X_{12})^2\right)\right)
\end{align*}
hence:
\begin{align*}
    &\frac{1}{N}\left(\sum_{i\neq j}X_{ij}(U_iU_j+V_{ij})-\sum_{i\neq j,k\neq i,j}\frac{U_i}{||U||}\frac{U_j}{||U||}X_{jk} (U_iU_k+V_{ik})\right)\\
    &\rightarrow_d 2 \frac{E(U_1^3)E(U_1)}{E(U_1^2)}E(X_{12})+\mathcal{N}\left(0, \sigma_V^2 \left( 2 E(X_{12}^2)-3 \frac{E(U_1)^2}{E(U_1^2)}E(X_{12}X_{23})+\frac{E(U_1)^4}{E(U_1^2)^2}E(X_{12})^2\right)\right) \\
\end{align*}
and by the Wold device, for a multivariate $X$: \\
\begin{align*}
    &\frac{1}{N}\left(\sum_{i\neq j}X_{ij}(U_iU_j+V_{ij})-\sum_{i\neq j,k\neq i,j}\frac{U_i}{||U||}\frac{U_j}{||U||}X_{jk} (U_iU_k+V_{ik})\right)\\
    &\rightarrow_d 2 \frac{E(U_1^3)E(U_1)}{E(U_1^2)}E(X_{12})+\mathcal{N}\left(0, \sigma_V^2 \left( 2 E(X_{12}X_{12}')-3 \frac{E(U_1)^2}{E(U_1^2)}E(X_{12}X_{23}')+\frac{E(U_1)^4}{E(U_1^2)^2}E(X_{12})E(X_{12})'\right)\right) \\
\end{align*}
therefore
{\footnotesize
\begin{equation*}
\begin{split}
    N(\hat{\mu}^*-\mu_0) \rightarrow_d & \left(E(X_{12}'X_{12})-\frac{E(U_1)^2}{E(U_1^2)}E(X_{12}'X_{32})\right)^{-1} \times\\
    & \left(2 \frac{E(U_1^3)E(U_1)}{E(U_1^2)}E(X_{12})+\mathcal{N}\left(0, \sigma_V^2 \left( 2 E(X_{12}X_{12}')-3 \frac{E(U_1)^2}{E(U_1^2)}E(X_{12}X_{23}')+\frac{E(U_1)^4}{E(U_1^2)^2}E(X_{12})E(X_{12})'\right)\right)\right) \\ 
\end{split}
\end{equation*}
}
\textbf{Step 2:}

Again, I will use the Wold device. Let $\eta\in \mathbb{R}^L$ and denote $X_{ij, \eta}=\eta X_{ij}'\in \mathbb{R}$ and $X_\eta:=(X_{ij, \eta})_{ij}\in \mathbb{R}^{N\times N}$.\\

Let's determine the asymptotic of 
\begin{equation}\label{Eq_step2}
\begin{split}
    &\sum_{i\neq j,k\neq i,j}\nu_i(\Tilde{\mu})\nu_j(\Tilde{\mu})X_{jk,\eta} (U_iU_k+V_{ik}) - \sum_{i\neq j,k\neq i,j}\frac{U_i}{||U||}\frac{U_j}{||U||}X_{jk,\eta}(U_iU_k+V_{ik})\\
&=\nu(\Tilde{\mu})'X_\eta M(\mu_0)\nu(\Tilde{\mu})- \nu(\Tilde{\mu})'diag\left(XM(\mu_0)\right)\nu(\Tilde{\mu})-\frac{1}{|| U||^2}U'X_\eta M(\mu_0)U+\frac{1}{|| U||^2}U'diag(X_\eta M(\mu_0))U\\
&=\nu(\Tilde{\mu})'X_\eta M(\mu_0)\nu(\Tilde{\mu})-\frac{1}{|| U||^2}U'X_\eta M(\mu_0)U-\left( \nu(\Tilde{\mu})'diag\left(X_\eta M(\mu_0)\right)\nu(\Tilde{\mu})-\frac{U'}{||U||}diag(X_\eta M)\frac{U'}{||U||}\right)
\end{split}
\end{equation}

\begin{itemize}
    \item \underline{Case 1: $E(U_i)\neq 0$}\\
On one side, note:\footnote{Remember that, by definition, $X_{ii,\eta}=0$ for all $i$.}
\begin{align*}
    \bigg|  \nu(\Tilde{\mu})'diag\left(X_\eta M(\mu_0)\right)\nu(\Tilde{\mu})&-\frac{U'}{||U||}diag(X_\eta M)\frac{U'}{||U||} \bigg| \leq \bigg| \bigg| \nu(\Tilde{\mu})-\frac{U}{||U||}\bigg| \bigg|  \max_k \big|\sum_i X_{ik,\eta}(U_i U_k +V_{ik}) \big|\\
    &\leq \bigg| \bigg| \nu(\Tilde{\mu})-\frac{U}{||U||}\bigg| \bigg|  \max_k  \sum_i \left( \big|X_{ik,\eta}(U_i U_k +V_{ik})\big|-E\left( \big|X_{ik,\eta}(U_i U_k +V_{ik})\big|\right)\right)\\
    &+N \bigg| \bigg| \nu(\Tilde{\mu})-\frac{U}{||U||}\bigg| \bigg| E\left( \big|X_{ik.\eta}(U_i U_k +V_{ik})\big|\right)
\end{align*}
I want to show that: $$\max_k  \sum_i \left( \big|X_{ik,\eta}(U_i U_k +V_{ik})\big|-E \big|X_{ik,\eta}(U_i U_k +V_{ik})\big|\right) =O_p\left(N\sqrt{N}\right) $$
Fix some $x>0$ and by a union bound:
\begin{align*}
    & \mathbb{P}\left(\frac{1}{N\sqrt{N}}\max_k  \sum_i \left( \big|X_{ik,\eta}(U_i U_k +V_{ik})\big|-E \big|X_{ik,\eta}(U_i U_k +V_{ik})\big| \right)\geq x \right)\\
    & \leq \sum_k \mathbb{P}\left(\frac{1}{N\sqrt{N}}  \sum_i \left( \big|X_{ik,\eta}(U_i U_k +V_{ik})\big|-E \big|X_{ik,\eta}(U_i U_k +V_{ik})\big| \right)\geq x \right)\\
    &= N \times \mathbb{P}\left(\frac{1}{N\sqrt{N}}  \sum_i \left( \big|X_{i1,\eta}(U_i U_1 +V_{i1})\big|-E \big|X_{i1,\eta}(U_i U_1 +V_{i1})\big|\right)\geq x \right)\\
    & \leq \frac{1}{N^2} \frac{Var\left(\sum_i \left( \big|X_{i1,\eta}(U_i U_1 +V_{i1})\big|-E \big|X_{i1,\eta}(U_i U_1 +V_{i1})\big|\right)\right)}{x^2}\\
    &\leq \frac{1}{x^2} \left( Var\left( \big|X_{12,\eta}(U_2 U_1 +V_{12})\big|\right) + Cov\left( \big|X_{12,\eta}(U_2 U_1 +V_{12})\big|, \big|X_{13,\eta}(U_3 U_1 +V_{13})\big|\right) \right)
\end{align*}
where the second inequality is Markov's. This implies:
$$\max_k  \sum_i \left( \big|X_{ik,\eta}(U_i U_k +V_{ik})\big|-E \big|X_{ik,\eta}(U_i U_k +V_{ik})\big|\right) =O_p\left(N\sqrt{N}\right) $$
as desired. Since:
$$ \bigg| \bigg| \nu(\Tilde{\mu})-\frac{U}{||U||}\bigg| \bigg| \leq \frac{1}{|| U||}\bigg| \bigg| v(\Tilde{\mu})-U \bigg|\bigg|+||v(\Tilde{\mu}) || \bigg| \frac{1}{||v(\Tilde{\mu})||}-\frac{1}{||U||} \bigg|=O_p\left( \frac{1}{\sqrt{N}}\right)$$
then:
\begin{align*}
  \nu(\Tilde{\mu})'diag\left(X_\eta M(\mu_0)\right)\nu(\Tilde{\mu})&-\frac{U'}{||U||}diag(X_\eta M(\mu_0))\frac{U'}{||U||} = O_p(N)
\end{align*}
and equation (\ref{Eq_step2}) becomes:
\begin{equation}\label{Eq2_Step2}
   \begin{split}
   \sum_{i\neq j,k\neq i,j}\nu_i(\Tilde{\mu})\nu_j(\Tilde{\mu})X_{jk,\eta} (U_iU_k+V_{ik}) - &\sum_{i\neq j,k\neq i,j}\frac{U_i}{||U||}\frac{U_j}{||U||}X_{jk,\eta}(U_iU_k+V_{ik})\\
&=\nu(\Tilde{\mu})'X_\eta M(\mu_0)\nu(\Tilde{\mu})-\frac{1}{|| U||^2}U'X_\eta M(\mu_0)U+O_p(N)\\
&=\nu(\Tilde{\mu})'X_\eta UU'\nu(\Tilde{\mu})-\frac{1}{|| U||^2}U'X_\eta UU'U\\
    &\; +\nu(\Tilde{\mu})'XV_\eta \nu(\Tilde{\mu})-\frac{1}{|| U||^2}U'X_\eta VU+O_p(N)\\
&=v(\Tilde{\mu})'X_\eta U-U'X_\eta U +\nu(\Tilde{\mu})'X_\eta V\nu(\Tilde{\mu})-\frac{1}{|| U||^2}U'X_\eta VU+O_p(N)\\
\end{split} 
\end{equation}

    On one side:
    $$ \bigg| \nu(\Tilde{\mu})'X_\eta V\nu(\Tilde{\mu})-\frac{1}{|| U||^2}U'X_\eta VU \bigg| \leq \bigg|\bigg| \nu(\Tilde{\mu})-\frac{U}{|| U||} 
 \bigg|\bigg| || X_\eta V|| =O_p\left( N\right) $$
On the other side:
\begin{align*}
v(\Tilde{\mu})'X_\eta U-U'X_\eta U &=U'X_\eta (v(\Tilde{\mu})-U)\\
&=-\frac{1}{\lambda_1(\Tilde{\mu})}\sum_{l=1}^L(\Tilde{\mu}_l-\mu_{0,l})U'X_\eta X_lU+\sum_{l=1}^L (\Tilde{\mu}_l-\mu_{0,l}) \frac{U'X_l U}{U'U}\frac{U'X_\eta U}{\lambda_1(\Tilde{\mu})}+O_p(N)\\
&=\sum_{l=1}^L (\Tilde{\mu}_l-\mu_{0,l})\left( \frac{U'X_l U}{U'U}\frac{U'X_\eta U}{\lambda_1(\Tilde{\mu})} - \frac{1}{\lambda_1(\Tilde{\mu})} U'X_\eta X_lU\right)+O_p(N)
\end{align*}

where the second equality is a consequence of equation (\ref{MainEq}) and from noting that $U'VX_\eta U=O_p(N^2)$ since, almost surely:
$$ Var\left(U'X_\eta VU \bigg|X, U\right)=\sigma_V^2 \sum_{j,k}U_k^2 \left( \sum_i X_{ij,\eta}U_i\right)^2=O(N^4)$$

Hence:
\begin{align*}
   \frac{\sqrt{N}}{N^2} \bigg(& \sum_{i\neq j,k\neq i,j}\nu_i(\Tilde{\mu})\nu_j(\Tilde{\mu})X_{jk,\eta} (U_iU_k+V_{ik}) - \sum_{i\neq j,k\neq i,j}\frac{U_i}{||U||}\frac{U_j}{||U||}X_{jk,\eta}(U_iU_k+V_{ik}) \bigg)\\
   &=\sqrt{N}\sum_{l=1}^L (\Tilde{\mu}_l-\mu_{0,l})\left( \frac{U'X_l U}{N U'U}\frac{U'X_\eta U}{N\lambda_1(\Tilde{\mu})} - \frac{N}{\lambda_1(\Tilde{\mu})} \frac{1}{N^2}U'X_\eta X_lU\right)+O_p\left(\frac{1}{\sqrt{N}}\right)\\
   &=\sqrt{N}\sum_{l=1}^L (\Tilde{\mu}_l-\mu_{0,l})\left(\frac{E(U_1)^4 }{E(U_1^2)^2} E(X_{12,l})E(X_{12,\eta})-\frac{E(U_1)^2}{E(U_1^2)}E(X_{12,\eta}X_{23,l})\right)+ O_p\left(\frac{1}{\sqrt{N}}\right)\\
\end{align*}
the previous equality holds for any fixed $\eta$, so:
\begin{align*}
   \frac{\sqrt{N}}{N^2} \bigg(& \sum_{i\neq j,k\neq i,j}\nu_i(\Tilde{\mu})\nu_j(\Tilde{\mu})X_{jk} (U_iU_k+V_{ik}) - \sum_{i\neq j,k\neq i,j}\frac{U_i}{||U||}\frac{U_j}{||U||}X_{jk}(U_iU_k+V_{ik}) \bigg)\\
   &=\sqrt{N}\sum_{l=1}^L (\Tilde{\mu}_l-\mu_{0,l})\left(\frac{E(U_1)^4 }{E(U_1^2)^2} E(X_{12,l})E(X_{12})-\frac{E(U_1)^2}{E(U_1^2)}E(X_{12}X_{23,l})\right)+ O_p\left(\frac{1}{\sqrt{N}}\right)\\
   &=\frac{E(U_1)^2}{E(U_1^2)}\sqrt{N}\sum_{l=1}^L (\Tilde{\mu}_l-\mu_{0,l})\left(\frac{E(U_1)^2 }{E(U_1^2)} E(X_{12,l})E(X_{12})-E(X_{12}X_{23,l})\right)+ O_p\left(\frac{1}{\sqrt{N}}\right)\\
   \end{align*}
since, by step 1: $N(\hat{\mu}^*-\mu_0)=O_p(1)$, which allows to conclude:
\begin{equation*}
\begin{split}
    \sqrt{N}(\hat{\mu}-\mu_0)=&\left(  E(X_{12}X_{12}')-\frac{E(U_1)^2}{E(U_1^2)}E(X_{12}X_{23}')\right)^{-1}\\
    &\times \frac{E(U_1)^2}{E(U_1^2)}\sum_{l=1}^L \sqrt{N}(\Tilde{\mu}_l-\mu_{0,l})\left(E(X_{12}X_{23,l})-\frac{E(U_1)^2 }{E(U_1^2)} E(X_{12,l})E(X_{12})\right)+ O_p\left(\frac{1}{\sqrt{N}}\right)\\
    &=\frac{E(U_1)^2}{E(U_1^2)} \left(  E(X_{12}X_{12}')-\frac{E(U_1)^2}{E(U_1^2)}E(X_{12}X_{23}')\right)^{-1}\\
    &\times  \left(E(X_{12}X_{23}')-\frac{E(U_1)^2 }{E(U_1^2)} E(X_{12})E(X_{12}')\right) \sqrt{N}(\Tilde{\mu}-\mu_{0})+ O_p\left(\frac{1}{\sqrt{N}}\right)
    \end{split}
\end{equation*}

\item \underline{Case 2: $E(U_i)=0$}\\

Let's prove the appropriate version of equation (\ref{Eq2_Step2}) for this case. On one side, note
{\small
\begin{align*}
    \bigg|  \nu(\Tilde{\mu})'diag\left(X_\eta M(\mu_0)\right)\nu(\Tilde{\mu})&-\frac{U'}{||U||}diag(X_\eta M)\frac{U'}{||U||} \bigg| \leq \bigg| \bigg| \nu(\Tilde{\mu})-\frac{U}{||U||}\bigg| \bigg|  \max_k \big|\sum_i X_{ik,\eta}(U_i U_k +V_{ik}) \big|\\
    &\leq \bigg| \bigg| \nu(\Tilde{\mu})-\frac{U}{||U||}\bigg| \bigg|  \max_k \big|\sum_i X_{ik,\eta}(U_i U_k +V_{ik}) -E\left( X_{ik,\eta}(U_i U_k +V_{ik})| X_k, U_k  \right) \big|\\
    &+ N \bigg| \bigg| \nu(\Tilde{\mu})-\frac{U}{||U||}\bigg| \bigg| \max_k  \big| E\left( X_{ik,\eta}(U_i U_k +V_{ik})| X_k, U_k  \right) \big| \\
    &\leq \bigg| \bigg| \nu(\Tilde{\mu})-\frac{U}{||U||}\bigg| \bigg|  \max_k \big| \sum_i  X_{ik,\eta}(U_i U_k +V_{ik}) -E\left( X_{ik,\eta}(U_i U_k +V_{ik})| X_k, U_k  \right) \big|
\end{align*}}
because under $E(U_1)=0$, $E\left( X_{ik,\eta}(U_i U_k +V_{ik})| X_k, U_k  \right)=0$. 
I want to show that: $$\max_k  \big| \sum_i \left(  X_{ik,\eta}(U_i U_k +V_{ik}) -E\left( X_{ik,\eta}(U_i U_k +V_{ik})| X_k, U_k  \right)  \right)\big|  =O_p\left(N\right) $$
Fix some $x>0$ and by a union bound:
{\small
\begin{align*}
    & \mathbb{P}\left(\frac{1}{N}\max_k  \big| \sum_i  X_{ik,\eta}(U_i U_k +V_{ik}) -E\left( X_{ik,\eta}(U_i U_k +V_{ik})| X_k, U_k  \right)\big| \geq x \right)\\
    & \leq \sum_k \mathbb{P}\left(\frac{1}{N}  \big|\sum_i  X_{ik,\eta}(U_i U_k +V_{ik}) -E\left( X_{ik,\eta}(U_i U_k +V_{ik})| X_k, U_k  \right) \big| \geq x \right)\\
    &= N \times \mathbb{P}\left(\frac{1}{N} \big| \sum_i  X_{ik,\eta}(U_i U_k +V_{ik}) -E\left( X_{ik,\eta}(U_i U_k +V_{ik})| X_k, U_k  \right)\big| \geq x \right)\\
    & \leq \frac{1}{N} \frac{Var\left(\sum_i \left( X_{ik,\eta}(U_i U_k +V_{ik})-E\left( X_{ik.\eta}(U_i U_k +V_{ik}) \big| X_k, U_k\right) \right)\right)}{x^2}\\
    &= \frac{1}{x^2}  Var\left( X_{12,\eta}(U_2 U_1 +V_{12})-E\left( X_{12.\eta}(U_1 U_2 +V_{12}) \big| X_1, U_1\right)\right)  
\end{align*}}
where the second inequality is Markov's and the last equality results from the fact that for a fixed $k$, the terms {\small $X_{ik,\eta}(U_i U_k +V_{ik})-E\left( X_{ik.\eta}(U_i U_k +V_{ik}) \big| X_k, U_k\right)$} are uncorrelated for different $i$'s, because they are centered and independent conditionally on $X_k, U_k$.

This implies:
$$\max_k  \big| \sum_i \left(  X_{ik,\eta}(U_i U_k +V_{ik}) -E\left( X_{ik,\eta}(U_i U_k +V_{ik})| X_k, U_k  \right)  \right)\big|  =O_p\left(N\right) $$
as desired. Since:
$$ \bigg| \bigg| \nu(\Tilde{\mu})-\frac{U}{||U||}\bigg| \bigg| =O_p\left( \frac{1}{\sqrt{N}}\right)$$
then:
\begin{align*}
  \nu(\Tilde{\mu})'diag\left(X_\eta M(\mu_0)\right)\nu(\Tilde{\mu})&-\frac{U'}{||U||}diag(X_\eta M(\mu_0))\frac{U'}{||U||} = O_p(\sqrt{N})
\end{align*}
and equation (\ref{Eq_step2}) becomes:
{\small
\begin{equation}\label{Eq3_Step2}
   \begin{split}
   \sum_{i\neq j,k\neq i,j}\nu_i(\Tilde{\mu})\nu_j(\Tilde{\mu})X_{jk,\eta} (U_iU_k+V_{ik}) - &\sum_{i\neq j,k\neq i,j}\frac{U_i}{||U||}\frac{U_j}{||U||}X_{jk,\eta}(U_iU_k+V_{ik})\\
&=\nu(\Tilde{\mu})'X_\eta M(\mu_0)\nu(\Tilde{\mu})-\frac{1}{|| U||^2}U'X_\eta M(\mu_0)U+O_p(\sqrt{N})\\
&=\nu(\Tilde{\mu})'X_\eta UU'\nu(\Tilde{\mu})-\frac{1}{|| U||^2}U'X_\eta UU'U\\
&\; +\nu(\Tilde{\mu})'XV_\eta \nu(\Tilde{\mu})-\frac{1}{|| U||^2}U'X_\eta VU+O_p(\sqrt{N})\\
&=v(\Tilde{\mu})'X_\eta U-U'X_\eta U +\nu(\Tilde{\mu})'X_\eta V\nu(\Tilde{\mu})-\frac{1}{|| U||^2}U'X_\eta VU+O_p(\sqrt{N})\\
\end{split} 
\end{equation}}
\newpage

Let's get back to the notation from earlier in step 2: Let $\eta\in \mathbb{R}^L$ and denote $X_{ij, \eta}=\eta X_{ij}'\in \mathbb{R}$ and $X_\eta:=(X_{ij, \eta})_{ij}\in \mathbb{R}^{N\times N}$. From equation (\ref{Eq3_Step2}):
\begin{equation*}
\begin{split}
   \sum_{i\neq j,k\neq i,j}\nu_i(\Tilde{\mu})\nu_j(\Tilde{\mu})X_{jk, \eta} (U_iU_k+V_{ik}) - &\sum_{i\neq j,k\neq i,j}\frac{U_i}{||U||}\frac{U_j}{||U||}X_{jk, \eta}(U_iU_k+V_{ik})\\
&=\nu(\Tilde{\mu})'X_\eta M(\mu_0)\nu(\Tilde{\mu})-\frac{1}{|| U||^2}U'X_\eta M(\mu_0)U+O_p(\sqrt{N})\\
&=\nu(\Tilde{\mu})'X_\eta M(\tilde{\mu})\nu(\Tilde{\mu})-\frac{1}{|| U||^2}U'X_\eta M(\tilde{\mu})U\\
&+\sum_{l=1}^L (\Tilde{\mu}_l-\mu_{0,l})\left(\nu(\tilde{\mu})'X_\eta X_l \nu(\tilde{\mu})-\frac{U'}{||U||}X_\eta X_l \frac{U}{||U||} \right)+ O_p(\sqrt{N})\\
\end{split} 
\end{equation*}
I next show two useful results: for any random matrix $X\in \mathbb{R}^{N\times N}$ such that $X$'s largest eigenvalue is at most of order  $N$ and $X_{ij}:=g(X_i, X_j)$ for some fixed function $g$, then: 1) $||XU||=O_p(N)$, and 2) $||Xv(\Tilde{\mu})||=||XU||+O_p(\sqrt{N})$.\\

Fix such a random matrix $X$, the proof of the two results goes as follows:
\begin{enumerate}
    \item Note that $||XU ||^2=U'X^2U=\sum_{ijk}U_iX_{ij}X_{jk}U_k$, and that:
    \begin{align*}
    Var\left(\sum_{ijk}U_iX_{ij}X_{jk}U_k\bigg| X \right)&=\sum_{i_1,k_1,i_2, k_2}E(U_{i_1}U_{i_2}U_{k_1}U_{k_2})\left(\sum_j X_{i_1j}X_{jk_1}\right)\left(\sum_j X_{i_2j}X_{jk_2}\right)\\
    &=N^4(c+o(1)) ; \mbox{ alomost surely}\\    \end{align*}
    for some real number $c$. Then:
    $$||XU||^2= O_p(N^2)$$
    as desired.
    \item By the equation (\ref{MainEq}):
    \begin{align*}
        \lambda_1(\Tilde{\mu})||XU-Xv(\Tilde{\mu})||\leq ||XM(\Tilde{\mu})r(\Tilde{\mu})||&+||XVU||+|\lambda_1(\Tilde{\mu})-U'U| \times ||XU|| \\
        &+ \sum_{l=1}^L |\Tilde{\mu}_l-\mu_{0,l}| \times||X X_lU|| +E(U_1^2)||XU|| 
    \end{align*}
    let's show that each term in the right hand side is $O_p(N\sqrt{N})$.
    \begin{enumerate}
        \item $||XM(\Tilde{\mu})r(\Tilde{\mu})||\leq \lambda_1(X) ||M(\Tilde{\mu})r(\Tilde{\mu}) ||=O_p(N\sqrt{N})$
        \item for the term $||XVU||$, note that:
        \begin{align*}
        U'VX^2VU&=\sum_{i,j,k,l,m}U_iV_{ij}X_{jk}X_{kl}V_{lm}U_l\\
        &=\sum_{i,j,k,l,m: \{i,j\}\neq \{l,m\}}U_iV_{ij}X_{jk}X_{kl}V_{lm}U_l\\
        &\; +\sum_{i,j,k}U_iV_{ij}X_{jk}X_{ki}V_{ij}U_j+ \sum_{i,j,k}U_iV_{ij}X_{jk}X_{kj}V_{ij}U_i\\
        &=\sum_{i,j,k,l,m, \{i,j\}\neq \{l,m\}}U_iV_{ij}X_{jk}X_{kl}V_{lm}U_l+O_p(N^3)\\
        \end{align*}
        
        almost surely:
    \begin{align*}
    Var\left(\sum_{i,j,k,l,m: \{i,j\}\neq \{l,m\}}U_iV_{ij}X_{jk}X_{kl}V_{lm}U_l\bigg|X, U\right)&= O(N^6)
    \end{align*}
        so $||XVU||^2=U'VX^2VU=O_p(N^3)$, or $||XVU||= O_p(N\sqrt{N}) $
        \item From proposition \ref{Prop4}$$ \lambda(\Tilde{\mu})=U'U+ \frac{\sum_{l=1}^L(\mu_{0,l}-\Tilde{\mu}_l)U'X_lU}{U'U}+O_p(1)$$
        when $E(U_1)=0$, then $\frac{\sum_{l=1}^L(\mu_{0,l}-\Tilde{\mu}_l)U'X_lU}{U'U}=O_p(1)$, $|\lambda(\Tilde{\mu})-U'U|=O_p(1)$, hence
        $$ |\lambda_1(\Tilde{\mu})-U'U| \times ||XU||= O_p(N\sqrt{N}) $$
        \item For every $l=1..L$: $||X X_l U||\leq \lambda_1(X) ||X_lU||=O_p(N^2)$, since $|\Tilde{\mu}_l-\mu_{0,l}|=O_p\left(\frac{1}{\sqrt{N}}\right),$ then $\sum_{l=1}^L |\Tilde{\mu}_l-\mu_{0,l}| \times||X X_lU|| =O_p(N\sqrt{N})$
        \item $E(U_1^2)||XU||= O_p(N)$
    \end{enumerate}
    In conclusion: $$  ||XU||-||Xv(\Tilde{\mu})|| \leq  ||XU -Xv(\Tilde{\mu})||=O_p(\sqrt{N})$$
    
\end{enumerate}
so equation (\ref{Eq2_Step2}) becomes:

\begin{equation*}
   \begin{split}
   \sum_{i\neq j,k\neq i,j}\nu_i(\Tilde{\mu})\nu_j(\Tilde{\mu})X_{jk,\eta} (U_iU_k+V_{ik}) - &\sum_{i\neq j,k\neq i,j}\frac{U_i}{||U||}\frac{U_j}{||U||}X_{jk,\eta}(U_iU_k+V_{ik})\\
&=\nu(\Tilde{\mu})'X_\eta M(\tilde{\mu})\nu(\Tilde{\mu})-\frac{1}{|| U||^2}U'X_\eta M(\tilde{\mu})U+ O_p(\sqrt{N})\\
&=\frac{\lambda_1(\tilde{\mu}) }{||v(\tilde{\mu})||^2}v(\tilde{\mu})'X_\eta v(\tilde{\mu})-\frac{\lambda_1(\tilde{\mu}) }{||U||^2}U'X_\eta v(\tilde{\mu})\\
&-\frac{1}{|| U||^2}U'X_\eta M(\tilde{\mu})r(\tilde{\mu})+ O_p(\sqrt{N})\\
&=-\frac{\lambda_1(\tilde{\mu}) }{||v(\tilde{\mu})||^2}v(\tilde{\mu})'X_\eta r(\tilde{\mu})+\lambda_1(\tilde{\mu}) \left( \frac{1}{||v(\tilde{\mu})||^2}-\frac{1}{||U||^2}\right)U'X_\eta v(\tilde{\mu})\\
&+O_p(\sqrt{N})\\
&=-\frac{\lambda_1(\tilde{\mu}) }{||v(\tilde{\mu})||^2}v(\tilde{\mu})'X_\eta r(\tilde{\mu})+O_p(\sqrt{N})\\
\end{split} 
\end{equation*} 
when $E(U_1)=0$, the equation (\ref{MainEq}) yields:
\begin{equation}
\begin{split}
\lambda_1(M(\Tilde{\mu}))v(\Tilde{\mu})'X_\eta r(\Tilde{\mu})&=v(\Tilde{\mu})'X_\eta M(\Tilde{\mu})r+v(\Tilde{\mu})'X_\eta VU+O_p(1)v(\Tilde{\mu})'X_\eta U \\
&+\sum_{l=1}^L(\tilde{\mu}_l-\mu_{0,l})v(\Tilde{\mu})'X_\eta X_l U+E(U_1^2)v(\Tilde{\mu})'X_\eta U\\
\end{split}
\end{equation}
I want to show that each of the terms in the right hand side is of order $O_p(N\sqrt{N})$:
\begin{enumerate}
    \item $|v(\Tilde{\mu})'X_\eta M(\Tilde{\mu})r|\leq ||v(\Tilde{\mu})'X_\eta || \times || M(\Tilde{\mu})r||=O_p(N)\times O_p(\sqrt{N})$
    \item for $ ||v(\Tilde{\mu})'X_\eta VU||$, first write: $v(\Tilde{\mu})'X_\eta VU=U'X_\eta VU-r(\Tilde{\mu})'X_\eta VU$  and note that  $|r(\Tilde{\mu})'X_\eta VU |\leq ||VU ||\times ||r(\Tilde{\mu})'X_\eta || = O_p(N)\times O_p(\sqrt{N}) $, since $||VU ||\leq \lambda_1(V) ||U|| $ and $||r(\Tilde{\mu})'X_\eta || =O_p(1)$ by the proof in bulletpoint (e) above. So let's examine the term $U'X_\eta VU$:
    \begin{align*}
        Var(U'X_\eta VU|U,X)&=\sigma_V^2\sum_{jk,\eta}\left(\sum_{i} U_iX_{ij, \eta }U_k\right)^2\\
        &=\sigma_V^2 \left( \sum_k U_k^2\right) \sum_j \left(\sum_{i} U_iX_{ij,\eta}\right)^2\\
        &=\sigma_V^2 ||U||^2 ||X_\eta U||^2
    \end{align*}
    so
    \begin{align*}
        Var\left(\frac{U'X_\eta VU}{||X_\eta U||}\bigg|U,X\right)&=\sigma_V^2 ||U||^2\\
        &=N\sigma_V^2 (E(U_1^2)+o(1)); \;\mbox{ almost surely}
    \end{align*}
    and $$ \frac{U'X_\eta VU}{||X_\eta U||}=O_p(\sqrt{N})$$
    since $||X_\eta U||=O_p(N)$, then
    $$ {U'X_\eta VU}=O_p(\sqrt{N})$$
    implying
    $$ ||v(\Tilde{\mu})'X_\eta VU||=O_p(\sqrt{N}) $$
    \item $|v(\Tilde{\mu})'X_\eta U|\leq ||v(\Tilde{\mu})|| \times ||X_\eta U||=O_p(N\sqrt{N}) $
    \item $|(\tilde{\mu}_l-\mu_{0,l})v(\Tilde{\mu})'X_\eta X_l U|\leq |\tilde{\mu}_l-\mu_{0,l}| \times ||v(\Tilde{\mu})'X_\eta ||\times  || X_l U||=O_p(N\sqrt{N}) $
\end{enumerate}
this allows to conclude:
$$\lambda_1(M(\Tilde{\mu}))v(\Tilde{\mu})'X_\eta r(\Tilde{\mu})=O_p(N\sqrt{N}) $$
and finally, the equation (\ref{Eq2_Step2}) becomes:

\begin{equation*}
   \begin{split}
   \sum_{i\neq j,k\neq i,j}\nu_i(\Tilde{\mu})\nu_j(\Tilde{\mu})X_{jk, \eta } (U_iU_k+V_{ik}) - &\sum_{i\neq j,k\neq i,j}\frac{U_i}{||U||}\frac{U_j}{||U||}X_{jk, \eta }(U_iU_k+V_{ik})
=O_p(\sqrt{N})\\
\end{split} 
\end{equation*}
so under the condition $E(U_1)=0$:
$$ N(\hat{\mu}^*-\hat{\mu})=O_p\left(\frac{1}{\sqrt{N}}\right)$$
and
$$N(\hat{\mu}-\mu_0)\rightarrow_d E(X_{12}'X_{12})^{-1} \times \mathcal{N}\left(0, 2 \sigma_V^2 E(X_{12}X_{12}')\right)\\ $$
\end{itemize}

\end{proof}

\subsection{Proof of proposition \ref{Keig}}\label{KeigProof}
\begin{proof}
     The case $E(U_1)=0$ is straightforward, let's prove the proposition for $E(U_1)\neq 0$. Note that:
    \begin{align*}
      &\frac{E(U_1)^2}{E(U_1^2)}\left(  E(X_{12}X_{12}')-\frac{E(U_1)^2}{E(U_1^2)}E(X_{12}X_{23}')\right)^{-1}   \left(E(X_{12}X_{23}')-\frac{E(U_1)^2 }{E(U_1^2)} E(X_{12})E(X_{12}')\right)\\
      &=\left(  E(X_{12}X_{12}')-\frac{E(U_1)^2}{E(U_1^2)}E(X_{12}X_{23}')\right)^{-1}   \left(\frac{E(U_1)^2}{E(U_1^2)}E(X_{12}X_{23}')-\frac{E(U_1)^4 }{E(U_1^2)^2} E(X_{12})E(X_{12}')\right)\\
      &=  \left(  E(U_1^2)^2 E(X_{12}X_{12}')-E(U_1^2)E(U_1)^2 E(X_{12}X_{23}')\right)^{-1}   \left(E(U_1^2)E(U_1) ^2 E(X_{12}X_{23}')-E(U_1)^4 E(X_{12})E(X_{12}')\right)\\
    \end{align*}
    denote:
    \begin{align*}
        A&:= E(U_1^2)E(U_1) ^2E(X_{12}X_{23}')-E(U_1)^4 E(X_{12})E(X_{12}')\\
        B&:=  E(U_1^2) ^2E(X_{12}X_{12}')-E(U_1^2)E(U_1)^2 E(X_{12}X_{23}')
    \end{align*}

    I begin by showing that $B-A$ is semi-definite positive. Denote:
    \begin{align*}
        \Tilde{X}_{12}&:=U_1 \times U_2 \times X_{12}\\
        f(X_1, U_1)&:=E(\Tilde{X}_{12}|X_{1},U_1)
    \end{align*}
    Write:
    \begin{align*}
        B-A&=\left(  E(U_1^2)^2 E(X_{12}X_{12}')-E(U_1^2)E(U_1)^2 E(X_{12}X_{23}')\right)\\
        & -\left(E(U_1^2)^2 E(X_{12}X_{23}')-E(U_1^2)E(U_1)^2 E(X_{12})E(X_{12}')\right)\\
        &= E(U_1X_{12}U_2U_1X_{12}'U_2)-E(U_1X_{12}U_2U_2X_{23}'U_3)\\
        &-E(U_1X_{12}U_2U_2X_{23}'U_3)+E(U_1X_{12}U_2)E(U_1X_{12}'U_2)\\
        &=E(U_1X_{12}U_2U_1X_{12}'U_2)-E(U_1X_{12}U_2U_2X_{24}'U_4)\\
        &-E(U_1X_{13}U_3U_1X_{12}'U_2)+E(U_1X_{12}U_2)E(U_2X_{24}'U_4)\\
        &=E\left(\left(\Tilde{X}_{12}-\Tilde{X}_{13} \right) \left(\Tilde{X}_{12}-\Tilde{X}_{24} \right)' \right)\\
        &=E\left(E\left(\left(\Tilde{X}_{12}-\Tilde{X}_{13} \right) \left(\Tilde{X}_{12}-\Tilde{X}_{24} \right)' \bigg|X_1, U_1, X_2, U_2\right)\right)\\
        &=E\left(\left(\Tilde{X}_{12}-E\left(\Tilde{X}_{13}|X_1, U_1 \right) \right)\left(\Tilde{X}_{12}-E\left(\Tilde{X}_{24} |X_{2}, U_2\right)\right)'\right)\\
        &=E\left(\left(\Tilde{X}_{12}-f(X_1, U_1) \right)\left(\Tilde{X}_{12}-f(X_2,U_2)\right)'\right)\\
        &=E\left(\left(\Tilde{X}_{12}-f(X_1, U_1) -f(X_2, U_2)+E(\Tilde{X}_{12}) \right)\left(\Tilde{X}_{12}-f(X_2,U_2)\right)'\right)\\
        &+E\left(\left(f(X_2, U_2)-E(\Tilde{X}_{12}) \right)\left(\Tilde{X}_{12}-f(X_2,U_2)\right)'\right)\\ &=E\left(\left(\Tilde{X}_{12}-f(X_1, U_1) -f(X_2, U_2)+E(\Tilde{X}_{12}) \right)\left(\Tilde{X}_{12}-f(X_2,U_2)\right)'\right)\\
        &=E\left(\left(\Tilde{X}_{12}-f(X_1, U_1) -f(X_2, U_2)+E(\Tilde{X}_{12}) \right)\left(\Tilde{X}_{12}-f(X_1, U_1) -f(X_2, U_2)+E(\Tilde{X}_{12}) \right)'\right)\\
        &+E\left(\left(\Tilde{X}_{12}-f(X_1, U_1) -f(X_2, U_2)+E(\Tilde{X}_{12}) \right)\left(f(X_1, U_1)-E(\Tilde{X}_{12}) \right)'\right)\\
        &=E\left(\left(\Tilde{X}_{12}-f(X_1, U_1) -f(X_2, U_2)+E(\Tilde{X}_{12}) \right)\left(\Tilde{X}_{12}-f(X_1, U_1) -f(X_2, U_2)+E(\Tilde{X}_{12}) \right)'\right)
    \end{align*}
so $B-A$ is semi definite positive. Moreover, assume there is some deterministic non null vector $\lambda$ such that:
$$\lambda'  (B-A)\lambda =0$$
then, almost surely:
$$ \lambda' \left(\Tilde{X}_{12}-f(X_1, U_1) -f(X_2, U_2)+E(\Tilde{X}_{12}) \right)=0$$
or
$$ \lambda' \Tilde{X}_{12}= \lambda'f(X_1, U_1) + \lambda'f(X_2, U_2)- \lambda' E(\Tilde{X}_{12}) $$
i.e.
$$ U_1U_2 \lambda'X_{12}=U_1 E(U_1)\lambda'E(X_{12}|X_1)+U_2 E(U_1)\lambda'E(X_{12}|X_2)-E(U_1)^2\lambda'E(X_{12})$$

Take an expectation conditional on $U_1$ and $U_2$:
$$ E(\lambda'X_{12})(U_1U_2-U_1 E(U_1)-U_2 E(U_1)+E(U_1)^2)=0$$

either:
\begin{enumerate}
 \item $E(\lambda'X_{12})=0$, then conditioning on $X_2, U_1, U_2$:
    $$ U_1 U_2 E(\lambda'X_{12}|X_2)=U_2 E(U_1)E(\lambda'X_{12}|X_2)$$
   Taking an expectation conditional on $U_2$ and $X_2$:
   $$ U_1 E(U_2)E(\lambda'X_{12}|X_2)=E(U_2)^2E(\lambda'X_{12}|X_2)$$
    or equivalently
    $$ E(\lambda'X_{12}|X_2)E(U_2) (U_1- E(U_1))=0$$
    $E(U_1)=0$ is ruled out by assumption, so either
    \begin{enumerate}
        \item $ U_1- E(U_1)=0 \; \mbox{a.s.}$ so $\sigma_U^2=0$, contradicts the assumptions of proposition \ref{MainTheorem}; or
        \item $ E(\lambda'X_{12}|X_2)=0; \;\mbox{a.s.}$ then $E(U_1U_2 \lambda'X_{12}X_{12}\lambda)=0$ , so $E(U_1)=0$ or $ \lambda'E(X_{12}X_{12}')\lambda=0$, a contradiction (with $E(U_1)\neq$ and $E(X_{12}X_{12}')$ being invertible.)
    \end{enumerate}
    or

    \item $U_1U_2-U_1 E(U_1)-U_2 E(U_1)+E(U_1)^2=0\; a.s. $, or equivalently $(U_1-E(U_1))(U_2-E(U_2))=0 \; a.s.$, by squaring the right hand side and then taking expectations: $\sigma_U^2=0$, which is again a contradiction.    
\end{enumerate}

This allows to conclude that $B-A$ is definite positive.

To prove that all of $B^{-1}A$'s eigenvalues are smaller than 1 in absolute value,  first note that both $A$ and $B$ are definite positive, since:
 \begin{align*}
        A&:= E(\Tilde{X}_{12}\Tilde{X}_{23}')-E(\Tilde{X}_{12})E(\tilde{X}_{12}')\\
        &=E\left((E(\tilde{X}_{12}|X_{1}, U_1)-E(\tilde{X}_{12}))(E(\tilde{X}_{12}|X_{1}, U_1)-E(\tilde{X}_{12}))'\right)\\
        &>0\\
        B&:=  E(\tilde{X}_{12}\tilde{X}_{12}')-E(\tilde{X}_{12}\tilde{X}_{23}')\\
        &=E(\tilde{X}_{12}\tilde{X}_{12}')-E(E(\tilde{X}_{12}|X_2, U_2)E(\tilde{X}_{12}|X_2, U_2)')\\
        &=E\left( \left(\tilde{X}_{12}-E(\tilde{X}_{12}|X_2, U_2) \right)\left(\tilde{X}_{12}-E(\tilde{X}_{12}|X_2, U_2) \right)' \right)\\
        &>0
    \end{align*}
    Let $\lambda$ be an eigenvalue of $B^{-1}A$. There exists some non null vector $x$ such that  $B^{-1}Ax=\lambda x$, so $x'Ax=\lambda x'Bx$, so $\lambda\in (0,1)$.
\end{proof}

\subsection{Proof of proposition 
\ref{Kmatrix}}\label{KmatrixProof}
\begin{proof}
    Write:
    \begin{equation*}
    \begin{split}
        K_N&= \frac{E(U_1)^2}{E(U_1^2)} \left(  E(X_{12}X_{12}')-\frac{E(U_1)^2}{E(U_1^2)}E(X_{12}X_{23}')\right)^{-1}  \left(E(X_{12}X_{23}')-\frac{E(U_1)^2 }{E(U_1^2)} E(X_{12})E(X_{12}')\right)\\
        &=: F\left(E(U_1), E(U_1^2), E(X_{12}X_{12}'), E(X_{12}X_{23}'), E(X_{12}) \right)
    \end{split}
    \end{equation*}
    for a function $F$ that is continuously differentiable at $x:=\left(E(U_1), E(U_1^2), E(X_{12}X_{12}'), E(X_{12}X_{23}'), E(X_{12}) \right)$. For any estimator $x_N$ of $x$:
    $$|F(x_N)-F(x)|\leq ||x_N-x ||\times \bigg|\bigg|\frac{\partial F}{\partial x} (\Bar{x}) \bigg|\bigg|$$
    where $||.||$ is the Euclidean norm and where $\Bar{x} $ is a convex combination of $x_N$ and $x$. So $||x_N-x ||=O_p\left(\frac{1}{\sqrt{N}}\right)$ implies $|F(x_N)-F(x)|=O_p\left(\frac{1}{\sqrt{N}}\right) $. 
    Therefore, it is enough to propose $\sqrt{N}$ consistent estimators for each of the elements $E(U_1),$ $E(U_1^2)$, $E(X_{12}X_{12}')$ ,  $E(X_{12}X_{23}')$ and $E(X_{12})$.

    Clearly, by the standard CLT: $\frac{\sum_{i=1 \leq N/2} X_{2i, 2i+1}X_{2i, 2i+1}'}{N/2}$, $\frac{\sum_{i=1 \leq N/3} X_{3i, 3i+1}X_{3i+1, 3i+2}'}{N/3}$ and $\frac{\sum_{i=1 \leq N/2} X_{2i, 2i+1}}{N/2}$ are $\sqrt{N}$- consistent for  $E(X_{12}X_{12}')$ ,  $E(X_{12}X_{23}')$ and $E(X_{12})$ respectively.
    
    For the parameters $E(U_1)$ and $E(U_1^2)$, lemma \ref{BiasCorrection} shows that the estimators $\frac{\sum_i \hat{U}_i}{N}$ and $\frac{\sum_i \hat{U}_i^2}{N}$ (Cf. lemma \ref{BiasCorrection} for the definitions) are enough for our purposes. 
    
Plugging the five estimators in the function $F$ yields the desired estimator:
 \begin{align*}
        \hat{K}_N:&=\frac{\left(  \sum_i \nu_i(\Tilde{\mu})\right)^2}{N}\left(  \frac{\sum_{i=1 \leq N/2} X_{2i, 2i+1}X_{2i, 2i+1}'}{N/2}-\frac{\left(  \sum_i \nu_i(\Tilde{\mu})\right)^2}{N}\frac{\sum_{i=1 \leq N/3} X_{3i, 3i+1}X_{3i+1, 3i+2}'}{N/3}\right)^{-1} \\
        &\times \left(\frac{\sum_{i=1 \leq N/3} X_{3i, 3i+1}X_{3i+1, 3i+2}'}{N/3}-\frac{\left(  \sum_i \nu_i(\Tilde{\mu})\right)^2}{N} \left( \frac{\sum_{i=1 \leq N/2} X_{2i, 2i+1}}{N/2}\right)\left( \frac{\sum_{i=1 \leq N/2} X_{2i, 2i+1}}{N/2}\right)'\right) \\
    \end{align*}
\end{proof}

\subsection{Proof of corollary \ref{Contraction}} \label{ContractionProof}
\begin{proof}
\begin{enumerate}
     \item The function $\mu \rightarrow |\lambda_{1}(M(\mu)^2)-\lambda_{2}(M(\mu)^2)|$ is continuous on the compact $B({\mu}_0, \frac{C}{\sqrt{N}})$. Let $\mu_N$ be a minimizer on $B({\mu}_0, \frac{C}{\sqrt{N}})$.  We show in the proof of proposition \ref{Prop4}  that $\lambda_{1}(M(\mu_N)^2)=O_p(N^2)$ and $\lambda_{2}(M(\mu_N)^2)=O_p({N}) $. So $|\lambda_{1}(M(\mu_N)^2)-\lambda_{2}(M(\mu_N)^2)|=O_p({N}^2) $. So with probability approaching 1, the largest eigenvalue of $M(\mu_N)$ in absolute value is simple on all of $B({\mu}_0, \frac{C}{\sqrt{N}})$. Theorem 1 in \cite{Magnus1985} allows to conclude that $\mu \rightarrow\nu(\mu)$ is infinitely continuously differentiable on  $B({\mu}_0, \frac{C}{\sqrt{N}})$. The proof of corollary \ref{Existence} (section \ref{ExistenceProof})
     \item Following 1), assume $f_N$ is continuously differentiable on $B({\mu}_0, \frac{2C}{\sqrt{N}})$. Let $ \mu_{\max}$ be a minimizer of $||f_N'(\mu)||$ on $B({\mu}_0, \frac{C}{\sqrt{N}})$. By equation (\ref{SingleIteration}):
     $\sqrt{N}(f(\mu_{\max})-\mu_0)=K\sqrt{N}(\mu_{\max}-\mu_0)+O_p\left( \frac{1}{\sqrt{N}}\right)$\\
     also
     $$\sqrt{N}(f(\mu_{\max}+\frac{1}{\sqrt{N}})-\mu_0)=K\sqrt{N}(\mu_{\max}+\frac{1}{\sqrt{N}}-\mu_0)+O_p\left( \frac{1}{\sqrt{N}}\right) $$
     taking the difference of the two last equations:
     $$\sqrt{N}\left(f\left(\mu_{\max}+\frac{1}{\sqrt{N}}\right)-f(\mu_{\max})\right)=K+O_p\left( \frac{1}{\sqrt{N}}\right) $$
     on the other side, by a Taylor expansion:
     $$ \sqrt{N}\left(f\left(\mu_{\max}+\frac{1}{\sqrt{N}}\right)-f(\mu_{\max})\right)=f'(\mu_{\max})+o_p\left( 1\right) $$
     hence
     $$f'(\mu_{\max})-K=o_p(1) $$
     with  a probability approaching 1:
     $$ ||f'(\mu_{\max})||=\sup_{\mu\in B({\mu}_0, \frac{C}{\sqrt{N}})}||f'(\mu)|| \leq \kappa $$
     for any $\kappa \in(\lambda_1(K), 1) $.
     \item Fix some $\kappa \in(\lambda_1(K), 1) $ and $\epsilon>0$. There exists $M>0$ such that for $N$ large enough, with probability at least $1-\epsilon$, $\hat{\mu}_1, \hat{\mu}_0 \in B({\mu}_0, \frac{M}{2\sqrt{N}})$ so that $|| \hat{\mu}_{1}- \hat{\mu}_0||\leq \frac{M}{\sqrt{N}}$ (let this be event $E_N$) . Assume $f_N$ is continuously differentiable on $B\left({\mu}_0, \frac{M}{\sqrt{N}}\left(1+\frac{1}{1-\kappa}\right)\right)$ (denote this event $F_N$) and that   $\sup_{\mu\in B\left({\mu}_0, \frac{M}{\sqrt{N}}\left(1+\frac{1}{1-\kappa}\right)\right)}||f'(\mu)|| \leq \kappa $ (let this be event $G_N$). Then for any $\mu, \mu' \in B({\mu}_0, \frac{M}{\sqrt{N}}(1+\frac{1}{1-\kappa})) $,   we have:
     $$||f_N(\mu)-f_N(\mu')||\leq \kappa ||\mu-\mu'|| $$\\
     By induction on $m$, assume $\hat{\mu}_0, ..., \hat{\mu}_m \in B\left({\mu}_0, \frac{M}{\sqrt{N}} \left(1+\frac{1}{1-\kappa}\right)\right)$
    $$ || \hat{\mu}_{m+1}-\hat{\mu}_m ||=|| f_N(\hat{\mu}_{m})-f_N(\hat{\mu}_{m-1}) ||\leq \kappa || \hat{\mu}_m- \hat{\mu}_{m-1}||$$
    and \begin{align*}
        || \hat{\mu}_{m+1}-\mu_0 ||&\leq  || \hat{\mu}_{m+1}- \hat{\mu}_m||+||\hat{\mu}_{m}- \hat{\mu}_{m-1}||+...+|| \hat{\mu}_{1}- \hat{\mu}_0||+||\hat{\mu}_{0}- {\mu}_{0}||\\
        &\leq \sum_{i=0}^m \kappa ^i || \hat{\mu}_{1}- \hat{\mu}_0|| +||\hat{\mu}_{0}- {\mu}_{0}||\\
        &\leq \frac{M}{\sqrt{N}} ( 1+\sum_{i=0}^m\kappa^i) \\
        &\leq \frac{M}{\sqrt{N}} \left( 1+\frac{1}{1-\kappa}\right)
    \end{align*} 
     so $\hat{\mu}_{m+1} \in B\left({\mu}_0, \frac{M}{\sqrt{N}}\left(1+\frac{1}{1-\kappa} \right)\right) $.\\
     So even though $f_N$ is not necessarily contracting on $B\left({\mu}_0, \frac{M}{\sqrt{N}}\left(1+\frac{1}{1-\kappa} \right)\right) $, because it may not preserve $B\left({\mu}_0, \frac{M}{\sqrt{N}}\left(1+\frac{1}{1-\kappa} \right)\right)$, we can follow the proof of the Banach  fixed point theorem for the specific sequence $\hat{\mu}$  (or in fact any sequence initiated in a way that the first two first elements are in $B\left({\mu}_0, \frac{M}{2\sqrt{N}}\right)$ and not just $B\left({\mu}_0, \frac{M}{\sqrt{N}}\left(1+\frac{1}{1-\kappa}\right)\right)$). First we show that the sequence $\hat{\mu}_m$ is a Cauchy sequence, let $p, q \in \mathbb{N}$, without loss of generality take $p>q$
    \begin{align*}
         ||\hat{\mu}_p-\hat{\mu}_q||\leq \frac{\kappa^q}{1-\kappa}||\hat{\mu}_1-\hat{\mu}_0||\rightarrow 0, \mbox{ as } q\rightarrow + \infty
     \end{align*} 
     so the sequence is a Cauchy sequence. Therefore, it has a limit in $B\left({\mu}_0, \frac{M}{\sqrt{N}}\left(1+\frac{1}{1-\kappa} \right)\right)$ that  can only be a fixed point of $f_N$. By lemma \ref{Existence}, the sequence converges to a minimizer.
     We have shown the following:
     $$ E_N, F_N \mbox{ and } G_N \Rightarrow  \mbox{The sequence converges to a minimizer}$$
     Which proves that with probability approching 1, the the sequence converges to a minimizer as desired.
 \end{enumerate}
 Finally, the last result along with lemma \ref{Existence} ensure that $\hat{\mu}^*$ is  a solution to the minimization problem \ref{Oracle} and is $\sqrt{N}$-consistent. Equation (\ref{SingleIteration}) yields
 $$(I-K)\sqrt{N}(\hat{\mu}^*-\mu_0)=O_p\left(\frac{1}{\sqrt{N}}\right) $$
 and finally
  $$(\hat{\mu}^*-\mu_0)=O_p\left(\frac{1}{\sqrt{N}}\right) $$
 \end{proof}

\subsection{Proof of proposition \ref{N rate version}} \label{Proof N rate version}
\begin{proof}

As for the proof of propositions \ref{MainTheorem} (section \ref{MainTheoremProof}), assume that $\delta=1$. The result for an unknown $\delta\in  \{-1, 1\}$ immediately follows  as described in  the the proof section \ref{MainTheoremProof} .

Again, I use the Wold device. Let $\eta\in \mathbb{R}^L$ and denote $X_{ij, \eta}=\eta X_{ij}'\in \mathbb{R}$ and $X_\eta:=(X_{ij, \eta})_{ij}\in \mathbb{R}^{N\times N}$.

Following equation (\ref{Eq2_Step2}):
\begin{align*}
    &\sum_{i\neq j}X_{ij,\eta}(U_iU_j+V_{ij})-\sum_{i\neq j,k\neq i,j}\nu_i(\Tilde{\mu})\nu_j(\Tilde{\mu})X_{jk,\eta}(U_iU_k+V_{ik})\\
    & = \sum_{i\neq j}X_{ij, \eta}(U_iU_j+V_{ij})-\sum_{i\neq j,k\neq i,j}\frac{U_i}{||U||}\frac{U_j}{||U||}X_{jk, \eta} U_iU_k \\
    &+\sum_{i\neq j,k\neq i,j}\frac{U_i}{||U||}\frac{U_j}{||U||}X_{jk, \eta} U_iU_k-\sum_{i\neq j,k\neq i,j}\nu_i(\Tilde{\mu})\nu_j(\Tilde{\mu})X_{jk, \eta}(U_iU_k+V_{ik})\\
    &= \sum_{i\neq j}X_{ij}(U_iU_j+V_{ij})-\sum_{i\neq j,k\neq i,j}\frac{U_i}{||U||}\frac{U_j}{||U||}X_{jk}( U_iU_k+V_{ik}) \\
    &+U'X(U-v(\tilde{\mu}))-\nu(\tilde{\mu})'X_\eta V\nu(\tilde{\mu})+\frac{1}{||U||^2} U' X_\eta VU+O_p(\sqrt{N})\\
    &=N \left(2 \frac{E(U_1^3)E(U_1)}{E(U_1^2)}E(X_{12,\eta})\right)+\sum_{i\neq j}X_{ij}V_{ij}-\sum_{i\neq j,k\neq i,j}\frac{U_i}{||U||}\frac{U_j}{||U||}X_{jk, \eta}V_{ik} \\
    &+U'X_\eta r(\tilde{\mu})+\frac{1}{||v(\tilde{\mu}))||^2}\left( U'X_\eta VU-v (\tilde{\mu})'X_\eta V v(\tilde{\mu}))\right)+U' X_\eta VU \frac{||v(\tilde{\mu})||^2-||U||^2}{||v(\tilde{\mu})||^2||U||^2}+O_p(\sqrt{N})
\end{align*}
where the second equality results from equation (\ref{Eq2_Step2}) and the third from equation (\ref{Eq1Step1}).

Note that 
\begin{enumerate}
    \item $$U'X_\eta VU-v (\tilde{\mu})'X_\eta Vv(\tilde{\mu})=- \frac{1}{\lambda_1 (\tilde{\mu})} U'X_\eta  V^2U+O_p(N\sqrt{N})$$
    to see that, observe that from equation (\ref{MainEq}):
    \begin{align*}
        Vv(\tilde{\mu})&=VU - \frac{1}{\lambda_1 (\tilde{\mu})}\bigg( VM(\Tilde{\mu})r(\tilde{\mu})-V^2U+O_p(1)VU +\sum_{l=1}^L(\tilde{\mu}_l-\mu_{0,l})VX_l U+E(U_1^2)VU\\
        &- \frac{\sum_{k}(\Tilde{\mu}_k-\mu_{0,k})U'X_kU}{U'U}VU\bigg)\\
        X_\eta v(\tilde{\mu})&=X_\eta U - \frac{1}{\lambda_1 (\tilde{\mu})}\bigg( X_\eta M(\Tilde{\mu})r(\tilde{\mu})-X_\eta VU+O_p(1)X_\eta U +\sum_{l=1}^L(\tilde{\mu}_l-\mu_{0,l})X_\eta X_l U+E(U_1^2)X_\eta U\\
        &- \frac{\sum_{k}(\Tilde{\mu}_k-\mu_{0,k})U'X_kU}{U'U}X_\eta U\bigg)
    \end{align*}
    combining both identities:
    \begin{align*}
        v(\tilde{\mu})'X_\eta Vv(\tilde{\mu})&=U'X_\eta VU- \frac{1}{\lambda_1 (\tilde{\mu})} U'X_\eta  \bigg( VM(\Tilde{\mu})r(\tilde{\mu})-V^2U+O_p(1)VU \\
        &+\sum_{l=1}^L(\tilde{\mu}_l-\mu_{0,l})VX_l U+E(U_1^2)VU - \frac{\sum_{k}(\Tilde{\mu}_k-\mu_{0,k})U'X_kU}{U'U}VU\bigg)\\
        &-\frac{1}{\lambda_1 (\tilde{\mu})} U'V \bigg( X_\eta M(\Tilde{\mu})r(\tilde{\mu})-X_\eta VU+O_p(1)X_\eta U \\
        & +\sum_{l=1}^L(\tilde{\mu}_l-\mu_{0,l})X_\eta X_l U+E(U_1^2)X_\eta U - \frac{\sum_{k}(\Tilde{\mu}_k-\mu_{0,k})U'X_kU}{U'U}X_\eta U\bigg)\\
        &+\frac{1}{\lambda_1^2 (\tilde{\mu})}\bigg( X_\eta M(\Tilde{\mu})r(\tilde{\mu})-X_\eta VU+O_p(1)X_\eta U \\
        & +\sum_{l=1}^L(\tilde{\mu}_l-\mu_{0,l})X_\eta X_l U+E(U_1^2)X_\eta U - \frac{\sum_{k}(\Tilde{\mu}_k-\mu_{0,k})U'X_kU}{U'U}X_\eta U\bigg)'\\
        & \times \bigg( VM(\Tilde{\mu})r(\tilde{\mu})-V^2U+O_p(1)VU      +\sum_{l=1}^L(\tilde{\mu}_l-\mu_{0,l})VX_l U+E(U_1^2)VU \\
        &- \frac{\sum_{k}(\Tilde{\mu}_k-\mu_{0,k})U'X_kU}{U'U}VU\bigg)\\
        &=U'X_\eta VU+ \frac{1}{\lambda_1 (\tilde{\mu})} U'X_\eta  V^2U+O_p(N\sqrt{N})
    \end{align*}
    \item Remark that $Var(U'X_\eta VU| X, U)=\sigma_V^2 \sum_{jk}\left(\sum_i U_iX_{ij}U_k \right)^2=O(N^4) $ almost surely, hence $$U'X_\eta VU=O_p(N^2) $$
\end{enumerate}
so:
\begin{align*}
    &\sum_{i\neq j}X_{ij, \eta}(U_iU_j+V_{ij})-\sum_{i\neq j,k\neq i,j}\nu_i(\Tilde{\mu})\nu_j(\Tilde{\mu})X_{jk, \eta}(U_iU_k+V_{ik})\\
    &=N \left(2 \frac{E(U_1^3)E(U_1)}{E(U_1^2)}E(X_{12,\eta})\right)+\sum_{i\neq j}X_{ij, \eta }V_{ij}-\sum_{i\neq j,k\neq i,j}\frac{U_i}{||U||}\frac{U_j}{||U||}X_{jk, \eta}V_{ik} \\
    &+U'X_\eta r(\tilde{\mu})-\frac{1}{||v(\tilde{\mu})||^2\lambda_1 (\tilde{\mu})} U'X_\eta  V^2U+O_p(\sqrt{N})
\end{align*}

Let's determine the asymptotic distribution of $U'X_
\eta r(\tilde{\mu})$. From the equation (\ref{MainEq})
{\small
\begin{equation*}
\begin{split}
\lambda_1(M(\Tilde{\mu}))U'X_\eta r(\Tilde{\mu})&=U'X_\eta M(\Tilde{\mu})r-U'X_\eta VU+(\lambda_1(M(\Tilde{\mu}))-U'U)U'X_\eta U +\sum_{l=1}^L(\tilde{\mu}_l-\mu_{0,l})U'X_\eta X_l U+E(U_1^2)U'X_\eta U
\end{split}
\end{equation*}
}
Also
{\small
\begin{equation*}
\begin{split}
\lambda_1(M(\Tilde{\mu}))U'X_\eta M(\Tilde{\mu})r(\Tilde{\mu})&=U'X_\eta M(\Tilde{\mu})^2r-U'X_\eta M(\Tilde{\mu})VU+(\lambda_1(M(\Tilde{\mu}))-U'U)U'X_\eta M(\Tilde{\mu})U \\
&+\sum_{l=1}^L(\tilde{\mu}_l-\mu_{0,l})U'X_\eta M(\Tilde{\mu})X_l U+E(U_1^2)U'X_\eta M(\Tilde{\mu})U\\
&=-U'X_\eta UU'VU-U'XV^2U-\sum_{l=1}^L(\mu_{0,l}-\title{\mu}_l)U'X_\eta X_l VU+(\lambda_1(M(\Tilde{\mu}))-U'U)U'X_\eta UU'U\\
&+(\lambda_1(M(\Tilde{\mu}))-U'U)U'X_\eta VU+(\lambda_1(M(\Tilde{\mu}))-U'U)\sum_{l=1}^L (\mu_0-\tilde{\mu})U'X_\eta X_l U\\
&+\sum_{l=1}^L(\tilde{\mu}_l-\mu_{0,l})U'X_\eta UU'X_l U+\sum_{l=1}^L(\tilde{\mu}_l-\mu_{0,l})U'X_\eta VX_l U\\
&+\sum_{k=1}^L(\mu_{0,k}-\Tilde{\mu}_k)\sum_{l=1}^L(\tilde{\mu}_l-\mu_{0,l})U'X_\eta X_k X_l U+E(U_1^2)U'X_\eta UU'U+E(U_1^2)U'X_\eta VU\\
&+E(U_1^2)\sum_{l=1}^L(\mu_{0,l}-\Tilde{\mu}_l)U'X_\eta X_l U+O_p(N^2\sqrt{N})\\
&=-U'X_\eta UU'VU-U'X_\eta V^2U+(\lambda_1(M(\Tilde{\mu}))-U'U)U'X_\eta UU'U\\
&+\sum_{l=1}^L(\tilde{\mu}_l-\mu_{0,l})U'X_\eta UU'X_l U+E(U_1^2)U'X_\eta UU'U+O_p(N^2\sqrt{N})\\
\end{split}
\end{equation*}
}
By proposition \ref{Prop4}, when $\Tilde{\mu}-\mu_0=\left( \frac{1}{N}\right)$ as we are assuming here, we get:
$$ \lambda_1(\Tilde{\mu})=U'U+ \frac{\sum_{k}(\mu_{0,k}-\Tilde{\mu}_k)U'X_kU}{U'U} -E(U_1^2)+\frac{U'V^2U}{(U'U)^2}+O_p\left(\frac{1}{\sqrt{N}} \right)$$
therefore:
{\small
\begin{equation*}
\begin{split}
\lambda_1(M(\Tilde{\mu}))U'X_\eta M(\Tilde{\mu})r(\Tilde{\mu})&=-U'X_\eta UU'VU-U'X_\eta V^2U+\frac{U'V^2U U'X_\eta U}{U'U}+O_p(N^2\sqrt{N})\\
\end{split}
\end{equation*}
}
plugging back in the expansion of $U'X_\eta r(\Tilde{\mu})$:
\begin{align*}
\lambda_1(M(\Tilde{\mu}))U'X_\eta r(\Tilde{\mu})&=-\frac{1}{\lambda_1(\Tilde{\mu})}U'X_\eta UU'VU-\frac{1}{\lambda_1(\Tilde{\mu})}U'X_\eta V^2U+\frac{U'V^2U U'X_\eta U}{\lambda_1(\Tilde{\mu}) U'U}-U'X_\eta VU\\
&+(\lambda_1(M(\Tilde{\mu}))-U'U)U'X_\eta U +\sum_{l=1}^L(\tilde{\mu}_l-\mu_{0,l})U'X_\eta X_l U+E(U_1^2)U'X_\eta U+O_p(N\sqrt{N})\\
&=-\frac{1}{\lambda_1(\Tilde{\mu})}U'X_\eta UU'VU-\frac{1}{\lambda_1(\Tilde{\mu})}U'X_\eta V^2U+\frac{U'V^2U U'X_\eta U}{\lambda_1(\Tilde{\mu}) U'U}-U'X_\eta VU\\
&+\frac{U'V^2U U'X_\eta U}{(U'U)^2} -\frac{\sum_{k}(\Tilde{\mu}_k-\mu_{0,k})U'X_kUU'X_\eta U}{U'U}+\sum_{l=1}^L(\tilde{\mu}_l-\mu_{0,l})U'X_\eta X_l U+O_p(N\sqrt{N})
\end{align*}
so
\begin{align*}
U'X_\eta r(\Tilde{\mu})&=-\frac{1}{(U'U)^2}U'X_\eta UU'VU-\frac{1}{(U'U)^2}U'X_\eta V^2U+2\frac{U'V^2U U'X_\eta U}{(U'U)^3}-\frac{1}{U'U}U'X_\eta VU\\
& -\frac{\sum_{k}(\Tilde{\mu}_k-\mu_{0,k})U'X_kUU'X_\eta U}{(U'U)^2}+\frac{1}{U'U}\sum_{l=1}^L(\tilde{\mu}_l-\mu_{0,l})U'X_\eta X_l U+O_p(\sqrt{N})
\end{align*}
plugging:
\begin{align*}
    &\sum_{i\neq j}X_{ij, \eta }(U_iU_j+V_{ij})-\sum_{i\neq j,k\neq i,j}\nu_i(\Tilde{\mu})\nu_j(\Tilde{\mu})X_{jk, \eta }(U_iU_k+V_{ik})\\
    &=N \left(2 \frac{E(U_1^3)E(U_1)}{E(U_1^2)}E(X_{12, \eta})\right)+\sum_{i\neq j}X_{ij, \eta}V_{ij}-\sum_{i\neq j,k\neq i,j}\frac{U_i}{||U||}\frac{U_j}{||U||}X_{jk, \eta}V_{ik} \\
    &-\frac{1}{(U'U)^2}U'X_\eta UU'VU-\frac{1}{(U'U)^2}U'X_\eta V^2U+2\frac{U'V^2U U'X_\eta U}{(U'U)^3}-\frac{1}{U'U}U'X_\eta VU\\
& -\frac{\sum_{k}(\Tilde{\mu}_k-\mu_{0,k})U'X_kUU'X_\eta U}{(U'U)^2}+\frac{1}{U'U}\sum_{l=1}^L(\tilde{\mu}_l-\mu_{0,l})U'X_\eta X_l U-\frac{1}{||v(\tilde{\mu})||^2\lambda_1 (\tilde{\mu})} U'X_\eta  V^2U+O_p(\sqrt{N})
\end{align*}
Notice that
$$U'V^2U =\sum_{i,j}U_i^2 V_{ij}^2+O_p(N\sqrt{N}) $$
 and
 $$U'XV^2U=\sum_{i,j,k}U_i X_{ij}V_{jk}^2U_j+ O_p(N^2\sqrt{N})$$
so that
$$\frac{1}{||v(\tilde{\mu})||^2\lambda_1 (\tilde{\mu})} U'X_\eta  V^2U- \frac{1}{(U'U)^2}U'X_\eta V^2U =O_p\left ( \frac{1}{\sqrt{N}}\right)$$
subsequently
\begin{align*}
    &\sum_{i\neq j}X_{ij, \eta }(U_iU_j+V_{ij})-\sum_{i\neq j,k\neq i,j}\nu_i(\Tilde{\mu})\nu_j(\Tilde{\mu})X_{jk, _\eta }(U_iU_k+V_{ik})\\
    &=N \left(2 \frac{E(U_1^3)E(U_1)}{E(U_1^2)}E(X_{12, \eta })\right)+\sum_{i\neq j}X_{ij, \eta}V_{ij}-\sum_{i\neq j,k\neq i,j}\frac{U_i}{||U||}\frac{U_j}{||U||}X_{jk, \eta}V_{ik} \\
    &-\frac{E(U_1)^2}{E(U_1^2)^2}E(X_{12, \eta})U'VU-\frac{1}{U'U}U'X_\eta VU\\
& -\frac{\sum_{k}(\Tilde{\mu}_k-\mu_{0,k})U'X_kUU'X_\eta U}{(U'U)^2}+\frac{1}{U'U}\sum_{l=1}^L(\tilde{\mu}_l-\mu_{0,l})U'X_\eta X_l U+O_p(\sqrt{N})\\
&=\sum_{l=1}^L(\tilde{\mu}_l-\mu_{0,l})\left(\frac{U'X_\eta X_l U}{U'U}-\frac{U'X_lU}{U'U}\frac{U'X_\eta U}{U'U}\right)\\
&+2N \frac{E(U_1^3)E(U_1)}{E(U_1^2)}E(X_{12, \eta })\\
&+\sum_{i\neq j}X_{ij, \eta }V_{ij}-2\sum_{i\neq j,k\neq i,j}\frac{U_i}{||U||}\frac{U_j}{||U||}X_{jk, \eta}V_{ik}-\frac{E(U_1)^2}{E(U_1^2)^2}E(X_{12, \eta })U'VU+O_p(\sqrt{N})\\
&=N^2\sum_{l=1}^L(\tilde{\mu}_l-\mu_{0,l})\left(\frac{E(U_1)^2}{E(U_1^2)}E(X_{12, \eta }X_{23,l})-\frac{E(U_1)^4 }{E(U_1^2)^2} E(X_{12,l})E(X_{12,\eta})\right)\\
&+2N \frac{E(U_1)E(U_1^3)}{E(U_1^2)}  E(X_{12, \eta })\\
&+\sum_{i\neq j}X_{ij, \eta }V_{ij}-2\sum_{i\neq j,k\neq i,j}\frac{U_i}{||U||}\frac{U_j}{||U||}X_{jk, \eta }V_{ik}-\frac{E(U_1)^2}{E(U_1^2)^2}E(X_{12, \eta})U'VU+O_p(\sqrt{N})\\
&=N^2\sum_{l=1}^L(\tilde{\mu}_l-\mu_{0,l})\left(\frac{E(U_1)^2}{E(U_1^2)}E(X_{12,\eta }X_{23,l})-\frac{E(U_1)^4 }{E(U_1^2)^2} E(X_{12,l})E(X_{12,\eta})\right)+R_{N,\eta}+O_p(\sqrt{N})
\end{align*}
where the residual $R_{N,\eta}$ is of order $O_p(N)$ and is given by:
\begin{align*}
    R_{N, \eta}:=&2 N \frac{E(U_1)E(U_1^3)}{E(U_1^2)} E(X_{12, \eta})\\
&+\sum_{i\neq j}X_{ij, \eta }V_{ij}-2\frac{1}{N E(U_1^2)} \sum_{i\neq j,k\neq i,j} U_i U_j X_{jk, \eta}V_{ik}-\frac{E(U_1)^2}{E(U_1^2)^2}E(X_{12, \eta})U'VU+O_p(\sqrt{N})\\
&=2N \frac{E(U_1)E(U_1^3)}{E(U_1^2)} E(X_{12, \eta})\\
&+\sum_{ij}V_{ij}\left( X_{ij, \eta }-2U_i \frac{1}{N E(U_1^2)} \sum_{k\neq i,j}  U_k X_{jk, \eta}-\frac{E(U_1)^2}{E(U_1^2)^2}E(X_{12, \eta}) U_i U_j\right)+O_p(\sqrt{N})
\end{align*}

we get: 
\begin{align*}
    Var(R_{N, \eta}|X, U)&=\sigma_V^2 \sum_{i<j} \bigg( X_{ij, \eta }-2U_i \frac{1}{N E(U_1^2)} \sum_{k\neq i,j}  U_k X_{jk, \eta}-\frac{E(U_1)^2}{E(U_1^2)^2}E(X_{12, \eta}) U_i U_j \\
    &\;\;+ X_{ij, \eta }-2U_j \frac{1}{N E(U_1^2)} \sum_{k\neq i,j}  U_k X_{ik, \eta}-\frac{E(U_1)^2}{E(U_1^2)^2}E(X_{12, \eta}) U_i U_j\bigg)^2 \\
    &=\sigma_V^2\sum_{i<j} \bigg( X_{ij, \eta }-2U_i \frac{1}{N E(U_1^2)} \sum_{k\neq i,j}  U_k X_{jk, \eta}-\frac{E(U_1)^2}{E(U_1^2)^2}E(X_{12, \eta}) U_i U_j \bigg)^2\\
    &+\sigma_V^2\sum_{i<j} \bigg(X_{ij, \eta }-2U_j \frac{1}{N E(U_1^2)} \sum_{k\neq i,j}  U_k X_{ik, \eta}-\frac{E(U_1)^2}{E(U_1^2)^2}E(X_{12, \eta}) U_i U_j\bigg)^2\\
    &+2\sigma_V^2\sum_{i<j} \bigg( X_{ij, \eta }-2U_i \frac{1}{N E(U_1^2)} \sum_{k\neq i,j}  U_k X_{jk, \eta}-\frac{E(U_1)^2}{E(U_1^2)^2}E(X_{12, \eta}) U_i U_j \bigg)\\
    & \; \; \; \; \; \times \bigg(X_{ij, \eta }-2U_j \frac{1}{N E(U_1^2)} \sum_{k\neq i,j}  U_k X_{ik, \eta}-\frac{E(U_1)^2}{E(U_1^2)^2}E(X_{12, \eta}) U_i U_j\bigg)\\
    &=\sigma_V^2\sum_{i\neq j} \bigg( X_{ij, \eta }-2U_i \frac{1}{N E(U_1^2)} \sum_{k\neq i,j}  U_k X_{jk, \eta}-\frac{E(U_1)^2}{E(U_1^2)^2}E(X_{12, \eta}) U_i U_j \bigg)^2\\
    &+2\sigma_V^2\sum_{i<j} \bigg( X_{ij, \eta }-2U_i \frac{1}{N E(U_1^2)} \sum_{k\neq i,j}  U_k X_{jk, \eta}-\frac{E(U_1)^2}{E(U_1^2)^2}E(X_{12, \eta}) U_i U_j \bigg)\\
    & \; \; \; \; \; \times \bigg(X_{ij, \eta }-2U_j \frac{1}{N E(U_1^2)} \sum_{k\neq i,j}  U_k X_{ik, \eta}-\frac{E(U_1)^2}{E(U_1^2)^2}E(X_{12, \eta}) U_i U_j\bigg)\\
    &=\sigma_V^2 \sum_{i\neq j} X_{ij, \eta }^2+4U_i^2 \frac{1}{N^2 E(U_1^2)^2} \left(\sum_{k\neq i,j}  U_k X_{jk, \eta}\right)^2+\frac{E(U_1)^4}{E(U_1^2)^4}E(X_{12, \eta})^2 U_i^2 U_j^2\\
    &-\sigma_V^2\frac{4}{N E(U_1^2)} \sum_{i\neq j,k\neq i,j} 
 X_{ij, \eta } U_i  U_k X_{jk, \eta}-2\frac{E(U_1)^2}{E(U_1^2)^2}E(X_{12, \eta})\sum_{i\neq j} 
 X_{ij, \eta }U_i U_j\\
 &+4 \sigma_V^2\frac{E(U_1)^2}{N E(U_1^2)^3} E(X_{12, \eta}) \sum_{i\neq j,k\neq i,j} U_i^2 U_j U_k  X_{jk, \eta} \\
 &+\sigma_V^2 \sum_{i\neq j}\bigg( X_{ij, \eta}^2-2U_j \frac{1}{N E(U_1^2)} \sum_{k\neq i,j}  U_k X_{ik, \eta}X_{ij, \eta}-\frac{E(U_1)^2}{E(U_1^2)^2}E(X_{12, \eta}) U_i U_j X_{ij,\eta} \bigg)\\
& +\sigma_V^2 \sum_{i\neq j}\bigg(-2U_i \frac{1}{N E(U_1^2)} \sum_{k\neq i,j}  U_k X_{jk, \eta}  X_{ij, \eta }+4U_i U_j \frac{1}{N^2 E(U_1^2)^2} \sum_{k,l\neq i,j}  U_kU_l X_{ik, \eta}X_{jl, \eta}\\
&+2U_i \frac{1}{N E(U_1^2)} \sum_{k\neq i,j}  U_k X_{jk, \eta}  \frac{E(U_1)^2}{E(U_1^2)^2}E(X_{12, \eta}) U_i U_j  \bigg)\\
&+\sigma_V^2 \sum_{i\neq j}\bigg(-\frac{E(U_1)^2}{E(U_1^2)^2}E(X_{12, \eta}) U_i U_j X_{ij, \eta }+ 2\frac{E(U_1)^2}{NE(U_1^2)^3}E(X_{12, \eta})  \sum_{k\neq i,j}  U_i U_j^2 U_k X_{ik, \eta}\\
&+ \frac{E(U_1)^4}{E(U_1^2)^4}E(X_{12, \eta})^2 U_i^2 U_j^2 \bigg)\\
 &=N^2\sigma_V^2 \bigg( E(X_{12, \eta }^2)+4\frac{E(U_1)^2}{ E(U_1^2)} E(X_{12,\eta}X_{23,\eta})+\frac{E(U_1)^4}{E(U_1^2)^2}E(X_{12, \eta})^2 \\
    &-\frac{4E(U_1)^2}{ E(U_1^2)} E(X_{12,\eta}X_{23, \eta})-2\frac{E(U_1)^4}{E(U_1^2)^2}E(X_{12, \eta})^2 \\
 &+4 \frac{E(U_1)^4}{E(U_1^2)^2} E(X_{12, \eta})^2 \\
 &+E( X_{12, \eta}^2)-2 \frac{E(U_1)^2}{ E(U_1^2)} E(X_{12, \eta}X_{23, \eta})-\frac{E(U_1)^4}{E(U_1^2)^2}E(X_{12, \eta})^2\\
& -2 \frac{E(U_1)^2}{E(U_1^2)} E(X_{12, \eta}X_{23, \eta})+4 \frac{E(U_1)^4}{ E(U_1^2)^2} E(X_{12, \eta})^2\\
&+2\frac{E(U_1)^4}{E(U_1^2)^2}  E(X_{12, \eta}) ^2 \\
&-\frac{E(U_1)^4}{E(U_1^2)^2}E(X_{12, \eta})^2+ 2\frac{E(U_1)^4}{E(U_1^2)^2}E(X_{12, \eta})^2\\
&+ \frac{E(U_1)^4}{E(U_1^2)^2}E(X_{12, \eta})^2 +o(1)\bigg) \; \mbox{ almost surely.}\\
 &= N^2 \sigma_V^2 \bigg( 2 E(X_{12, \eta }^2)+10\frac{E(U_1)^4}{ E(U_1^2)^2} E(X_{12,\eta})^2-4\frac{E(U_1)^2}{ E(U_1^2)} E(X_{12,\eta}X_{23,\eta})+o(1)\bigg) \; \mbox{ almost surely.}\\
\end{align*}

clearly, the Lyapunov condition is met and by
the Lyapunov CLT 
\begin{align*}
    \frac{1}{N}R_{N,\eta}\rightarrow_d \mathcal{N}\left( 0,\sigma_V^2 \eta \Sigma \eta' \right)
\end{align*}
for 
\begin{align*}
    \Sigma &:= \bigg( 2 E(X_{12 }X_{12 }')+10\frac{E(U_1)^4}{ E(U_1^2)^2} E(X_{12,\eta})E(X_{12,\eta})'-4\frac{E(U_1)^2}{ E(U_1^2)} E(X_{12}X_{23}')\bigg) 
\end{align*}

Finally:

\begin{align*}
    N(\hat{\mu}-\mu_0)&= \left(E(X_{12}'X_{12})- \frac{E(U_1)^2}{E(U_1^2)}E(X_{12}'X_{32})\right)^{-1} \\
    &\times \frac{1}{N}\left(\sum_{i\neq j}X_{ij}'(U_iU_j+V_{ij})-\sum_{i\neq j,k\neq i,j}\nu_i(\Tilde{\mu})\nu_j(\Tilde{\mu})X_{jk}' (U_iU_k+V_{ik}) \right)\\
    &= \left(E(X_{12}'X_{12})- \frac{E(U_1)^2}{E(U_1^2)}E(X_{12}'X_{32})\right)^{-1} \\
    &=\frac{E(U_1)^2}{E(U_1^2)} \left(  E(X_{12}X_{12}')-\frac{E(U_1)^2}{E(U_1^2)}E(X_{12}X_{23}')\right)^{-1}\\
    &\times  \left(E(X_{12}X_{23}')-\frac{E(U_1)^2 }{E(U_1^2)} E(X_{12})E(X_{12}')\right) N (\Tilde{\mu}-\mu_{0})\\
    &+ R_N + O_p\left(\frac{1}{\sqrt{N}}\right)\\
    &=K_N N (\Tilde{\mu}-\mu_{0}) +R_N + O_p\left(\frac{1}{\sqrt{N}}\right)
\end{align*}

with
\begin{align*}
    K_N&:=\frac{E(U_1)^2}{E(U_1^2)} \left(  E(X_{12}X_{12}')-\frac{E(U_1)^2}{E(U_1^2)}E(X_{12}X_{23}')\right)^{-1}  \left(E(X_{12}X_{23}')-\frac{E(U_1)^2 }{E(U_1^2)} E(X_{12})E(X_{12}')\right)\\
    R_N& \rightarrow_d 2 \frac{E(U_1)}{E(U_1^2)}\left( E(U_1^3)+\frac{\sigma_V^2E(U_1)}{E(U_1^2)}\right)\left(E(X_{12}X_{12}')- \frac{E(U_1)^2}{E(U_1^2)}E(X_{12}X_{32}')\right)^{-1} E(X_{12})\\
&\; \; \; + \left(E(X_{12}X_{12}')- \frac{E(U_1)^2}{E(U_1^2)}E(X_{12}X_{32}')\right)^{-1}  \mathcal{N}\left( 0,\sigma_V^2 \Sigma \right)
\end{align*}

\end{proof}

\subsection{Proof of lemma \ref{BiasCorrection}}\label{BiasCorrectionProof}
\begin{proof}
    
    When $\delta=1$, by proposition \ref{Prop4}, with probability approaching 1  $\hat{\delta}=\frac{\lambda_1(\tilde{\mu})}{|\lambda_1(\tilde{\mu})|}$ and  $\frac{\lambda_1(\tilde{\mu})}{|\lambda_1(\tilde{\mu})|}=1$.
    
    When $\delta=-1$, 
    \begin{align*}
    \hat{\delta}:&= \frac{-\lambda_N(-M(\Tilde{\mu}))\mathbbm{1}\{-\lambda_N(-M(\Tilde{\mu}))>\lambda_1(-M(\Tilde{\mu})))\} -\lambda_1(-M(\Tilde{\mu}))\mathbbm{1}\{-\lambda_N(-M(\Tilde{\mu}))<\lambda_1(-M(\Tilde{\mu}))\}}{\max_i |\lambda_i(\tilde{\mu})|} 
    \end{align*}
    so with probability approaching 1:
    \begin{align*}
    \hat{\delta}:&=-\frac{\lambda_N(-M(\Tilde{\mu}))\mathbbm{1}\{|\lambda_N(-M(\Tilde{\mu}))|>\lambda_1(-M(\Tilde{\mu})))\} +\lambda_1(-M(\Tilde{\mu}))\mathbbm{1}\{|\lambda_N(-M(\Tilde{\mu}))|<\lambda_1(-M(\Tilde{\mu}))\}}{\max_i |\lambda_i(\tilde{\mu})|}
    \end{align*}
    by the same reasonning as for the case $\delta=1$, $\frac{\lambda_N(-M(\Tilde{\mu}))\mathbbm{1}\{|\lambda_N(-M(\Tilde{\mu}))|>\lambda_1(-M(\Tilde{\mu})))\} +\lambda_1(-M(\Tilde{\mu}))\mathbbm{1}\{|\lambda_N(-M(\Tilde{\mu}))|<\lambda_1(-M(\Tilde{\mu}))\}}{\max_i |\lambda_i(\tilde{\mu})|}=1$ with probability approaching 1. so $\hat{\delta}=-1$  with probability approaching 1.\\    
    
     I begin by proving the third approximation. Note that when $\delta=1$:
    \begin{align*}
        ||U-\sqrt{\lambda_1(\Tilde{\mu})} \nu(\tilde{\mu})||_2&\leq ||U-v(\Tilde{\mu})||_2+||v(\Tilde{\mu})-\sqrt{\lambda_1(\Tilde{\mu})} \nu_i(\tilde{\mu})||_2\\
        &=||U-v(\Tilde{\mu})||_2 +|\sqrt{\lambda_1(\tilde{\mu})}-|| v(\Tilde{\mu})||_2 |\\
        &=||U-v(\Tilde{\mu})||_2 + \frac{\lambda_1(\tilde{\mu})-|| v(\Tilde{\mu})||_2^2}{\sqrt{\lambda_1(\tilde{\mu})}+|| v(\Tilde{\mu})||_2}\\
        &=||U-v(\Tilde{\mu})||_2 + \frac{\lambda_1(\tilde{\mu})-||U||_2^2+||U||_2^2-|| v(\Tilde{\mu})||_2^2}{\sqrt{\lambda_1(\tilde{\mu})}+|| v(\Tilde{\mu})||_2}
    \end{align*}

    By proposition \ref{Prop4}, $||U-v(\Tilde{\mu})||_2=O_p(1)$, $\lambda_1(\tilde{\mu})-||U||_2^2=O_p(\sqrt{N})$ and $||U||_2^2-|| v(\Tilde{\mu})||_2^2=(||U||_2-|| v(\Tilde{\mu})||_2)(||U||_2+|| v(\Tilde{\mu})||_2)=O_p(\sqrt{N})$, therefore $||U-\sqrt{\lambda_1(\Tilde{\mu})} \nu_(\tilde{\mu})||_2=O_p(1)$. Likewise, when $\delta=-1$, the same reasonning applies to the matrix $-M(\Tilde{\mu})$ and we get that $||U+\sqrt{|\lambda_N(\Tilde{\mu})|} \nu_(\tilde{\mu})||_2=O_p(1)$

Combining both cases, we establish that $||U-\delta \sqrt{\max_i |\lambda_i(\Tilde{\mu})|} \nu(\tilde{\mu})||_2=O_p(1)$. Hence: 
\begin{align*}
||U-\hat{U}||&\leq ||U-\delta \sqrt{\max_i |\lambda_i(\Tilde{\mu})|} \nu(\tilde{\mu})||_2+||\delta \sqrt{\max_i |\lambda_i(\Tilde{\mu})|} \nu(\tilde{\mu})- \hat{\delta} \sqrt{\max_i |\lambda_i(\Tilde{\mu})|} \nu(\tilde{\mu})||_2\\
&= ||U-\delta \sqrt{\max_i |\lambda_i(\Tilde{\mu})|} \nu(\tilde{\mu})||_2+|\delta -\hat{\delta}|\times \sqrt{\max_i |\lambda_i(\Tilde{\mu})|} \nu(\tilde{\mu})||_2\\
&=O_p(1)
\end{align*}
    By the Cauchy-Schwartz inequality:
    \begin{align*}
        \bigg| \sum_i U_i-\hat{U}_i \bigg| &\leq \sum_i |U_i-\hat{U}_i|\\
        &\leq \sqrt{N}   ||U-\hat{U}||_2
    \end{align*}
    which prove the first point.

    Finally, note:
    $$ \sum_i ( U_i-\hat{U}_i)^3= \sum_i U_i^3-\hat{U}_i^3 - 3\sum_i U_i \hat{U}_i (U_i - \hat{U}_i)  $$
    then
    \begin{align*}
        \bigg|  \sum_i U_i^3-\hat{U}_i^3 \bigg| &\leq \sum_i | U_i-\hat{U}_i|^3 +3 \sum_i |U_i| (U_i-\hat{U}_i)^2+3 \sum_i U_i^2 |U_i-\hat{U}_i|\\
        &\leq \sum_i | U_i-\hat{U}_i|^3 +3 \left( \sum_i U_i^2\right)^{1/2}\left( \sum_i  (U_i-\hat{U}_i)^4\right)^{1/2}+3\left( \sum_i U_i^4\right)^{1/2}\left( \sum_i  (U_i-\hat{U}_i)^2\right)^{1/2}\\
        &\leq ||U-\hat{U}||_3^3+3 \left( \sum_i U_i^2\right)^{1/2} ||U-\hat{U}||_4^2+3\left( \sum_i U_i^4\right)^{1/2} ||U-\hat{U}||_2\\
        &\leq ||U-\hat{U}||_2^{3}+3 \left( \sum_i U_i^2\right)^{1/2} ||U-\hat{U}||_2^2+3\left( \sum_i U_i^4\right)^{1/2} ||U-\hat{U}||_2 \end{align*}
Proposition \ref{Prop4} allows to conclude that $\bigg|  \sum_i U_i^3-\hat{U}_i^3 \bigg|=O_p(\sqrt{N})$
\end{proof}
\end{document}